# COMPLEMENTARITY AND ENTANGLEMENT IN QUANTUM INFORMATION THEORY

## BY

## TRACEY EDWARD TESSIER

B.S., Computer Science, University of Massachusetts, Amherst, 1993
M.S., Physics, Creighton University, 1997

DISSERTATION

Submitted in Partial Fulfillment of the
Requirements for the Degree of

**Doctor of Philosophy**
**Physics**

The University of New Mexico
Albuquerque, New Mexico

**December, 2004**

# Dedication

*To my family.*

# Acknowledgments

First and foremost I would like to thank my advisor, Ivan Deutsch, for his guidance and for giving me the freedom to investigate topics that truly inspired me. Thanks Ivan! Thanks also to Carl Caves for helpful discussions and guidance. Many thanks to all the members of the information physics group including Paul Alsing, René Stock, Andrew Silberfarb, Bryan Eastin, Kiran Manne, Clark Highstrete, Steve Flammia, Iris Rappert, Colin Trail, Seth Merkel, Nick Menicucci, Animesh Datta, and Aaron Denney. Special thanks also to former members Joe Renes, Sonja Daffer, Shohini Ghose, Gavin Brennen, John Grondalski, Andrew Scott, Mark Tracy, Pranaw Rungta, and especially Aldo Delgado and Ivette Fuentes-Guridi for stimulating discussions.

I am also grateful to those at other institutions, especially Robert Raussendorf, Dave Bacon, Michael Nielsen, Tobias Osborne, and Chris Fuchs, each of whom gave me helpful advice, and to the remaining members of my dissertation committee, Sudhakar Prasad and Christoper Moore, for their scrutiny of this work and for the valuable time it takes.

Finally, I would like to thank N. David Mermin and William K. Wootters whose work I find so inspiring, my family for their love and support, and Lori for making every day special.

# COMPLEMENTARITY AND ENTANGLEMENT IN QUANTUM INFORMATION THEORY

BY

TRACEY EDWARD TESSIER

# Complementarity and Entanglement in Quantum Information Theory

by

**Tracey Edward Tessier**

B.S., Computer Science, University of Massachusetts, 1993

M.S., Physics, Creighton University, 1997

Doctor of Philosophy, Physics, University of New Mexico, 2004

## Abstract

This research investigates two inherently quantum mechanical phenomena, namely complementarity and entanglement, from an information-theoretic perspective. Beyond philosophical implications, a thorough grasp of these concepts is crucial for advancing our understanding of foundational issues in quantum mechanics, as well as in studying how the use of quantum systems might enhance the performance of certain information processing tasks. The primary goal of this thesis is to shed light on the natures and interrelationships of these phenomena by approaching them from the point of view afforded by information theory. We attempt to better understand these pillars of quantum mechanics by studying the various ways in which they govern the manipulation of information, while at the same time gaining valuable insight into the roles they play in specific applications.

The restrictions that nature places on the distribution of correlations in a multipartite quantum system play fundamental roles in the evolution of such systems and yield vital insights into the design of protocols for the quantum control of ensembles with potential applications in the field of quantum computing. By augmenting the existing formalism for quantifying entangled correlations, we show how this entanglement sharing behavior may be studied in increasingly complex systems of both theoretical and experimental significance. Further, our results shed light on the dynamical generation and evolution of multipartite entanglement by demonstrating that individual members of an ensemble of identical systems coupled to a common probe can become entangled with one another, even when they do not interact directly.

The phenomenon of entanglement sharing, as well as other unique features of entanglement, e.g. the fact that maximal information about a multipartite quantum system does not necessarily entail maximal information about its component subsystems, may be understood as specific consequences of the phenomenon of complementarity extended to composite quantum systems. The multi-qubit relations which we derive imply that quantum mechanical systems possess the unique ability to encode information directly in entangled correlations, without the need for the correlated subsystems to possess physically meaningful values.

We present a local hidden-variable model supplemented by an efficient amount of classical communication that reproduces the quantum-mechanical predictions for the entire class of Gottesman-Knill circuits. The success of our simulation provides strong evidence that the power of quantum computation arises not directly from entanglement, but rather from the nonexistence of an efficient, local realistic description of the computation, even when augmented by an efficient amount of nonlocal, but classical communication. This conclusion is fully consistent with our generalized complementarity relations and implies that the unique ability of quan-tum systems to support directly encoded correlations is a necessary ingredient for performing truly quantum computation. Our results constitute further progress towards the information-theoretic goal of identifying the minimal classical resources required to simulate the correlations arising in an arbitrary quantum circuit in order to determine the roles played by complementarity and entanglement in achieving an exponential quantum advantage in computational efficiency.

The findings presented in this thesis support the conjecture that Hilbert space dimension is an objective property of a quantum system since it constrains the number of valid conceptual divisions of the system into subsystems. These arbitrary observer-induced distinctions are integral to the theory since they determine the possible forms which our subjective information may take. From this point of view the phenomenon of complementarity, which limits the in-principle types and amounts of information that may simultaneously exist about different conceptual divisions of the system, may be identified as that part of quantum mechanics where objectivity and subjectivity intersect.

# Contents

# List of Figures

# List of Tables

# Chapter 1

# Introduction

This research investigates two inherently quantum mechanical phenomena, namely complementarity and entanglement, from an information-theoretic perspective. Beyond philosophical implications, a thorough grasp of these concepts is crucial for advancing our understanding of foundational issues in quantum mechanics [1], as well as in studying how the use of quantum systems might enhance the performance of certain information processing tasks [2]. The primary goal of this thesis is to shed light on the natures and interrelationships of these phenomena by approaching them from the point of view afforded by information theory. That is, we attempt to better understand these pillars of quantum mechanics by studying various ways in which they govern the manipulation of information while at the same time gaining valuable insight into the roles they play in specific applications.

Debates about how to properly interpret quantum mechanics have raged ever since the inception of the theory [1], and continue to this day [3]. For the most part, the plethora of seemingly distinct interpretations of quantum mechanics are all variants on a theme; an attempt to come to grips with the phenomena of complementarity and entanglement, so far removed from our everyday experience. We thus begin by reviewing these concepts, focusing on the fundamental differences between them and the classical ideas that have been so successful in explaining most macroscopic phenomena. After that, we briefly survey several fundamental results in the field of quantum information; these illustrate some of the counterintuitive implications of entanglement in the context of the manipulation and processing of information encoded in physical systems. The last section gives an overview of the specific issues explored in this thesis.

## 1.1 Complementarity

Complementarity is perhaps the most important phenomenon distinguishing systems that are inherently quantum mechanical from those that may accurately be treated classically. Niels Bohr introduced this term, as part of what is now known as the Copenhagen interpretation of quantum mechanics, to refer to the fact that information about a quantum object obtained under different experimental arrangements cannot always be comprehended within a single causal picture [4]. The results of experiments designed to probe different aspects of a quantum system are complementary to one another in the sense that only the totality of the potentially observable attributes exhausts the *possible* information that may be obtained about the system.

An alternative statement of complementarity, which makes no reference to experimental arrangements or measurements, states that a quantum system may possess properties that are equally real, but mutually exclusive [4]. This is an enormous departure from the observed behavior of everyday objects like automobiles and billiard balls where the implicit assumption, correct to a high degree of accuracy, is that physical systems possess properties, and that all of these properties may be simultaneously ascertained to an arbitrary degree of accuracy via an appropriate measurement procedure. Indeed, we identify the classical world with precisely those

*1.1. Complementarity*

systems and processes for which it is possible to unambiguously combine the space-time coordinates of objects with the dynamical conservation laws that govern their mutual interactions. However, in the more general setting of quantum mechanics, complementarity precludes the existence of such a picture. It was this insight that led Bohr to consider complementarity to be the natural generalization of the classical concept of causality [4].

Viewed in this context, the famed Heisenberg uncertainty relation [5] between two Hermitian operators $A$ and $B$

$$\left\langle (\Delta A)^2 \right\rangle \left\langle (\Delta B)^2 \right\rangle \geq \frac{1}{4} \left| \langle [A, B] \rangle \right|^2 , \qquad (1.1)$$

where $\Delta A \equiv A - \langle A \rangle$, $\left\langle (\Delta A)^2 \right\rangle = \langle A^2 \rangle - \langle A \rangle^2$, and $\langle A \rangle$ is the quantum expectation value of the observable $A$, is seen to be one specific consequence of complementarity. The existence of this relation is essential for ensuring the consistency of quantum theory by defining the limits within which the use of classical concepts belonging to two complementary pictures, e.g., the wave-particle duality exhibited by a photon in a double-slit experiment [6], the tradeoff between the uncertainties in the position and momentum of a subatomic particle [7], etc., may be applied without contradiction.

Bohr also described three related implications of complementarity that have no logical counterparts in the classical world. The first, known as indivisibility, expresses the idea that the 'interior' of a quantum phenomenon is physically unknowable. This form of 'quantum censorship' is, according to Bohr, inextricably linked with an aspect of the measurement process known as closure. The occurrence of a definite physical event (or classically knowable result) brought on by an "irreversible act of amplification" yielding a classical outcome, 'closes' a quantum phenomenon with a certain probability distribution for the different possible outcomes [4]. Thus, until a measurement yields a definite outcome corresponding to the value of some physical property, it is inconsistent to associate that property, or indeed any property for which there is no physical evidence, to the measured system. Failure to respect this proviso leads to seemingly paradoxical results [8].

Finally, Bohr pointed out that complementarity implies the "impossibility of any sharp separation between the behavior of atomic objects and their interaction with the measuring instruments which serve to define the conditions under which the phenomena appear" [6]. The import of this statement is often taken to be that quantum mechanics does not provide a mechanism via which to understand the observed existence of the macroscopic world, since in the end any system, no matter how large or complex, is governed by the laws of quantum mechanics. Indeed, great bodies of research have been performed on the so-called quantum to classical transition (see, e.g., [9]), as well as on the related measurement problem [1]. We will not concern ourselves with these questions here. Rather, we note that Bohr's observation also implies an unavoidable necessity for the development of correlations in any attempt to determine the 'properties' of a quantum object. Of course, the possible types of correlation associated with a quantum system are not limited to correlations with a macroscopic measuring apparatus. Correlations between atomic systems and the environment lead to the whole field of decoherence [10, 9], and quantum correlations among multiple atomic systems provide interesting examples of entanglement.

## 1.2 Correlations

A *correlation* is a relation between two or more variables. Generally speaking, the ultimate goal of all scientific inquiry is discovering correlations, i.e. uncovering the relations that exist between distinct physical properties. The philosophy of science teaches us that there is no other way of representing 'meaning' except in terms of these relations between different quantities or qualities, while information theory [11] teaches us that these relations contain information that pertains to the correlated

*1.2. Correlations*

entities.

Consider for example, two random variables representing the weights and heights, respectively, of men over thirty. If we restrict our attention to men over six feet tall then we find that, on average, these men weigh more than the average adult male. This is an example of a correlation between two properties, the weight and height, of a single physical system, in this case an adult male.

Correlations can also arise between distinct physical systems. For example, suppose that two parties, whom we refer to as Alice and Bob, each have a fair coin in their possession. If these two parties toss their coins and compare their results, they will find that their outcomes are either the same, i.e., both heads or both tails, referred to as perfectly correlated, or different (perfectly anti-correlated). In either case, they are now in a position to communicate information about the correlations that exist in their joint system to a third party (Charlie) without having to disclose any information about the outcomes of either coin toss. As a result, given the knowledge of whether the two coins are correlated or anti-correlated, and subsequently being told the outcome of e.g. Alice's coin toss, Charlie can correctly infer the outcome of Bob's coin toss.

This simple example highlights an important feature of correlations arising in classical systems; classical correlations are secondary quantities in the sense that there always exist properties possessed by individual subsystems from which these correlations may in principle be inferred. The fundamental quantities granted 'physical reality' in the above example are the results of each individual coin toss. One of the aims of this thesis is to demonstrate the primacy of information stored in entangled correlations which cannot be inferred, even in principle, from information about the correlated entities since these distinct types of information share a complementary existence with one another.

One of the main conceptual departures of quantum mechanics from the everyday 'classical' description of reality results from the fact, codified by John Bell [12] and verified experimentally [13, 14], that entangled quantum systems exhibit stronger correlations than are achievable with any local hidden variable model. Here, locality is taken to mean that the result of a measurement performed on one system is unaffected by any operations performed on a space-like separated system with which it has interacted in the past. The goal of any LHV model is to account for the statistical predictions of quantum mechanics in terms of averages over more well-defined states, the complete knowledge of which would yield deterministic predictions, in the same way that the values of thermodynamic variables are defined by averaging over the various possible microstates in a classical statistical ensemble [15, 16]. The specific values of the local variables in such models are assumed to be 'hidden' since, if it were possible to ascertain these values, then the status of quantum mechanics would be trivially reduced to that of an incomplete theory.

If we assume that locality is respected by quantum systems, then the violation of Bell-type inequalities demonstrates the in-principle failure of LHV models to account for all of the predictions of quantum mechanics. This implies that nature does not respect the constraints either of locality or of realism, where realism in this context means that a physical system possesses definite values for properties that exist independent of observation. Bohmian mechanics [17, 18], for example, is a highly nonlocal theory which purports, at all times, to yield a precise, rational, and objective description of individual systems at a quantum level of accuracy. The price one pays in adopting this point of view is the acceptance of superluminal action-at-a-distance in physical processes [19], the existence of which flies in the face of the relativistic lesson that no signal can propagate at a speed faster than that of light [20].

An alternative approach to trying to understand the implications of Bell's result is

*1.2. Correlations*

to (i) accept quantum mechanics as it is or, perhaps more correctly, as it purports to be, i.e., as a complete theory that contains an unavoidable element of randomness at a fundamental level, and (ii) assume that locality is respected by quantum mechanics, and see where these two assumptions lead. According to the Ithaca interpretation of quantum mechanics [21], the conclusion is this: correlations have physical reality; that which they correlate does not. More generally, we show that the presence of entanglement in a composite quantum system precludes, to the degree that it exists, the simultaneous existence of information about the individual subsystems to which these correlations refer. This, in turn, suggests that inherently bipartite (or in general multipartite) entangled correlations share a complementary relationship with the existence of information normally associated with individual systems, as well as with one another. As a result, many of the bizarre implications of entanglement can be understood as specific consequences of complementarity in composite quantum systems.

Finally we mention the Bayesian interpretation of quantum mechanics, which considers the quantum state to be a representation of our subjective knowledge about a quantum system [16], rather than a description of its physical properties. One advantage of this interpretation is that the collapse of the wave function [22] is viewed not as a real physical process, but simply represents a change in our state of knowledge. This is an important point of view for our purposes since we are inquiring about the implications of quantum mechanics for information theory. However, it is unclear exactly what the knowledge encoded by the quantum state pertains to since, from this perspective, we are generally prohibited from associating objective properties with individual systems. Our results shed some light on this question and suggest that a constructive approach might be to merge the Bayesian, Ithaca, and Copenhagen interpretations into a single interpretation that treats the information encoded in both individual subsystems and in quantum correlations as fundamental elements of quantum theory, while at the same time recognizing that the in-principle existence of each of these distinct types of information is constrained by the phenomenon of complementarity.

## 1.3 Quantum Information Theory

Quantum information is the study of information processing tasks that can be accomplished using physical systems that must be described according to the laws of quantum mechanics. The goal of this section is not to give a comprehensive overview of this vast subject, but to introduce some of the additional resources that become available when information is encoded in quantum rather than classical systems, and to give simple examples of their usefulness in enhancing the performance of various tasks. The reader is referred to [2] for a thorough treatment of the fields of quantum information and computation.

The quantum bit, or qubit [23], is the fundamental unit of quantum information. A qubit may be physically implemented by any two-state quantum system such as a spin-1/2 particle or two energy levels in an atom. Designating the orthogonal states of a qubit to be $|0\rangle$ and $|1\rangle$, representing the Boolean possibilities of a classical bit, the most general pure state $|\psi\rangle$ of a single qubit is given by a coherent linear superposition of the basis states

$$|\psi\rangle = \alpha |0\rangle + \beta |1\rangle, \tag{1.2}$$

where $\alpha$ and $\beta$ are complex numbers satisfying $|\alpha|^2 + |\beta|^2 = 1$.

As discussed in the previous section, the Bayesian interpretation of quantum mechanics considers a quantum state to be a representation of the information that we possess about a quantum system [24]. From this point of view, we are justified in asking information-theoretic questions about these states. Schumacher's quantum noiseless channel coding theorem [23] is one example of the efficacy of this sort of

*1.3. Quantum Information Theory*

approach. This theorem establishes the qubit as a resource for performing quantum communication by quantifying the number of qubits, transmitted from sender to receiver, that are asymptotically necessary and sufficient to faithfully transmit unknown pure quantum states randomly selected from an arbitrary, but known, source ensemble. Schumacher's result generalizes Shannon's noiseless channel coding theorem [25], which quantifies the minimum number of bits (in an asymptotic sense) required to reliably encode the output of a given classical information source, to the quantum case.

The superposition principle illustrated in Eq. (1.2), coupled with the tensor product structure of Hilbert space [26], implies that two qubits $A$ and $B$ may become correlated with one another such that they cannot be written in the form

$$|\psi_{AB}\rangle = |\psi_A\rangle \otimes |\psi_B\rangle, \tag{1.3}$$

where $|\psi_{A(B)}\rangle$ is a pure state describing the first (second) qubit, respectively. States of the form (1.3) are referred to as *product states*. A pure state of two qubits which cannot be written in product form contains entanglement. For example, the *singlet state*

$$\left|\psi_{AB}^{(s)}\right\rangle \equiv \frac{1}{\sqrt{2}}(|01\rangle - |10\rangle) \tag{1.4}$$

is a maximally entangled state of two qubits. The fact that entangled states cannot be factored into states representing individual subsystems suggests that the presence of entanglement precludes, to some degree, the existence of single particle information. We quantify this intuition in terms of a tradeoff between bipartite and single-qubit properties in Chapter 4.

The relationship between entanglement and complementarity alluded to above is not limited to tension between the existence of single particle properties and bipartite entanglement, but also manifests in the form of entanglement sharing in multipartite, i.e., tripartite or higher, systems. The concept of entanglement sharing [27, 28] refers to the fact that entanglement cannot be freely distributed among subsystems in a multipartite system. Rather, the distribution of entanglement in these systems is subject to certain constraints. As a simple example, consider a tripartite system of three qubits $A$, $B$, and $C$. Suppose that qubits $A$ and $B$ are known to be in a maximally entangled pure state such as the singlet state. In this case, it is obvious that the overall system $ABC$ is constrained such that no entanglement may exist either between $A$ and $C$ or between $B$ and $C$. Otherwise, tracing over subsystem $C$ would necessarily result in a *mixed* marginal density operator for $AB$ in contradiction to the known purity of the state in Eq. (1.4).

The restriction of the correlations that several systems may share with one another is unique to quantum mechanics since a classical random variable may be correlated, to an arbitrary degree, with an arbitrary number of other random variables. Expanding on our earlier examples, one finds that the weight of an adult male is correlated not only with his height, but also with his average daily caloric intake, the heights of his parents, his level of physical activity, etc. Similarly, it is clear that there is nothing to prevent the results of an arbitrarily large number of coin tosses from being, e.g., perfectly correlated with one another. One might therefore expect the study of this purely quantum effect to yield new insights into the nature of entanglement and its usefulness for information processing. Accordingly, we extend the analysis of entanglement sharing to a system of both theoretical and experimental interest in Chapter 3, and demonstrate that this phenomenon is a manifestation of complementarity in tripartite systems in Chapter 4.

Entangled quantum systems also provide a new resource for performing information processing tasks. Quantum superdense coding [29] and teleportation [30] are two examples of processes which make use of entanglement as a resource for communication. The fundamental unit of entanglement, defined as the amount of entanglement in a maximally entangled state of two qubits, e.g. the singlet state, is referred to as

*1.3. Quantum Information Theory*

an *ebit* [31]. Suppose that Alice and Bob each possess one of the two qubits in the state given by Eq. (1.4), i.e., they share one ebit of entanglement. Superdense coding utilizes this shared entanglement to enhance the ability of Alice to communicate classical information to Bob and vice-versa. Specifically, Alice can communicate two bits of information to Bob by (i) performing one of the four operations $\{I, X, Y, Z\}$ corresponding, respectively, to the identity operation and the three Pauli rotations, to the qubit in her possession and (ii) sending her modified qubit to Bob. Since the four different possible two-qubit pure states resulting from the procedure described in (i) are all mutually orthogonal, Bob can perform a single joint measurement in the so-called Bell basis [2] to determine which operation Alice performed on her qubit. Thus, Alice can transmit two bits of classical information to Bob by sending him just a single physical qubit that is one-half of a maximally entangled pair.

Alternatively, Alice may use an ebit that she shares with Bob to transmit quantum information. Suppose that Alice possesses an additional qubit in an unknown quantum state $|\phi\rangle$ that she wishes to communicate to Bob, who is at some remote location unknown to Alice. (This latter condition prevents Alice from simply sending the qubit to Bob directly.) Briefly, the teleportation protocol requires that Alice (i) allow the two qubits in her possession (the qubit to be sent and her half of the singlet state) to interact and become entangled, (ii) measure the qubits in the logical basis thereby obtaining two classical bits of information, and (iii) transmitting these two bits of information to Bob.[1] Depending on the classical message received, Bob performs one of the four operations $\{I, X, Y, Z\}$ on his qubit, after which the qubit is in the desired state $|\phi\rangle$.

The success of the above protocol is in some sense surprising since, even if Alice knew the state of the qubit to be teleported and the location of Bob, it would take an infinite amount of classical communication to describe the state precisely since $|\phi\rangle$ takes on values in a continuous space. Further, this example illuminates certain relationships between the different physical resources involved. Specifically, we see that qubits are more powerful than ebits since the transmission of a single qubit that is also one-half of a maximally entangled pair is sufficient to create one ebit of shared entanglement, but an ebit (or many ebits) is by itself insufficient to teleport an arbitrary state of a qubit. To accomplish this one must also send classical information [31]. The teleportation protocol therefore implies that one qubit is at least equivalent to one ebit of entanglement and two bits of classical communication. Uncovering relationships such as these between the different available resources is one of the main goals of quantum information theory.

---

[1]In the standard protocol, Alice performs a coherent two-qubit measurement in the Bell basis on her entangled pair. Here, we consider an equivalent protocol employing measurements in the logical basis since they are more straightforward to implement physically.

Two other major topics studied in quantum information theory, in addition to communication, are cryptography and computation. Quantum cryptography [32] relies on the indeterminism inherent in quantum phenomena to perform secure communication. This application exploits the fact that quantum theory forbids physical measurements from yielding enough information to enable nonorthogonal quantum states to be reliably distinguished [33]. Accordingly, information encoded and transmitted in nonorthogonal states is secure since any attempt by an eavesdropper to intercept and measure such a signal necessarily results in a detectable disturbance. We will not study cryptography in any detail in this thesis. However, we do conjecture that the complementarity relations presented in Chapter 4 will be useful in extending the discussion of information vs. disturbance tradeoff relations, on which the various cryptographic protocols are based, to composite quantum systems.

The final topic, quantum computation [2], refers to the manipulation and processing of the quantum information stored in qubits, in much the same way that classical computation is concerned with the manipulation and processing of bits of information. The enormous amount of interest in this field stems mainly from the fact that quantum algorithms exist for certain problems which outperform the fastest known classical algorithms. The most well-known example of this is Shor's algorithm [34] capable of factoring numbers in polynomial rather than exponential time. This has obvious applications in the field of cryptography where most encryption schemes are based on the presumed difficulty of factoring large numbers. Our focus, however, will not be on quantum algorithms, but rather on identifying the resources generally required to achieve such an exponential quantum advantage in computational efficiency. Specifically, we investigate the fundamental properties of composite quantum systems that enable a pure state quantum computer to operate outside the constraints imposed by local realism (and obeyed by classical computers) to a degree sufficient for yielding an exponential speedup. Beyond questions of efficiency, our progress in this area also has implications for foundational issues in quantum mechanics.

## 1.4  Overview of Thesis

This research investigates the roles played by complementarity and entanglement in certain information processing tasks. The primary goals of this programme are (i) to augment the formalism that currently exists for quantifying entanglement, (ii) to extend the discussion of entanglement sharing to larger and more complex systems, (iii) to illuminate the role played by entanglement in performing pure state quantum computation, and (iv) to demonstrate that the bizarre implications of entanglement and entanglement sharing may be understood, in a larger context, as specific consequences of the phenomenon of complementarity. Finally, we hope that the insights gained here will shed some light on the problem of interpreting quantum mechanics.

In Chapter 2 we begin by briefly reviewing the formalism that currently exists for quantifying quantum mechanical correlations using entanglement monotones [35].

Several examples of different monotones, motivated by various physical and information theoretic principles, are presented. We then derive a new family of analytic entanglement monotones that provides a global structure illustrating certain relationships between several different measures of entanglement. These functions possess analytic forms that are computable in the most general cases, an important feature since the evaluation of most entanglement monotones entails solving a notoriously difficult minimization problem.

Chapter 3 presents a detailed analytic and numerical study of the phenomenon of entanglement sharing in the Tavis-Cummings model, a system of both theoretical and experimental interest. Our results indicate that individual members of an ensemble of identical systems coupled to a common probe can become entangled with one another, even when they do not interact directly. We investigate how this type of multipartite entanglement is generated in the context of a system consisting of $N$ two-level atoms resonantly coupled to a single mode of the electromagnetic field. In the case $N = 2$, the dynamical evolution is studied in terms of the entanglements in the different bipartite divisions of the system, as quantified by an entanglement monotone known as the I-tangle [36]. We also propose a generalization of the so-called residual tangle [27] that quantifies the inherent three-body correlations in this tripartite system. This enables us to completely characterize the phenomenon of entanglement sharing in the case of the two-atom Tavis-Cummings model. Finally, we gain some insight into the behavior of larger ensembles by employing the results of Section 2.2. Specifically, we find that one member of our family of entanglement monotones constitutes a lower bound on the I-tangle of an arbitrary bipartite system, and can be computed in cases when the I-tangle has no known analytic form.

Chapter 4 presents two novel complementarity relations that govern the bipartite and individual subsystem properties possessed by systems of qubits. The first relation shows that the amount of information that an individual qubit may encode

*1.4. Overview of Thesis*

is constrained solely by the amount of entanglement which that qubit shares with the remaining $N-1$ qubits when the entire system is in an overall pure state. One immediate implication of this result is that the phenomenon of entanglement sharing may be understood as a consequence of complementarity in multipartite systems.

The second expression illustrates the complementary nature of the relationship between entanglement, a quantity which we dub the separable uncertainty, and the single particle properties possessed by an arbitrary state of two qubits, pure or mixed. The separable uncertainty is shown to be a natural measure of ignorance about the individual subsystems, and may be used to completely characterize the relationship between entanglement and mixedness in two-qubit systems. Our results yield a geometric picture in which the root mean square values of local subsystem properties act like coordinates in the space of density matrices, and suggest possible insights into the problem of interpreting quantum mechanics.

Chapter 5 investigates the nature of certain types of entanglement and the role that such correlations play in performing pure state quantum computation. Specifically, we present a local hidden variable model supplemented by classical communication that reproduces the quantum-mechanical predictions for measurements of all products of Pauli operators on two classes of globally entangled states: the $N$-qubit GHZ states [37] (also known as "cat states"), and the one- or two-dimensional cluster states [38] of $N$ qubits. In each case the simulation is efficient since the required amount of communication scales linearly with the number of qubits.

The results for the $N$-qubit GHZ states are somewhat surprising when one considers that Bell-type inequalities exist for these states for which the amount of violation grows exponentially with $N$. However, the results for the cluster states are even more enlightening. The structure of our model yields insight into the Gottesman-Knill theorem [39], a result which goes a long way toward clarifying the role that global entanglement plays in pure state quantum computation. Specifically, we show that the correlations in the set of nonlocal hidden variables represented by the stabilizer generators [2, 39] that are tracked in the Gottesman-Knill theorem are captured by an appropriate set of local hidden variables augmented by $N-2$ bits of classical communication. This fact has profound consequences for our understanding of the necessary ingredients for achieving an exponential quantum advantage in computational efficiency. These implications are fully discussed towards the end of the chapter.

Finally, we summarize our research and draw certain conclusions in Chapter 6. Throughout this work we point out possible directions for further investigation where appropriate. Much of the research presented in this dissertation has been published or submitted for publication. Table 1.1 lists the chapters and the corresponding articles in which this material appears.

*1.4. Overview of Thesis*

| Chapter 2 | A. P. Delgado and T. E. Tessier, "Family of analytic entanglement monotones." e-print quant-ph/0210153, 2002. Submitted to Physical Review A (Rapid Communications). |
|---|---|
| Chapter 3 | T. E. Tessier, I. H. Deutsch, A. P. Delgado, and I. Fuentes-Guridi, "Entanglement sharing in the two-atom Tavis-Cummings model," *Phys. Rev. A*, Vol. 68, pp. 062316/1-10, 2003. <br><br> T. E. Tessier, I. H. Deutsch, and A. P. Delgado, "Entanglement sharing in the Tavis-Cummings model," in *Proceedings to SPIE*, Vol. 5105, (Orlando, FL), Aerosense 2003, 2003. |
| Chapter 4 | T. E. Tessier, "Complementarity relations for multi-qubit systems," *Found. Phys. Lett.*, Vol. 18(2), pp. 107-121, 2005. |
| Chapter 5 | T. E. Tessier, I. H. Deutsch, and C. M. Caves, "Efficient classical-communication-assisted local simulation of N-qubit GHZ correlations." e-print quant-ph/0407133, 2004. Submitted to Physical Review Letters. <br><br> T. E. Tessier, C. M. Caves, and I. H. Deutsch, "Efficient classical-communication-assisted local simulation of the Gottesman-Knill circuits." In preparation. |

Table 1.1: List of chapters in this dissertation and the corresponding published, submitted, or in progress papers.

# Chapter 2

# Measures of Entanglement

## 2.1 Entanglement Montones

As illustrated by the examples presented in Section 1.3, a great deal of quantum information theory is concerned with answering the following question: In what way, if any, does the potential use of entanglement enhance the performance of a given classical information processing task? In this context, where quantum mechanical correlations are viewed as a resource, it is important to have a consistent way of quantifying entanglement.

The sole requirement for a function of a multipartite quantum state to be a good measure of entanglement is that it be non-increasing, on average, under the set of local quantum operations and classical communication (LOCCs) [35]. The most general local quantum operation on an arbitrary quantum state (represented by a density operator $\rho$) is described by a set $\{\mathcal{K}_i\}$ of completely positive linear maps [2] satisfying

$$\rho'_i = \frac{\mathcal{K}_i(\rho)}{p_i}. \tag{2.1}$$

Here, $p_i \equiv \text{Tr}\left[\mathcal{K}_i\left(\rho\right)\right]$, $0 \leq p_i \leq 1$, is the probability that the system is left in the state $\rho'_i$ after the operation. Mathematically, a function $E\left(\rho\right)$ is a so-called *entanglement monotone* if and only if it satisfies the conditions [35]:

$$E\left(\rho\right) \geq \sum_i p_i E\left(\rho'_i\right) \tag{2.2}$$

for all local operations $\{\mathcal{K}_i\}$ and

$$\sum_k p_k E\left(\rho_k\right) \geq E\left(\rho\right) \tag{2.3}$$

for all ensemble decompositions $\rho = \sum_k p_k \rho_k$.

Consider, for example, two spatially separated observers,[1] Alice and Bob, each in possession of one member of a pair of qubits that have interacted in the past and so may share some entanglement. Due to the inherent nonlocality of quantum correlations one intuitively expects that, on average, these two should not be able to increase the entanglement between the qubits if they are only allowed to perform local operations and to communicate with one another over an ordinary channel. Of course, allowing Alice and Bob to communicate enables Alice to condition her local interventions on the outcomes obtained by Bob and vice-versa, which implies that it is possible for Alice and Bob to increase the classical correlations between their respective qubits. Thus, entanglement monotones are specifically designed to detect and quantify only the quantum mechanical correlations in a composite system. In this context, Eq. (2.2) ensures monotonicity, on average, for any individual local operation, and hence for a general LOCC protocol. The second condition, Eq. (2.3), states that $E\left(\rho\right)$ is a convex function which ensures that monotonicity is also preserved under mixing, i.e., when some of the information about the results of local operations is forgotten or is not communicated to the other party.

In general, any multipartite quantum state with $l$ subsystems, described by den-

---

[1]This bipartite example can easily be generalized to multipartite quantum systems.

2.1. *Entanglement Montones*

sity operators $\rho^{(l)}$, that can be written in the form

$$\rho_{\text{sep}} = \sum_i \omega_i \rho_i^{(1)} \otimes \rho_i^{(2)} \otimes \cdots \otimes \rho_i^{(l)}; \qquad \omega_i \geq 0, \qquad \sum_i \omega_i = 1 \qquad (2.4)$$

contains no entanglement and is referred to as a *separable* state; otherwise, the state is entangled. Indeed, any state of the form (2.4) can be constructed according to some LOCC protocol which implies that an entanglement monotone must assign the same value (which can always be taken to be zero) to all separable states. This leads to the additional positivity requirement

$$E(\rho) \geq 0; \qquad E(\rho_{\text{sep}}) = 0, \qquad \forall \rho_{\text{sep}}. \qquad (2.5)$$

Finally, we note that an entanglement monotone must remain invariant under the action of all reversible LOCC protocols, one specific subclass of which is the set of local unitary transformations. This observation yields the intuitive result that the entanglement in a system is independent of the choice of local bases used to describe the subsystems.

The remainder of this section is devoted to introducing some of the existing measures of entanglement that we will have occasion to use and the relationships between them. We will consider both pure and mixed state quantities, but will limit our discussion to bipartite measures. Considerations specifically related to multipartite entanglement will be held off until Chapter 3 and the discussion of entanglement sharing. The reader is referred to [40] for a comprehensive review of the most commonly used measures of entanglement.

We begin by defining the entropy of entanglement, the fiducial measure of entanglement for bipartite pure states. The relationship of this quantity to asymptotic conversion rates between different pure states leads naturally to a measure of mixed state entanglement known as the entanglement of formation. In general, calculating the entanglement of formation involves performing a difficult minimization procedure. Accordingly, we also discuss two related quantities, the concurrence and the tangle, that have known analytic solutions in certain cases. Finally, we mention the negativity, an entanglement monotone that, while not directly related to the entanglement of formation, can be evaluated in the most general situations.

### 2.1.1 Entropy of entanglement

Consider a bipartite system $AB$ with Hilbert space dimension $D_A \times D_B$ in an overall pure state $|\psi\rangle$. The quantum state of one of the subsystems is obtained by performing a partial trace over the other subsystem such that $\rho_{A(B)} \equiv \text{Tr}_{B(A)}(|\psi\rangle\langle\psi|)$. If $|\psi\rangle$ is an entangled state, i.e., if it cannot be written in the form (1.3), then the marginal density operators will be mixed signaling the presence of entanglement. The entropy of entanglement $E_S$ [31] makes use of this fact. It is defined as the *von Neumann entropy* [22],

$$S(\rho) = -\text{Tr}\rho\log_2\rho, \tag{2.6}$$

of the marginal density operator associated with either subsystem $A$ or subsystem $B$,

$$E_S(\psi) = S(\rho_A) = S(\rho_B). \tag{2.7}$$

This quantity enjoys a privileged position among measures of the entanglement in bipartite pure states because of its relationship to thermodynamics [41] and classical information theory. This connection is best illustrated by writing $|\psi\rangle$ in its Schmidt decomposition [2, 42]

$$|\psi\rangle = \sum_{i=1}^{d} c_i |\alpha_i\rangle \otimes |\beta_i\rangle, \tag{2.8}$$

where $d = \min\{D_A, D_B\}$, the expansion coefficients $c_i$ are real and positive, and the sets $\{|\alpha_i\rangle\}$ and $\{|\beta_i\rangle\}$ form orthonormal bases for subsystems $A$ and $B$, respectively.

*2.1. Entanglement Montones*

According to Eq. (2.7),

$$E_S(\psi) = -\text{Tr}\rho_A \log_2 \rho_A = -\text{Tr}\rho_B \log_2 \rho_B = -\sum_{i=1}^{d} c_i^2 \log_2 c_i^2, \qquad (2.9)$$

which shows that the entropy of entanglement of a bipartite pure state is equivalent to the classical Shannon entropy [11] of the squares of its Schmidt coefficients. Since the Shannon entropy of a classical probability distribution is a measure of the information contained in the distribution, Eq. (2.9) provides a first glimpse of the relationship between quantum mechanical correlations and information, a connection which we endeavor to illuminate throughout this work.

### 2.1.2 Entanglement of formation

Additional justification for using the entropy of entanglement as the fiducial measure of pure state entanglement is provided by two asymptotic results [43] concerning the interconversion of an arbitrary pure state $|\psi\rangle$ and a maximally entangled state of two qubits such as the spin singlet state given by Eq. (1.4). The first states that the entanglement in $n$ non-maximally entangled pure states can be concentrated or "distilled" into $m$ singlet states via an optimal LOCC protocol with a yield $m/n$ that approaches $E_S(\psi)$ as $n \to \infty$. A measure of mixed state entanglement known as the *distillable entanglement* [31, 44] is based on this observation.

Conversely, two separated observers supplied with an entanglement resource of $n$ shared singlets can prepare $m$ arbitrarily good copies of an arbitrary pure state $|\psi\rangle$ with an optimal asymptotic yield $m/n$ that approaches $1/E_S(\psi)$ as $n \to \infty$ [43]. Extending these results to an arbitrary bipartite mixed state $\rho$ with the pure state decomposition

$$\rho = \sum_k p_k |\psi_k\rangle \langle\psi_k|, \qquad (2.10)$$

one finds that the number of singlets needed to create this particular decomposition of $\rho$ is given by

$$n = m \sum_k p_k E_S(\psi_k). \tag{2.11}$$

Of course, a general density matrix has an infinite number of decompositions of the form (2.10). The *entanglement of formation $E_F$* [31, 44], which quantifies the minimum number of singlets required to create $\rho$, is therefore defined as the average entropy of entanglement, minimized over all pure state decompositions of $\rho$, i.e.

$$E_F \equiv \min_{\{p_k, \psi_k\}} \sum_k p_k E_S(\psi_k). \tag{2.12}$$

The generalization of the entropy of entanglement, defined only for pure states, to the entanglement of formation, which is defined for both pure and mixed states, is a specific example of a convex-roof extension [45]. More generally, any pure state entanglement monotone $E(\psi)$ can be extended to mixed states by finding the minimum average value of the measure over all pure state ensemble decompositions of $\rho$ [35]

$$E(\rho) \equiv \min_{\{p_k, \psi_k\}} \sum_k p_k E(\psi_k), \tag{2.13}$$

where the resulting function $E(\rho)$ is the largest convex function of $\rho$ that agrees with $E(\psi)$ on all pure states. Vidal [35] demonstrated that any such function automatically satisfies conditions (2.2) and (2.3).

Unfortunately the above minimization procedure is notoriously difficult [46]. Accordingly, closed forms for the entanglement of formation exist in only a very limited number of cases [47, 48, 49]. We therefore turn now to a discussion of two additional measures of entanglement, known respectively as the concurrence and the tangle, that are related to the entanglement of formation and have proven useful for deriving analytic expressions quantifying the entanglement in certain classes of bipartite systems.

## 2.1.3 Concurrence and tangle

Wootters [47] derived a closed-form expression for the entanglement of formation of a pair of qubits in an arbitrary state by introducing a related quantity known as the concurrence. For a pure state of two qubits, the *concurrence* $C_2(\psi)$ is given by

$$C_2(\psi) \equiv \left|\left\langle \psi | \widetilde{\psi} \right\rangle\right|, \tag{2.14}$$

where $\left|\widetilde{\psi}\right\rangle \equiv \sigma_y \otimes \sigma_y |\psi^*\rangle$ represents the 'spin-flip' of $|\psi\rangle$, $\sigma_y$ is the usual Pauli operator, and the '*' denotes complex conjugation in the standard basis. The spin-flip operation maps the state of each qubit to its corresponding orthogonal state. Thus, the concurrence of any product state of the form (1.3), is equal to zero as expected. Conversely, performing the spin-flip operation on a maximally entangled state such as the singlet state in Eq. (1.4) leaves the state invariant (up to an overall phase), demonstrating that the concurrence achieves its maximum value for the maximally entangled states.

More generally, the following relationship holds between the concurrence and the entropy of entanglement [50]

$$E_S(\psi) = \varepsilon(C_2(\psi)), \tag{2.15}$$

where the function $\varepsilon$ is defined by

$$\varepsilon(C_2) \equiv h\left(\frac{1+\sqrt{1-C_2^2}}{2}\right) \tag{2.16}$$

and

$$h(x) \equiv -x \log_2 x - (1-x) \log_2 (1-x) \tag{2.17}$$

is the binary entropy of the parameter $x$. That the concurrence satisfies the requirements for being an entanglement monotone follows immediately from the observation that $\varepsilon(C_2)$ is a monotonically increasing function of $C_2$ and vice-versa.

The generalization of the concurrence to a mixed state of two qubits proceeds by taking the convex-roof extension according to Eq. (2.13). In this way,

$$C_2(\rho) \equiv \min_{\{p_k,\psi_k\}} \sum_k p_k C_2(\psi_k) = \min_{\{p_k,\psi_k\}} \sum_k p_k \left|\langle \psi_k | \widetilde{\psi}_k \rangle\right|. \tag{2.18}$$

The analytic solution to this minimization procedure involves finding the eigenvalues of the nonHermitian operator $\rho\widetilde{\rho}$, where the tilde again denotes the spin-flip of the quantum state, i.e., $\tilde{\rho} \equiv \sigma_y \otimes \sigma_y \rho^* \sigma_y \otimes \sigma_y$. Specifically, the closed form solution for the concurrence of a mixed state of two qubits is given by

$$C_2(\rho) = \max\{0, \lambda_1 - \lambda_2 - \lambda_3 - \lambda_4\}, \tag{2.19}$$

where the $\lambda_i$'s are the square roots of the eigenvalues of $\rho\tilde{\rho}$ and are ordered in decreasing order [47]. Since there always exists an optimal decomposition of $\rho$ for a pair of qubits in which all of the pure states comprising the decomposition have the same entanglement, Wootters was able to show the following relationship between the entanglement of formation and the concurrence [47]

$$E_F(\rho) = \varepsilon(C_2(\rho)). \tag{2.20}$$

Rungta, *et. al.,* extended the above formalism by deriving an analytic form for the concurrence of a bipartite system $AB$, with *arbitrary* dimensions $D_A$ and $D_B$, in an overall pure state by generalizing the spin-flip operation to apply to higher dimensional systems [51]. The resulting quantity, dubbed the *I-concurrence*, is given by

$$C(\psi) = \sqrt{2\nu_A\nu_B[1 - \text{Tr}(\rho_A^2)]}, \tag{2.21}$$

where $\nu_A$ and $\nu_B$ are arbitrary scale factors. The convex-roof extension of this quantity to mixed states is then given by

$$\begin{aligned} C(\rho) &\equiv \min_{\{p_k,\psi_k\}} \sum_k p_k C(\psi_k) \\ &= \min_{\{p_k,\psi_k\}} \sum_k p_k \sqrt{2\nu_A\nu_B\left[1 - \text{Tr}\left(\rho_A^{(k)}\right)^2\right]}, \end{aligned} \tag{2.22}$$

*2.1. Entanglement Montones*

where we have used Eq. (2.21) for the pure state I-concurrence with $\rho_A^{(k)}$ as the marginal state of subsystem $A$ for the $k^{th}$ term in the ensemble decomposition.

The tangle $\tau_2$, another entanglement monotone applicable to a system of two qubits, is defined as the square of the concurrence in Eq. (2.19),

$$\tau_2(\rho) \equiv [C_2(\rho)]^2 = \max\{0, \lambda_1 - \lambda_2 - \lambda_3 - \lambda_4\}^2. \qquad (2.23)$$

This quantity was introduced in order to simplify investigations into the phenomenon of entanglement sharing [27]. Extending this definition to the I-concurence given by Eq. (2.21) yields an analytic form for the *I-tangle* [36] $\tau$ of a bipartite system in a pure state with arbitrary subsystem dimensions,

$$\tau(\psi) \equiv C^2(\psi) = 2\nu_A \nu_B \left[1 - \text{Tr}\left(\rho_A^2\right)\right] \qquad (2.24)$$

and corresponding convex-roof extension to bipartite mixed states

$$\begin{aligned}\tau(\rho) &\equiv \min_{\{p_k, \psi_k\}} \sum_k p_k C^2(\psi_k) \\ &= 2\nu_A \nu_B \min_{\{p_k, \psi_k\}} \sum_k p_k \left\{1 - \text{Tr}\left[\left(\rho_A^{(k)}\right)^2\right]\right\}.\end{aligned} \qquad (2.25)$$

At this point we note that the scale factors $\nu_A$ and $\nu_B$ in the definitions of the I-concurrence and the I-tangle, which may in general depend on the dimensions of the subsystems $D_A$ and $D_B$ respectively, are usually set to one so that agreement with the two qubit case is maintained, and so that the addition of extra unused Hilbert space dimensions has no effect on the value of the concurrence [51]. We will find in Section 3.4, when we attempt our own further generalization of the tangle formalism, that it is useful to take advantage of this scale freedom. For now, however, we adopt the usual convention both for the sake of clarity and to demonstrate exactly where in our proposed generalization this freedom is required. We also note that there exists no clear resource-based or information-theoretic interpretation for the tangle (or for the concurrence) such as we have for the entanglement of formation. We attempt to remedy this situation in our discussion of complementarity in bipartite systems in Chapter 4.

Using the definition of the I-tangle given by Eq. (2.25), Osborne derived an analytic form for $\tau(\rho)$ in the case where the rank of $\rho$ is no greater than two,

$$\tau(\rho) = \text{Tr}(\rho\tilde{\rho}) + 2\lambda_{min}\left[1 - \text{Tr}(\rho^2)\right], \qquad (2.26)$$

where $\tilde{\rho}$ now represents the universal inversion [51] of $\rho$, and $\lambda_{min}$ is the smallest eigenvalue of the so-called $M$ matrix defined by Osborne [52]. The details of the procedure to evaluate this quantity are quite involved and will not be discussed here. The important point for our purposes is that Eq. (2.26) yields a closed form for a certain class of bipartite mixed states which, as we will see in Section 3.3.3, corresponds to a specific bipartite partition of the two-atom Tavis-Cummings model.

Given the inherent difficulty of minimizing the average value of a pure state entanglement monotone over all possible ensemble decompositions, closed-form solutions for convex-roof extensions of pure state measures exist for only a limited number of classes of bipartite mixed states. In addition to the cases mentioned above, analytic expressions are known for the entanglement of formation of both the Werner states [48] and the isotropic states [49], while the I-concurrence and I-tangle have been calculated for the isotropic states [36]. These derivations rely on the high degree of symmetry possessed by these two classes of $d \times d$ dimensional (two qudit) states; the Werner states are invariant under all transformations of the form $U \otimes U$ while the isotropic states are invariant under the set of transformations $U \otimes U^*$, where in each case $U$ represents an arbitrary unitary operation on a $d$-dimensional quantum system.

The relative dearth of closed-form expressions for various bipartite systems of interest has prompted investigation into the existence of entanglement monotones that are easily computable in the most general cases. The negativity is one such measure. While this quantity also lacks a clear resource-based interpretation, it does illustrate a fundamental connection between the separability of a density operator and the theories of positive and completely positive maps [53].

### 2.1.4 Negativity

The negativity is a measure of entanglement that relies on the following essential difference between the dynamical behaviors of classically correlated (or separable) systems and entangled systems: a positive map applied to one subsystem of a composite system in a separable state always yields another valid quantum state, whereas the same operation applied to an entangled state does not, in general, result in a valid density operator. This is because a positive map $\mathcal{O}$ that maps positive operators to positive operators, does not behave like a completely positive map in the presence of entanglement. A completely positive map takes positive operators acting on a given subspace and all of its extensions to tensor-product spaces $\mathcal{O} \otimes I$, to positive operators on the extended subspace [2, 53]. Here $I$ is the identity operator on the appended space. In general, the operator resulting from a positive map applied to a quantum state may possess one or more negative eigenvalues, signaling the presence of entanglement.

The positive map used in the definition of the negativity is the partial transpose operation, and corresponds to a local permutation of the basis vectors describing the transposed subsystem. The *partial transpose* $\rho^{T_A}$ of $\rho$ with respect to subsystem $A$ is defined to have matrix elements given by

$$\left\langle i_A, j_B \left| \rho^{T_A} \right| k_A, l_B \right\rangle \equiv \left\langle k_A, j_B \left| \rho \right| i_A, l_B \right\rangle, \tag{2.27}$$

for arbitrary orthonormal basis sets $\{|i_A\rangle\}$, and $\{|j_B\rangle\}$. The positivity of $\rho^{T_A}$ is both a necessary and sufficient condition for the separability of $\rho$ in the cases of $2 \times 2$ and $2 \times 3$ dimensional systems, while for higher dimensional systems a positive partial transpose is only a necessary condition for $\rho$ to be separable [53]. Note that none of the results in what follows would change if we instead took the partial transpose with respect to subsystem $B$. This fact highlights a fundamental symmetry of entangled correlations.

The *trace norm* of an Hermitian operator $H$ is defined as [54]

$$\|H\|_1 \equiv \text{Tr}\sqrt{H^\dagger H}, \tag{2.28}$$

and is equal to the sum of the absolute values of the eigenvalues of $H$. In the case where $H = \rho$ is a density matrix, Eq. (2.28) reduces to the normalization condition $\text{Tr}(\rho) = 1$. Conversely, since $\rho^{T_A}$ may in general have negative eigenvalues, its trace norm reads

$$\|\rho^{T_A}\|_1 = \sum_i |\lambda_i| = \sum_i \lambda_i^{(+)} + \left|\sum_i \lambda_i^{(-)}\right|, \tag{2.29}$$

where $\lambda_i^{(\pm)}$ represents the $i^{th}$ positive (negative) eigenvalue of $\rho^{T_A}$. Using the fact that $\text{Tr}(\rho^{T_A}) = 1$,

$$1 = \sum_i \lambda_i = \sum_i \lambda_i^{(+)} + \sum_i \lambda_i^{(-)} \tag{2.30}$$

or

$$\sum_i \lambda_i^{(+)} = 1 + \left|\sum_i \lambda_i^{(-)}\right|, \tag{2.31}$$

Eq. (2.29) becomes

$$\|\rho^{T_A}\|_1 = 1 + 2\left|\sum_i \lambda_i^{(-)}\right| = 1 + 2\mathcal{N}(\rho), \tag{2.32}$$

where the *negativity* $\mathcal{N}(\rho)$

$$\mathcal{N}(\rho) \equiv \frac{\|\rho^{T_A}\|_1 - 1}{2} = \sum_i \left|\lambda_i^{(-)}\right| \tag{2.33}$$

is defined as the sum of the absolute values of the negative eigenvalues of $\rho^{T_A}$, and quantifies the degree to which $\rho^{T_A}$ fails to be a positive operator [55].

The negativity satisfies the conditions (2.2) and (2.3) for being an entanglement monotone, and has the added benefit of being computable for any mixed state of an arbitrary bipartite system [55]. However, it does not reduce to the entropy of entanglement for bipartite pure states, and so lacks a clear information-theoretic interpretation since it has no discernible connection to the entanglement of formation or concurrence formalisms. The next section addresses this issue by introducing a new class of computable entanglement monotones that provides a global structure highlighting some of the relationships between these quantities.

## 2.2 A New Family of Analytic Entanglement Monotones

This section presents a new family of entanglement monotones [56] based on the positive partial transpose criterion for separability [57, 53] and the theory of majorization [54]. Each is a simple function of the negative eigenvalues generated via the partial transposition operation, and may be evaluated with any standard linear algebra package. One member of this family is shown to be equivalent to the negativity, while two others constitute computable lower bounds on the I-concurrence and on the I-tangle, respectively. In order to estimate the quality of these functions as lower bounds, we compare their values with the values of the I-concurrence and the I-tangle on the family of isotropic states.

The construction given here is based on the theory of majorization. This formalism has been successfully used to characterize the necessary and sufficient conditions under which the process of entanglement transformation is possible [58], and has led to new insights into the operation of quantum algorithms [59, 60] and in the problem of optimal Hamiltonian simulation [55]. The majorization relation between the global and local eigenvalue vectors of a bipartite system has also been used to shed light on the phenomenon of bound entanglement [61].

The following is a brief review of the main tenets of majorization theory. The reader is referred to [54] for extensive background on the subject. Given two $d$-dimensional vectors $x$ and $y$ in $\mathbb{R}^d$, we say that $x$ is majorized by $y$, denoted by the expression $x \prec y$, when the following two conditions hold:

$$\sum_{i=1}^{k} x_i^{\downarrow} \leq \sum_{i=1}^{k} y_i^{\downarrow}; \qquad \forall k = 1, \ldots, d \tag{2.34}$$

$$\sum_{i=1}^{d} x_i^{\downarrow} = \sum_{i=1}^{d} y_i^{\downarrow}. \tag{2.35}$$

Here, the symbol $\downarrow$ indicates that the vector coefficients are arranged in decreasing order.

The mathematical construct of majorization is naturally connected with the comparative disorder between two vectors [54]. In fact, $x \prec y$ if and only if there exists a doubly stochastic matrix[2] $D$ such that

$$x = Dy. \tag{2.36}$$

According to Birkhoff's theorem [54] a $d \times d$ matrix is doubly stochastic if and only if it can be written as a convex combination of permutation matrices $P_j$ such that

$$D = \sum_{j} p_j P_j. \tag{2.37}$$

Taken together Eqs. (2.36) and (2.37) imply that $x \prec y$ if and only if $x$ can be written as a convex mixture of permutations of $y$; it is in this sense that $x$ is more

---

[2]A matrix $D$ is doubly stochastic if its coefficients $d_{ij}$ are non-negative and $\sum_i d_{ik} = \sum_j d_{kj} = 1, \forall k$.

## 2.2. A New Family of Analytic Entanglement Monotones

disordered than $y$ [2]. Further, if we consider $x$ and $y$ to be probability distributions, then the fact that $x$ is majorized by $y$ expresses the idea that $x$ is more disordered, in an entropic or information-theoretic sense, than $y$.

In the case that only condition (2.34) holds, we say that $x$ is weakly submajorized by $y$. This is denoted by the expression $x \prec_w y$. We will make use of the following two results concerning weak submajorization [54]:

$$x \prec_w y \in \mathbb{R}^d \Rightarrow x^+ \prec_w y^+ \tag{2.38}$$

$$x \prec_w y \in \mathbb{R}^d_+ \Rightarrow x^p \prec_w y^p; \qquad \forall p \geq 1, \tag{2.39}$$

where the operations $x^p$ and $x^+$ act on each component of $x$ individually. The $x^+$ operation simply converts each of the negative entries in $x$ into a zero.

The following relation enables us to construct a useful family of convex functions of the negative eigenvalues of an Hermitian matrix. Given two Hermitian matrices $Q$ and $R$,

$$\lambda^\downarrow(Q+R) \prec \lambda^\downarrow(Q) + \lambda^\downarrow(R), \tag{2.40}$$

where $\lambda(Q)$ denotes the vector whose elements are the eigenvalues of $Q$ [54]. Let us now define the vectors $\tilde{\lambda}(Q) = -\lambda(Q) = \lambda(-Q)$, such that the negative coefficients in $\lambda(Q)$ become positive in $\tilde{\lambda}(Q)$. Clearly Eq. (2.40) also holds for the vectors $\tilde{\lambda}(Q)$, i.e.,

$$\tilde{\lambda}^\downarrow(Q+R) \prec \tilde{\lambda}^\downarrow(Q) + \tilde{\lambda}^\downarrow(R). \tag{2.41}$$

Recognizing that the coefficients of $\tilde{\lambda}(Q)$ belong to $\mathbb{R}^d$ and applying property (2.38), Eq. (2.41) becomes

$$\left(\tilde{\lambda}^\downarrow(Q+R)\right)^+ \prec_w \left(\tilde{\lambda}^\downarrow(Q) + \tilde{\lambda}^\downarrow(R)\right)^+. \tag{2.42}$$

The coefficients of the vectors in Eq. (2.42) are, by definition, members of $\mathbb{R}_+^d$. Thus, using property (2.39) we obtain

$$\left[\left(\tilde{\lambda}^{\downarrow}(Q+R)\right)^+\right]^p \prec_w \left[\left(\tilde{\lambda}^{\downarrow}(Q)+\tilde{\lambda}^{\downarrow}(R)\right)^+\right]^p. \tag{2.43}$$

Applying condition (2.34) for $k=d$ then yields

$$\sum_{i=1}^d \left(\tilde{\lambda}_i^+(Q+R)\right)^p \leq \sum_{i=1}^d \left[\left(\tilde{\lambda}_i^{\downarrow}(Q)+\tilde{\lambda}_i^{\downarrow}(R)\right)^+\right]^p, \tag{2.44}$$

where we have removed the ordering of the vector on the left hand side. The term inside square brackets on the right hand side of Eq. (2.44) can be bounded from above by $\left[\tilde{\lambda}_i^{\downarrow}(Q)\right]^+ + \left[\tilde{\lambda}_i^{\downarrow}(R)\right]^+$, yielding

$$\sum_{i=1}^d \left(\tilde{\lambda}_i^+(Q+R)\right)^p \leq \sum_{i=1}^d \left\{\left[\tilde{\lambda}_i^{\downarrow}(Q)\right]^+ + \left[\tilde{\lambda}_i^{\downarrow}(R)\right]^+\right\}^p. \tag{2.45}$$

Finally, using *Minkowski's inequality* [62]

$$\left[\sum_{i=1}^d (x_i+y_i)^p\right]^{1/p} \leq \left(\sum_{i=1}^d x_i^p\right)^{1/p} + \left(\sum_{i=1}^d y_i^p\right)^{1/p}, \tag{2.46}$$

which shows that the $p^{\text{th}}$ root of the quantity on the right hand side of Eq. (2.45) satisfies the triangle inequality, we obtain

$$\left[\sum_{i=1}^d \left(\tilde{\lambda}_i^+(Q+R)\right)^p\right]^{1/p} \leq \left[\sum_{i=1}^d \left(\tilde{\lambda}_i^+(Q)\right)^p\right]^{1/p} + \left[\sum_{i=1}^d \left(\tilde{\lambda}_i^+(R)\right)^p\right]^{1/p}. \tag{2.47}$$

The terms in square brackets on the right hand side of Eq. (2.47) are the sums of the positive coefficients of $\tilde{\lambda}(Q)$ ($\tilde{\lambda}(R)$) to the $p^{th}$ power, or equivalently, to the sums of the absolute values of the negative coefficients of $\lambda(Q)$ ($\lambda(R)$) to the $p^{th}$ power. Thus, we see that the quantities

$$\mathcal{M}_p(Q) \equiv \bigg(\sum_{\lambda(Q)<0} |\lambda(Q)|^p\bigg)^{1/p}; \qquad \forall p \geq 1 \tag{2.48}$$

## 2.2. A New Family of Analytic Entanglement Monotones

obey the triangle inequality on the set of Hermitian matrices. In particular, they are convex functions since

$$\mathcal{M}_p(\alpha Q + \beta R) \leq \alpha \mathcal{M}_p(Q) + \beta \mathcal{M}_p(R) \tag{2.49}$$

for $\alpha$ and $\beta$ in the interval $[0, 1]$ such that $\alpha + \beta = 1$. A similar result also holds for the set of functions

$$\mathcal{N}_p(Q) \equiv [\mathcal{M}_p(Q)]^p = \sum_{\lambda(Q)<0} |\lambda(Q)|^p; \qquad \forall p \geq 1. \tag{2.50}$$

Using Eqs. (2.48) and (2.50) we now define two related sets of functions

$$\mathcal{M}_p^{\mathrm{T_A}}(\rho) \equiv \left( \sum_{\lambda(\rho^{\mathrm{T_A}})<0} \left|\lambda\left(\rho^{\mathrm{T_A}}\right)\right|^p \right)^{1/p}; \qquad \forall p \geq 1 \tag{2.51}$$

and

$$\mathcal{N}_p^{\mathrm{T_A}}(\rho) \equiv \left[\mathcal{M}_p^{\mathrm{T_A}}(\rho)\right]^p = \sum_{\lambda(\rho^{\mathrm{T_A}})<0} \left|\lambda\left(\rho^{\mathrm{T_A}}\right)\right|^p; \qquad \forall p \geq 1 \tag{2.52}$$

and prove that $\mathcal{M}_p^{\mathrm{T_A}}(\rho)$ is an entanglement monotone (with an analogous proof holding for $\mathcal{N}_p^{\mathrm{T_A}}(\rho)$). By Eq. (2.49) and the fact that the partial transpose operation defined by Eq. (2.27) is linear, it follows that $\mathcal{M}_p^{\mathrm{T_A}}(\rho)$ is a convex function so that condition (2.3) is satisfied. In order to show that condition (2.2) is also satisfied, we define

$$\rho' \equiv \sum_i p_i \rho'_i = \sum_i \mathcal{K}_i(\rho) = \sum_i (I_A \otimes K_i) \rho \left(I_A \otimes K_i^\dagger\right) \tag{2.53}$$

according to Eq. (2.1), where the last equality represents the Kraus decomposition (also known as the operator-sum representation) [2] of this operation with Kraus operators $K_i$. Since an arbitrary LOCC protocol can be decomposed into an equivalent protocol where only one party performs operations on their local subsystem, and since $\mathcal{M}_p^{\mathrm{T_A}}(\rho)$ is invariant under permutation of the parties, we may restrict our attention to quantum operations having Kraus decompositions with just a single term [63] acting nontrivially only on subsystem $B$ [64]. According to Eqs. (2.51) and (2.53) we then have,

$$\sum_i p_i \mathcal{M}_p^{\mathrm{T_A}}(\rho_i') = \sum_i \left\{ \sum_{\lambda(\rho^{\mathrm{T_A}})<0} \left| \lambda \left[ (I_A \otimes K_i) \rho \left( I_A \otimes K_i^\dagger \right) \right]^{\mathrm{T_A}} \right|^p \right\}^{1/p}. \qquad (2.54)$$

Recognizing that we may consistently interchange the order in which the Kraus operator and the partial transposition are applied then yields

$$\sum_i p_i \mathcal{M}_p^{\mathrm{T_A}}(\rho_i') = \sum_i \left\{ \sum_{\lambda(\rho^{\mathrm{T_A}})<0} \left| \lambda \left[ (I_A \otimes K_i) \rho^{\mathrm{T_A}} \left( I_A \otimes K_i^\dagger \right) \right] \right|^p \right\}^{1/p}. \qquad (2.55)$$

Next, we use the following relations [54]

$$S_j(QR) \leq ||Q|| S_j(R) \qquad S_j(QR) \leq ||R|| S_j(Q), \qquad (2.56)$$

where $S_j(Q)$ denotes the $j^{\mathrm{th}}$ singular value of $Q$, and $||Q||$ denotes the *operator norm* of $Q$, defined as the largest singular value of $Q$, i.e., $||Q|| \equiv \max_j S_j(Q)$. For Hermitian $Q$, the singular values are given by $S(Q) = \lambda\left(\sqrt{Q^\dagger Q}\right) = |\lambda(Q)|$. It then follows that

$$\left| \lambda \left[ (I_A \otimes K_i) \rho^{\mathrm{T_A}} \left( I_A \otimes K_i^\dagger \right) \right] \right|^p \leq \left|\left| I_A \otimes K_i \right|\right|^p \left|\left| I_A \otimes K_i^\dagger \right|\right|^p \left| \lambda\left(\rho^{\mathrm{T_A}}\right) \right|^p. \qquad (2.57)$$

Applying Eq. (2.57) to Eq. (2.55) then yields

$$\sum_i p_i \mathcal{M}_p^{\mathrm{T_A}}(\rho_i') \leq \sum_i \left( \left|\left| I_A \otimes K_i \right|\right| \left|\left| I_A \otimes K_i^\dagger \right|\right| \right) \mathcal{M}_p^{\mathrm{T_A}}(\rho), \qquad (2.58)$$

where we have made use of Eq. (2.51) in identifying $\mathcal{M}_p^{\mathrm{T_A}}(\rho)$. Finally, due to the normalization condition $\sum_i K_i^\dagger K_i \leq I_B$ [2], it follows that $||I_A \otimes K_i|| \leq 1$ and $\left|\left| I_A \otimes K_i^\dagger \right|\right| \leq 1$. Combined with Eq. (2.58) this implies that

$$\sum_i p_i \mathcal{M}_p^{\mathrm{T_A}}(\rho_i') \leq \mathcal{M}_p^{\mathrm{T_A}}(\rho), \qquad (2.59)$$

## 2.2. A New Family of Analytic Entanglement Monotones

demonstrating that the functions $\mathcal{M}_p^{T_A}(\rho)$ also satisfy condition (2.2) and are therefore entanglement monotones. A similar argument can be given for the monotonicity of the functions $\mathcal{N}_p^{T_A}(\rho)$.

Note that the negativity given by Eq. (2.33) is one member of this new family of entanglement monotones since $\mathcal{N}(\rho) = \mathcal{M}_1^{T_A}(\rho)$. Further, when we restrict our attention to the two-qubit case ($d = 4$) the partial transpose of $\rho$ has at most one negative eigenvalue, implying that $\mathcal{M}_2^{T_A}(\rho)$ also reduces to the negativity in this situation. For this special case it was shown that twice the negativity, referred to here as the *scaled negativity* (defined so as to take values between zero and one for pure states of two qubits), is a lower bound on the concurrence [65].

Our results may be used to generalize this last relationship by showing that the quantity

$$\mathcal{L}_C(\rho) \equiv 2\mathcal{M}_2^{T_A}(\rho) \tag{2.60}$$

is a lower bound on the I-concurrence given by Eq. (2.22), i.e.,

$$\mathcal{L}_C(\rho) \leq C(\rho) \tag{2.61}$$

for a bipartite system with arbitrary subsystem dimensions. We begin by writing the pure state I-concurrence in terms of the Schmidt coefficients given in Eq. (2.8),

$$C(\psi) = 2\Big(\sum_{i<j} c_i^2 c_j^2\Big)^{\frac{1}{2}} \tag{2.62}$$

and noting that the quantities $\sqrt{c_i^2 c_j^2}$ are the absolute values of the negative eigenvalues of the partial transpose of $|\psi\rangle$. This connection shows that our generalization of the scaled negativity $\mathcal{L}_C(\rho)$ and the I-concurrence agree on pure states, i.e.,

$$\mathcal{L}_C(\psi) = C(\psi). \tag{2.63}$$

It then follows that the relation in Eq. (2.61) holds since the scaled negativity and the I-concurrence are convex functions on the space of density matrices that agree on the extreme points (the pure states) of this space, while the I-concurrence is by definition the largest of all such functions.

A similar argument shows that the function

$$\mathcal{L}_\tau(\rho) \equiv [\mathcal{L}_C(\rho)]^2, \tag{2.64}$$

is a lower bound on the mixed-state I-tangle,

$$\mathcal{L}_\tau(\rho) \leq \tau(\rho). \tag{2.65}$$

These bounds, which are entanglement monotones in their own right, may be evaluated for a bipartite system with subsystems of arbitrary dimensions in a straightforward manner with the help of a standard linear algebra package.

It has been shown that the positive partial transposition criterion employed here is a necessary and sufficient condition for separability for $d \leq 6$ [53]. In higher dimensions, positivity under partial transposition is a necessary, but not sufficient, condition for separability. However, this is not a serious drawback for the usefulness of the quantities introduced above. Indeed, theoretical and numerical investigations have shown that the volume of the set of density operators with positive partial transpose decreases exponentially with the dimension $d$ of the Hilbert space [66].

Consider now the following example application of the lower bounds for the I-concurrence and I-tangle given by Eqs. (2.60) and (2.64) respectively, to the isotropic states. The isotropic states $\rho_F$ describe a quantum system composed of two subsystems of equal dimension $d$. They are mixtures formed from the convex combination of a maximally mixed state and a maximally entangled pure state,

$$\rho_F = (1-\omega)\frac{1}{d^2}I_d \otimes I_d + \omega|\Psi^+\rangle\langle\Psi^+|; \qquad 0 \leq \omega \leq 1. \tag{2.66}$$

Here, $I_d$ is the identity operator acting on a $d$-dimensional Hilbert space, and $|\Psi^+\rangle$ is the state given by

$$|\Psi^+\rangle = \sum_{i=1}^{d} \frac{1}{\sqrt{d}}|i\rangle \otimes |i\rangle. \tag{2.67}$$

## 2.2. A New Family of Analytic Entanglement Monotones

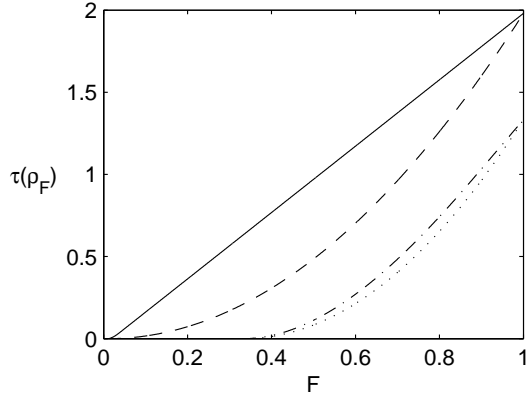

Figure 2.1: Comparison of $\tau(\rho_F)$ and $\mathcal{L}_\tau(\rho_F)$ as functions of the fidelity $F$ for different dimensions $d$. Solid line: $\tau(\rho_F)$ for $d = 100$. Dashed line: $\mathcal{L}_\tau(\rho_F)$ for $d = 100$. Dot-dashed line: $\tau(\rho_F)$ for $d = 3$. Dotted line: $\mathcal{L}_\tau(\rho_F)$ for $d = 3$.

The parameter $\omega$ in Eq. (2.66) can be related to the fidelity $F$

$$F \equiv \langle \Psi^+ | \rho_F | \Psi^+ \rangle \in [0, 1] \tag{2.68}$$

of $\rho_F$ with respect to the state $|\Psi^+\rangle$, via the relation

$$\omega = \frac{d^2 F - 1}{d^2 - 1}. \tag{2.69}$$

It has been shown that the isotropic states are separable for $F \leq 1/d$ [67], while the exact values of the I-tangle for the isotropic states $\rho_F$ were analytically calculated in [36].

The value of the lower bounds $\mathcal{L}_C(\rho_F)$ and $\mathcal{L}_\tau(\rho_F)$ on the isotropic states can be evaluated for arbitrary $d$. Since the partial transposition operation is linear, and since the identity operator is invariant under this operation, the eigenvalues of $\rho_F^{T_A}$ are readily obtained. They are given by $(1-\omega)/d^2 \pm \omega/d$ with multiplicity $d(d \pm 1)/2$, respectively. The negative eigenvalues $(1-\omega)/d^2 - \omega/d$ become positive when $\omega \leq 1/(d+1)$, or equivalently, when $F \leq 1/d$ such that

$$\mathcal{L}_C(\rho_F) = \begin{cases} \frac{2}{d}\left(\frac{\omega-1}{d} + \omega\right)\sqrt{\frac{d(d-1)}{2}} & \omega > 1/(d+1) \\ 0 & \omega \leq 1/(d+1) \end{cases} \tag{2.70}$$

The behaviors of the I-tangle $\tau(\rho_F)$ and of $\mathcal{L}_\tau(\rho_F)$ for the isotropic states are depicted in Fig. 2.1. For $d = 2$, the two functions assume the same values, while for larger dimensions and constant fidelity, the difference between the lower bound and the I-tangle increases. In the limit $d \to \infty$, $\tau(\rho_F)$ and $\mathcal{L}_\tau(\rho_F)$ behave as $\sqrt{2}F$ and $2F^2$, respectively. Similarly, the I-concurrence $C(\rho_F)$ and $\mathcal{L}_C(\rho_F)$ may be calculated analytically. Here we find that the two quantities agree over the isotropic states for any dimension $d$, demonstrating that the isotropic states saturate our lower bound.

## 2.3 Summary

The conditions for a function of a quantum state to be a good measure of entanglement are relatively straightforward; it must be non-increasing, on average, under the set of LOCC protocols and under mixing. Any pure state entanglement monotone satisfying these criteria may be readily extended to mixed states via the convex-roof formalism. The evaluation of such functions is, however, computationally intractable for many cases of interest. Accordingly, there is much interest in identifying quantities, such as the negativity, that are computable in the most general situations, even though they may lack a clear resource-based interpretation.

The entanglement measures derived in Section 2.2 comprise classes of monotones based on the positive partial transpose criterion for separability, and on the connection between the theory of majorization and comparative disorder. In this larger context, the negativity is seen to be one specific example of such a function. Other instances (with appropriate scaling) yield lower bounds on the I-concurrence and the I-tangle, providing useful tools for investigations of quantum information theoretic concepts and fundamental quantum mechanics. Apart from offering a larger structure from which to view these different entanglement measures, each member in this new family of functions also possesses an analytic form that may be evaluated in the

## 2.3. Summary

most general situations. In fact, as we will see in the next chapter, our results enable the quantification of entanglement in the context of a multipartite system of both theoretical and experimental significance for which there are no known closed-form solutions for any of the convex roof-based measures.

# Chapter 3

# Entanglement Sharing in the Tavis-Cummings Model

## 3.1 Introduction

The development of a mathematically rigorous theory of entanglement is highly desirable for investigating foundational issues in quantum mechanics, as well as for analyzing specific entanglement-enhanced information processing tasks. The previous chapter makes it clear that, while a consistent method of quantifying entanglement under the most general circumstances has not yet been formulated, progress has been made in certain specific cases. In fact, it turns out that the current state of the theory of entanglement is capable of yielding an essentially complete analysis of the quantum correlations arising in the two-atom Tavis-Cummings model (TCM), a system of both theoretical and experimental significance [68]. This chapter presents a detailed analysis of the different types of entanglement (corresponding to the different possible partitions of the system into subsystems) evolving in the two-atom TCM as a function of time.

This investigation is motivated by the opportunity to employ several different results from the theory of entanglement in order to study the dynamical evolution of quantum correlations in a nontrivial, yet experimentally realizable system. One specific goal of this work is to lay a foundation for future study of the relationship between the degree of entanglement in a system and the measurement backaction, or information-disturbance tradeoff [69, 70, 71], that occurs when measuring one subsystem in order to gain information about a second, correlated subsystem in the context of the TCM. This is seen as a necessary step toward being able to perform feedback and control of atomic ensembles with possible applications in the field of quantum computation.

The control of quantum systems through active measurement and feedback has been developing at a rapid pace. In a typical scenario, a single atom is monitored indirectly through its coupling to a traveling probe such as a laser beam. The scattered beam and the system become correlated, and a subsequent measurement of the probe leads to backaction on the system. A coherent drive applied to the system can then be made conditional on the measurement record, leading to a closed-loop control model [72, 73]. Such a protocol has been implemented to control a single mode electromagnetic field in a cavity [74], and has been envisioned for controlling a variety of systems such as the state of a quantum dot in a solid [75], the state of an atom coupled to a cavity mode [76], and the motion of a micro-mechanical resonator coupled to a Cooper pair charge box [77].

A common theme in the examples given above is that measurements are made on *single* copies of the quantum system of interest. However, in many situations one does not have access to an individually addressable system. In a gas, for example, preparing and/or addressing individual atoms is extremely difficult. In situations such as this, it is useful to think of the entire ensemble as a single many-body system. Indeed, recent experiments [78, 79] and theoretical proposals [80] have explored the

*3.1. Introduction*

control of such ensembles from the point of view of the Dicke model [81], where a collection of $N$ two-level atoms is treated as a pseudo-spin with $J = N/2$.

Measurement backaction on the pseudo-spin can lead to squeezing of the quantum fluctuations [78, 79, 80], which may be enhanced through active closed-loop control [72, 73]. This squeezing can reduce the quantum fluctuations of an observable as in, for example, the reduction of "projection noise" leading to enhanced precision measurements in an atomic clock [82]. Moreover, spin-squeezing is related to quantum entanglement between the atomic members of the ensemble [83]. This entanglement arises not through direct interaction between the atoms, but through their coupling to a common "quantum bus" in the form of an applied probe.

Measures of entanglement associated with these spin-squeezed states have been studied by Stockton, *et. al.,* [81] under the assumption that all of the atoms in the ensemble are symmetrically coupled to the bus. However, completely quantifying multipartite entanglement in the most general cases is extremely difficult, and as yet, an unsolved problem [46]. Here we consider the simplest possible ensemble consisting of two two-level atoms. Although at first sight this might appear trivial, when such a system is coupled to a quantum bus a rich structure emerges. Again, we consider the simplest realization of the bus – a single mode quantized electromagnetic field. The resulting physical system then corresponds to the two-atom Tavis-Cummings model [84]. A thorough understanding of the dynamical evolution of the TCM has obvious implications for the performance of quantum information processing [2, 85, 86], as well as for our understanding of fundamental quantum mechanics [2, 87]. Bipartite entanglement has been investigated in this system for the one-atom case, known as the Jaynes-Cummings model, for initial pure states [88] and mixed states [89, 90] of the field.

The two-atom TCM consists of two two-level atoms, or qubits, coupled to a single mode of the electromagnetic field in the dipole and rotating-wave approximations [84]. This model admits a total of five different possible partitionings, each of which may contain varying degrees of entanglement as a function of time. These five different partitions are (a) the bipartite division of the system into a subsystem consisting of the field and a second subsystem consisting of the ensemble of two atoms, (b) the bipartite division of the system into a subsystem consisting of a single two-level atom and a second subsystem containing the remainder, i.e., the remaining atom and the field, (c) the two atoms treated as separate subsystems, with the field being traced over, (d) a single atom and the field treated as separate subsystems, with the remaining atom being traced over, and (e) a single partition containing the entire tripartite system (capable of supporting irreducible three-body correlations).

Taken as a whole, the two-atom TCM in an overall pure state constitutes a tripartite quantum system in a Hilbert space with tensor product structure $2 \otimes 2 \otimes \infty$. Entanglement in tripartite systems has been studied by Coffman, *et. al.*, [27] for the case of three qubits. They found that such quantum correlations cannot be arbitrarily distributed amongst the subsystems; the existence of three-body correlations constrains the distribution of the bipartite entanglement which remains after tracing over any one of the qubits. For example, in a GHZ-state [37],

$$|\text{GHZ}\rangle = \frac{1}{\sqrt{2}}(|000\rangle + |111\rangle), \tag{3.1}$$

tracing over any one qubit results in a maximally mixed state containing no entanglement between the remaining two qubits. In contrast, for a W-state,

$$|\text{W}\rangle = \frac{1}{\sqrt{3}}(|001\rangle + |010\rangle + |100\rangle), \tag{3.2}$$

the average remaining bipartite entanglement is maximal [64]. Coffman, *et. al.,* analyzed this phenomenon of entanglement sharing [27], using the tangle between two qubits as defined in Eq. (2.23). In order to complete this analysis, they also introduced a quantity known as the residual tangle to quantify the irreducible tripartite correlations in a three qubit system [27].

Here, we extend the analysis of entanglement sharing to the case of the two-atom TCM [68]. This has implications for the study of quantum control of ensembles. For example, if we imagine that the quantum bus is measured, e.g., the field leaks out of the cavity and is then detected, then the degree of correlation between the field and one of the atoms determines the degree of backaction on one atom. We can then quantify the degree to which one can perform quantum control on a single member of an ensemble even when one couples only collectively to the entire ensemble. We accomplish this by extending the residual tangle formalism of Coffman, *et. al.*, to our $2 \otimes 2 \otimes \infty$ system.

The remainder of this chapter is organized as follows. First, the important features of the TCM are reviewed in Section 3.2. Using the formalism introduced in Section 2.1.3, we then calculate the tangle for each of the bipartite partitions of this tripartite system in Section 3.3. We will find an approximate analytic expression for the tangle between the field and the ensemble in the limit of large average photon number and in the Markoff approximation which provides further insight into these results. In Section 3.4, we study the irreducible tripartite correlations in the system using our proposed generalization of the residual tangle. Finally, we summarize our results and suggest possible directions for further research in Section 3.5.

## 3.2 The Tavis-Cummings Model

The Tavis-Cummings model (TCM) [84] (or "Dicke model"[91]) describes the simplest fundamental interaction between a single mode of the quantized electromagnetic field and a collection of $N$ atoms under the two-level and rotating wave approxima-tions [92]. The two-atom ($N = 2$) TCM is governed by the Hamiltonian

$$\begin{aligned} H &= H_0 + H_{int} \\ &= \hbar\omega\left(a^\dagger a + \frac{1}{2}\sigma_z^{(1)} + \frac{1}{2}\sigma_z^{(2)}\right) \\ &\quad + \hbar g\left[\left(\sigma_-^{(1)} + \sigma_-^{(2)}\right)a^\dagger + \left(\sigma_+^{(1)} + \sigma_+^{(2)}\right)a\right], \end{aligned} \qquad (3.3)$$

where $\sigma_\pm^{(i)}$ and $\sigma_z^{(i)}$ display a local $SU(2)$ algebra for the $i^{th}$ atom in the two-dimensional subspace spanned by the ground and excited states $\{|g\rangle, |e\rangle\}$, and $a$ ($a^\dagger$) are bosonic annihilation (creation) operators for the monochromatic field. The Hilbert space $\mathcal{H}$ of the joint system is given by the tensor product $\mathcal{H}_{A_1} \otimes \mathcal{H}_{A_2} \otimes \mathcal{H}_F$ where $\mathcal{H}_{A_1}$ ($\mathcal{H}_{A_2}$) denotes the Hilbert space of atom one (two) and $\mathcal{H}_F$ is the Hilbert space of the electromagnetic field.

The total number of excitations $K = a^\dagger a + \frac{1}{2}(\sigma_z^{(1)} + \sigma_z^{(2)} + 2)$ is a conserved quantity which allows one to split the Hilbert space $\mathcal{H}$ into a direct sum of subspaces, i.e., $\mathcal{H} = \sum_{K=0}^{\infty} \oplus \Omega_K$, with each subspace $\Omega_K$ spanned by the eigenvectors $\{|ee, k-2\rangle, |eg, k-1\rangle, |ge, k-1\rangle, |gg, k\rangle\}$ of $K$ with eigenvalue $k$. The analytic form for the time evolution operator within a subspace $\Omega_K$ is given by [93]

$$U(k,t) = \begin{pmatrix} \frac{1}{\delta}[\gamma c(\beta) + k] & \frac{i}{\sqrt{2}}\sqrt{\frac{\gamma}{\delta}}s(\beta) & \frac{i}{\sqrt{2}}\sqrt{\frac{\gamma}{\delta}}s(\beta) & -\frac{\sqrt{k\gamma}}{\delta}[1-c(\beta)] \\ -\frac{i}{\sqrt{2}}\sqrt{\frac{\gamma}{\delta}}s(\beta) & \frac{c(\beta)+1}{2} & \frac{c(\beta)-1}{2} & \frac{i}{\sqrt{2}}\sqrt{\frac{k}{\delta}}s(\beta) \\ -\frac{i}{\sqrt{2}}\sqrt{\frac{\gamma}{\delta}}s(\beta) & \frac{c(\beta)-1}{2} & \frac{c(\beta)+1}{2} & \frac{i}{\sqrt{2}}\sqrt{\frac{k}{\delta}}s(\beta) \\ -\frac{\sqrt{k\gamma}}{\delta}[1-c(\beta)] & -\frac{i}{\sqrt{2}}\sqrt{\frac{k}{\delta}}s(\beta) & -\frac{i}{\sqrt{2}}\sqrt{\frac{k}{\delta}}s(\beta) & \frac{1}{\delta}[kc(\beta) + \gamma] \end{pmatrix} \qquad (3.4)$$

where $c(x) \equiv \cos(x)$, $s(x) \equiv \sin(x)$, $\beta \equiv \sqrt{2\delta}gt$, $\gamma \equiv k-1$, and $\delta \equiv 2k-1$.

It is assumed throughout that the initial state of the TCM system is pure. Furthermore, we consider only the effects of the unitary evolution generated by Eq. (3.3), i.e., we do not include the effects of measurement, nor of mixing due to environment-induced decoherence [10], so that our system remains in an overall pure state at all times. Finally, by assuming an identical coupling constant $g$ between each of the atoms and the field, the Hamiltonian is symmetric under atom-exchange. This invariance under the permutation group was used by Stockton, *et. al.,* [81] to analyze the entanglement properties of very large ensembles. We will also make use of this fact in order to reduce the number of different partitioning schemes that one needs to consider when studying entanglement sharing in the two-atom TCM.

## 3.3 Bipartite Tangles in the Two-Atom TCM

Let the two atoms in the ensemble be denoted by $A_1$ and $A_2$, respectively, and the field, or quantum bus, by $F$. Because of the assumed exchange symmetry, there are four nonequivalent partitions of the two-atom TCM into tensor products of bipartite subsystems: (i) the field times the two-atom ensemble, $F \otimes (A_1 A_2)$, (ii) one atom times the remaining atom and the field, $A_1 \otimes (A_2 F) \equiv A_2 \otimes (A_1 F)$, (iii) the two atoms taken separately, having traced over the field, $A_1 \otimes A_2$, and (iv) one of the atoms times the field, having traced over the other atom, $A_1 \otimes F \equiv A_2 \otimes F$. We calculate how the tangle for each of these partitions evolves as a function of time under TCM Hamiltonian evolution using the formalism reviewed in Section 3.2. Taking the initial state to be a pure product state of the field with the atoms, we capture the key features of the tangle evolution by considering three classes of initial state vectors,

$$|e\rangle_{A_1} \otimes |e\rangle_{A_2} \otimes |n\rangle_F \equiv |ee, n\rangle, \tag{3.5a}$$

$$|ee, \alpha\rangle \text{ or } |gg, \alpha\rangle, \tag{3.5b}$$

and

$$\frac{1}{\sqrt{2}} (|eg\rangle + |ge\rangle) \otimes |\alpha\rangle \text{ or } \frac{1}{\sqrt{2}} (|gg\rangle + |ee\rangle) \otimes |\alpha\rangle, \tag{3.5c}$$

where $|g(e)\rangle$ denotes the ground (excited) state of the atom, $|n\rangle$ denotes a Fock state field with $n$ photons, and $|\alpha\rangle$ denotes a coherent state field with an average number of photons given by $\langle n \rangle$. The alternatives in Eqs. (3.5b) and (3.5c) arise from the fact that, in the limit of large $\langle n \rangle$, the evolution of all of the tangles are found to be identical for the two different initial atomic conditions, as shown below.

### 3.3.1 Field-ensemble and one atom-remainder tangles

Under the assumption that the system is in an overall pure state, we may calculate the tangles in partitions (i) and (ii) above by applying Eq. (2.24), with $\nu_A = \nu_B = 1$. Specifically,

$$\tau_{F(A_1 A_2)} = 2 \left[ 1 - \text{Tr} \left( \rho_F^2 \right) \right] = 2 \left[ 1 - \text{Tr} \left( \rho_{A_1 A_2}^2 \right) \right], \tag{3.6}$$

and

$$\tau_{A_1(A_2 F)} = 2 \left[ 1 - \text{Tr} \left( \rho_{A_1}^2 \right) \right] = 2 \left[ 1 - \text{Tr} \left( \rho_{A_2 F}^2 \right) \right], \tag{3.7}$$

where we have used the fact that the (nonzero) eigenvalue spectra of the two marginal density operators for a bipartite division of a pure state are identical [2, 42] in obtaining the rightmost equalities. These tangles have implications for the quantum control of atomic ensembles. Because the overall system is pure, any correlation between the field and the ensemble is necessarily in one-to-one correspondence with the amount of entanglement between these two subsystems. The quantum backaction on the ensemble due to measurement of the field is thus quantified by Eq. (3.6). Alternatively, a measurement of one atom leads to backaction on the remaining subsystem as described by Eq. (3.7).

The time evolutions for each of the different tangles, corresponding to the initial conditions given by Eqs. (3.5a) - (3.5c), are shown in Figs. 3.1(a) - 3.3(a) respectively. Figs. 3.1(b) - 3.3(b) show the time evolution of the atomic inversion, defined as the probability of finding both atoms in the excited state minus the probability of finding both atoms in the ground state, for reference purposes. Appendix A contains the

## 3.3. Bipartite Tangles in the Two-Atom TCM

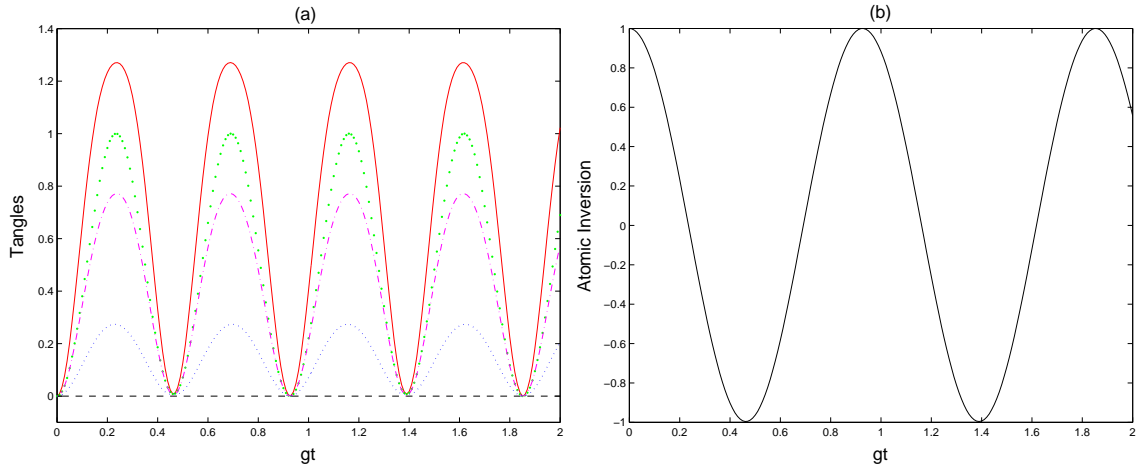

Figure 3.1: TCM evolution for both atoms initially in the excited state and the field in an initial Fock state with $n = 10$. (a) Solid curve (red): Field-ensemble tangle $\tau_{F(A_1A_2)}$; Large-dotted curve (green): One atom-remainder tangle $\tau_{A_1(A_2F)}$; Dashed curve (black): Atom-atom tangle $\tau_{A_1A_2}$. (Note that the atom-atom tangle is always zero for the given initial condition.); Dot-dashed curve (pink): Single atom-field tangle $\tau_{A_1F}$; Dotted curve (blue): Residual tangle $\tau_{A_1A_2F}$. (b) Atomic inversion of the ensemble.

Mathematica code used to numerically evaluate the evolution of all of the different quantities shown in these figures.

We find, under certain conditions, that the two stretched states in Eq. (3.5b) lead to identical evolution for the tangles in all of the bipartite partitions of the system, corresponding to the curves shown in Fig. 3.2(a). Similarly, the two symmetric states given in Eq. (3.5c) both yield the curves shown in Fig. 3.3(a). This behavior can be derived under a set of highly accurate approximations. In the limit of large average photon number, an initial coherent state field with zero phase will remain approximately separable from the atomic ensemble in an eigenstate of $J_x \equiv J_+ + J_-$ up to times on the order of $\langle n \rangle /g$ where, in the pseudospin picture, $J_\pm \equiv \sigma_\pm^{(1)} + \sigma_\pm^{(2)}$. This follows immediately from the time evolution operator generated by $H_{int}$ in Eq. (3.3) in the interaction picture. The key observation is that, for a macroscopic field, the removal or addition of a single photon has a negligible effect. This allows one to approximate the time evolution operator by $\exp^{-iH_{int}t/\hbar} \approx \exp^{-ig\sqrt{\langle n \rangle}J_x t}$. Thus, the eigenstates of $J_x$ form a convenient basis to use in describing the state of the atomic ensemble. This approach was taken by Gea-Banacloche in analyzing the behavior of the single atom Jaynes-Cummings model [94] and the generation of macroscopic superposition states [95], and extended to the multi-atom TCM by Chumakov, *et. al.*, [96, 97, 98].

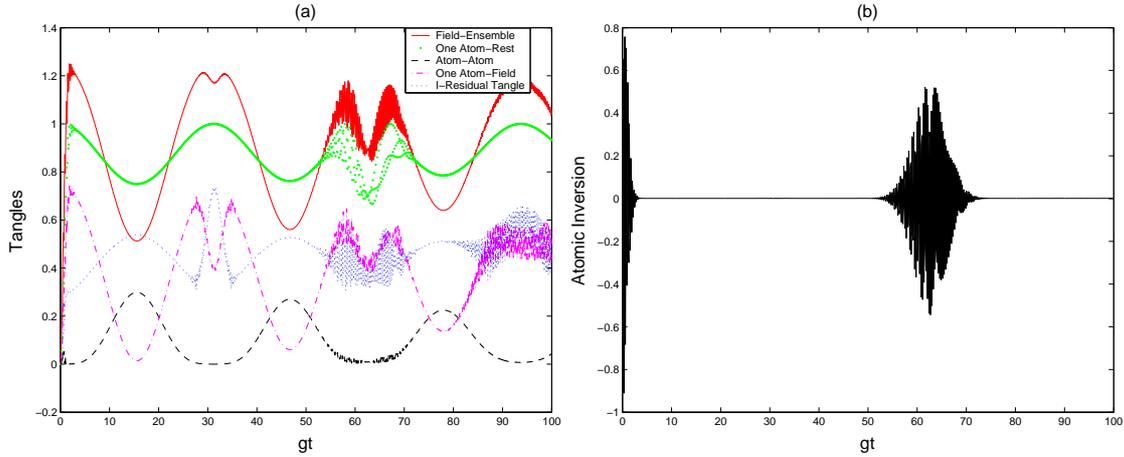

Figure 3.2: TCM evolution for both atoms initially in a stretched state and the field in an initial coherent state with $\langle n \rangle = 100$. (a) Solid curve (red): Field-ensemble tangle $\tau_{F(A_1A_2)}$; Large-dotted curve (green): One atom-remainder tangle $\tau_{A_1(A_2F)}$; Dashed curve (black): Atom-atom tangle $\tau_{A_1A_2}$; Dot-dashed curve (pink): Single atom-field tangle $\tau_{A_1F}$; Dotted curve (blue): I-Residual tangle $\tau_{A_1A_2F}$. (b) Atomic inversion of the ensemble.

We take as the appropriate basis the three symmetric eigenstates of $J_x$, which we label by $m = -1, 0,$ and $1$; the singlet state, $J = 0$, is a dark state and thus does not couple to the field. Writing the initial state of the system as

$$|\psi(0)\rangle = \sum_{m=-1}^{1} d_m |m\rangle \otimes |\alpha\rangle, \qquad (3.8)$$

and using the factorization approximation [96], we find that the state of the system

## 3.3. Bipartite Tangles in the Two-Atom TCM

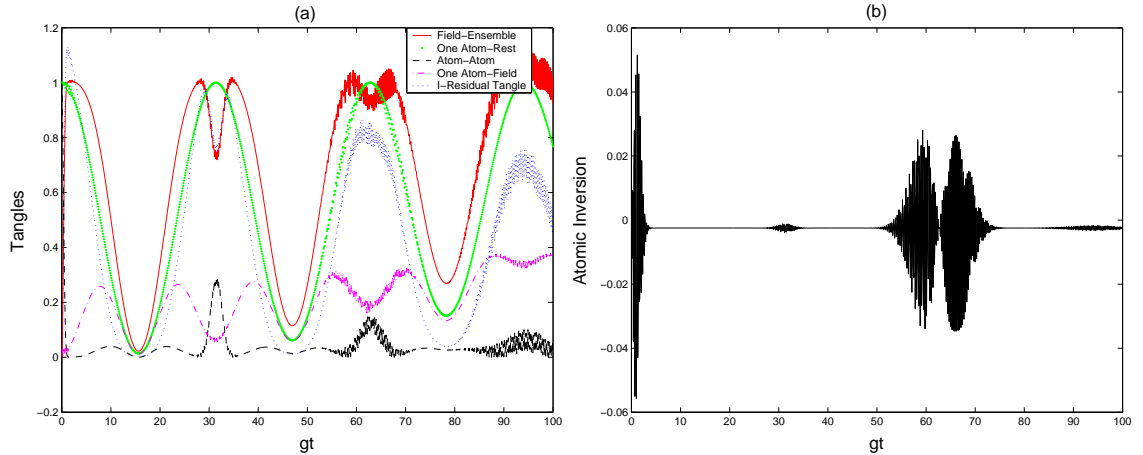

Figure 3.3: TCM evolution for the atoms initially in a symmetric state and the field in an initial coherent state with $\langle n \rangle = 100$. (a) Solid curve (red): Field-ensemble tangle $\tau_{F(A_1A_2)}$; Large-dotted curve (green): One atom-remainder tangle $\tau_{A_1(A_2F)}$; Dashed curve (black): Atom-atom tangle $\tau_{A_1A_2}$; Dot-dashed curve (pink): Single atom-field tangle $\tau_{A_1F}$; Dotted curve (blue): I-Residual tangle $\tau_{A_1A_2F}$. (b) Atomic inversion of the ensemble.

up to times on the order of $\langle n \rangle /g$ is given by

$$|\psi(t)\rangle \approx \sum_{m=-1}^{1} d_m |A_m(t)\rangle \otimes |\phi_m(t)\rangle, \tag{3.9}$$

where $|A_m(t)\rangle$ and $|\phi_m(t)\rangle$ are the time-evolved atomic and field states, respectively. The marginal density operator for the two atoms is then

$$\rho_{A_1A_2}(t) \approx \sum_{l,m} d_l^* d_m |A_m(t)\rangle \langle A_l(t)| f_{ml}(gt, \langle n \rangle), \tag{3.10}$$

where $f_{ml}(gt, \langle n \rangle) \equiv \sum_n \langle n | \phi_m(t) \rangle \langle \phi_l(t) | n \rangle$. We find that this function has "memory" only for $t \ll \sqrt{\langle n \rangle}/g$, and behaves very much like a delta function for longer time scales. Effectively, the large dimensional Hilbert space of the field acts as a broadband reservoir for the atoms – the generalization of the familiar "collapse" phenomenon in the Jaynes-Cummings model. This "Markoff" approximation is valid up to times on the order of $2\pi\sqrt{\langle n \rangle}/g$, corresponding to the well-known revival time in the Jaynes-Cummings Model [94]. Making this approximation in Eq. (3.10), the states $|A_m(t)\rangle$ act effectively as a "pointer basis" for decoherence [10] of the atomic density matrix, i.e.,

$$\rho_{A_1 A_2}(t) \approx \sum_m |d_m|^2 |A_m(t)\rangle \langle A_m(t)|. \tag{3.11}$$

Substituting this formula into Eq. (3.6) yields

$$\tau_{F(A_1 A_2)}(t) \approx 2\left\{1 - \frac{1}{4}[c - h(t')]\right\}, \tag{3.12}$$

where

$$\begin{aligned} c &\equiv 4\left(|d_{-1}|^4 + |d_0|^4 + |d_1|^4\right) + 2|d_0|^2 |d_1|^2 \\ &+ |d_{-1}|^2 \left(2|d_0|^2 + 3|d_1|^2\right) - 4|d_{-1}|^2 |d_1|^2 \end{aligned} \tag{3.13}$$

$$h(t') \equiv 2|d_0|^2 \left(|d_{-1}|^2 + |d_1|^2\right) \cos(4t') + |d_{-1}|^2 |d_1|^2 \cos(8t'), \tag{3.14}$$

and

$$t' \equiv \frac{gt}{2\sqrt{\langle n\rangle - \frac{N}{2} + \frac{1}{2}}}. \tag{3.15}$$

Under the factorization and Markoff approximations, the field-ensemble tangle is given by a constant term $c$ that depends only on the initial probabilities to find the atomic ensemble in each of the $J_x$ eigenstates, and a time-dependent piece $h(t')$. These probabilities depend solely on the absolute squares of the expansion coefficients of the atomic state given by Eq. (3.8). It is now clear why certain initial atomic conditions result in identical evolution for the different tangles. For example, the atomic states $|gg\rangle$ and $|ee\rangle$ both satisfy

$$|d_{-1}| = |d_1| = \frac{1}{2} \qquad \text{and} \qquad |d_0| = \frac{1}{\sqrt{2}}, \tag{3.16}$$

3.3. *Bipartite Tangles in the Two-Atom TCM*

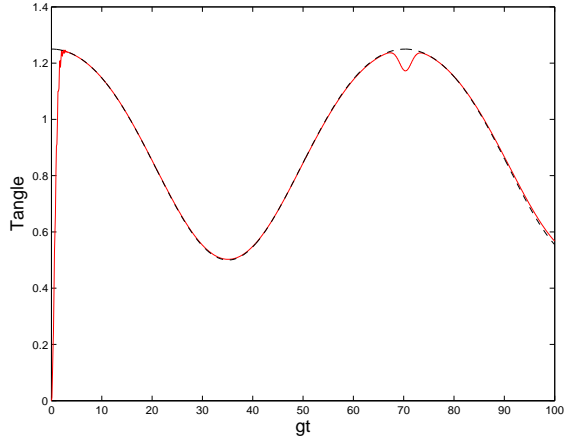

Figure 3.4: Exact field-ensemble tangle: Solid (red) curve, and approximate formula: Dashed (black) curve for an initial stretched atomic state and coherent state field with $\langle n \rangle = 500$.

corresponding to identical evolution for all of the tangles shown in Fig. 3.2(a). Similarly, the atomic states $1/\sqrt{2}\left(|eg\rangle + |ge\rangle\right)$ and $1/\sqrt{2}\left(|ee\rangle + |gg\rangle\right)$ both satisfy

$$|d_{-1}| = |d_1| = \frac{1}{\sqrt{2}} \qquad \text{and} \qquad |d_0| = 0, \qquad (3.17)$$

corresponding to the curves shown in Fig. 3.3(a). More generally, this property holds for any class of initial states $\left|\psi^{(i)}(0)\right\rangle$ having the form of Eq. (3.8) such that $\left|d_m^{(i)}\right| = \left|d_m^{(j)}\right|$, $m \in \{-1, 0, 1\}$. One immediate consequence of this result is that the relative phase information encoded in the initial state of the atomic system is irrelevant to the evolution of the field-ensemble tangle.

The field-ensemble tangle calculated according to Eq. (3.6) and the approximation given by Eq. (3.12) for an initial stretched atomic state and an initial coherent state field with $\langle n \rangle = 500$ are shown by the solid (red) and dashed (black) curves in Fig. 3.4, respectively. The approximation is seen to track the exact evolution extremely well over the range of its validity. The discrepancy at very small times is explained by the fact that at these times the Markoff approximation breaks down. It is also seen that our approximate solution does not capture the small dip in the field-ensemble tangle occurring at $t = \pi\sqrt{\langle n \rangle}/g$. The absence of this feature can be explained by noting that in making the Markoff approximation we have effectively wiped out any information regarding the initial coherence between the $|m = -1\rangle$ and $|m = 1\rangle$ states. The presence of this dip is then seen to be dependent upon the existence of this coherence. This is borne out by the fact that the dip in the field-ensemble tangle in Fig. 3.2(a) is much shallower than that in Fig. 3.3(a), where the initial atomic expansion coefficients are given by Eqs. (3.16) and (3.17), respectively.

### 3.3.2 Atom-atom tangle

Given an initial state, we time-evolve the system according to the dynamics governed by Eq. (3.3), and then trace over the field subsystem. The tangle of the two-atom mixed state $\rho_{A_1 A_2}(t)$ may then be calculated according to Eq. (2.23). The resulting atom-atom tangles corresponding to the initial conditions in Eqs. (3.5a) - (3.5c) are depicted by the dashed (black) curves in Figs. 3.1(a) - 3.3(a), respectively. These curves yield direct insight into the state of the atomic ensemble as a function of time. Specifically, the atom-atom tangle quantifies the degree to which the ensemble behaves as a collective entity, rather than as two individual particles.

It is somewhat surprising that for the initial condition given by Eq. (3.5a), i.e., when the field is initially in a Fock state with any value for $n$, the atom-atom tangle remains zero at all times, whereas the evolution of the atom-atom tangle resulting from an initial coherent state field is nontrivial and, in general, nonzero. In order to better understand these observations, we have performed a preliminary investigation into the evolution of the atom-atom tangle for initial conditions other than those given by Eqs. (3.5a) - (3.5c). When the field is initially in a Fock state and both atoms start in the ground state, the loss of an excitation in the field can result in the

*3.3. Bipartite Tangles in the Two-Atom TCM*

creation of an excitation in the atomic ensemble. This produces entanglement between the field and the ensemble and in the single atom-field and one atom-remainder partitions. Since it is not possible to distinguish in which atom the excitation is created, the two atoms become entangled with each other as well. It is found that the atom-atom entanglement falls off as $1/n^2$ so that, in the limit of a highly excited Fock state, these initial conditions yield results reminiscent of those found in Fig. 3.1(a). Specifically, we find that the entanglement in all of the different subsystem partitions always oscillate in phase at twice the Rabi frequency, and that the atom-atom tangle approaches zero as $n$ becomes large.

Next, we considered the case when both atoms initially reside in a stretched state, and the initial field state consists of a coherent superposition of two neighboring Fock states. We find, on a time scale much longer than that given by the inverse of the associated Rabi frequencies, that the overall behavior again closely resembles the evolution seen for an initial field consisting of a single Fock state. Specifically, we find that the general features of all of the different bipartite tangles oscillate in phase with one another. However, on much shorter time-scales, the effects of dephasing between the two Rabi frequencies become apparent, yielding the first clues regarding how the observed coherent state behavior arises in terms of initial Fock state superpositions. At the revival time, when there is a partial rephasing of the Rabi frequencies, all of the bipartite tangles decrease simultaneously, while at other times the tangles in certain bipartite partitions may be completely out of phase with one another. It seems likely that a detailed investigation of these initial findings will provide a quantitative explanation for the large scale Fock-like behavior seen in Fig. 3.2.

### 3.3.3  Single atom-field tangle

The final bipartite partition of the two-atom TCM is that consisting of a single atom, say $A_1$, as one subsystem and the field $F$ as the second subsystem. Again by exchange symmetry $\tau_{A_1F} = \tau_{A_2F}$, so we need calculate only one of these quantities. Because the tripartite system is in an overall pure state, the Schmidt decomposition theorem [2, 42] implies that the marginal density operator $\rho_{A_1F}$ has at most rank two. The rank of the reduced density matrix is set by the dimension of the smallest subsystem, which in this case is a two-level atom. This is exactly the scenario envisioned by Osborne [52], as described in Section 2.1.3. The tangle corresponding to this partition, $A_1 \otimes F$, is computed by first tracing over the state of the remaining atom, $A_2$, and then applying Eq. (2.26). Employing this procedure,

$$\tau_{A_1F} = \text{Tr}\left(\rho_{A_1F}\widetilde{\rho}_{A_1F}\right) + 2\lambda_{min}^{(A_1F)}\left[1 - \text{Tr}\left(\rho_{A_1F}^2\right)\right], \tag{3.18}$$

where $\lambda_{min}^{(A_1F)}$ represents the minimum eigenvalue of the Osborne $M$ matrix [52] generated from the marginal density operator $\rho_{A_1F}$. The dot-dashed (pink) curves in Figs. 3.1(a) - 3.3(a) give the time evolution of the single atom-field tangle for the different initial conditions considered.

We are now in possession of closed forms for the tangles of all bipartite partitions of the two-atom TCM. Any other entanglement that the system may possess must necessarily be in the form of irreducible three-body quantum correlations. In Section 3.4 we review the residual tangle formalism introduced by Coffman, *et. al.*, in order to quantify this type of tripartite entanglement in a system of three qubits. We then propose a generalization of this quantity that is applicable to a $2 \otimes 2 \otimes D$ system in an overall pure state. This extension of the tangle formalism allows us to study the phenomenon of entanglement sharing in the two-atom TCM.

## 3.4 Entanglement Sharing and the Residual Tangle

Coffman, *et. al.*, analyze the phenomenon of entanglement sharing for a system of three qubits $A$, $B$, and $C$ in an overall pure state in full generality by introducing a quantity known as the residual tangle [27]. This definition is motivated by the observation that the tangle of $A$ with $B$ plus the tangle of $A$ with $C$ cannot exceed the tangle of $A$ with the joint subsystem $BC$, i.e.,

$$\tau_{AB} + \tau_{AC} \leq \tau_{A(BC)}. \tag{3.19}$$

Here, $\tau_{AB}$ and $\tau_{AC}$ are calculated according to Eq. (2.23), and $\tau_{A(BC)}$ may be obtained from Eq. (2.24) with $\nu_A = \nu_B = 1$.

The original proof [27] of the inequality in Eq. (3.19), which forms the heart of the phenomenon of entanglement sharing for the case of three qubits, may be substantially simplified by making use of certain results due to Rungta, *et. al.* Specifically, we note that [51]

$$\mathrm{Tr}\left(\rho_{xy}\widetilde{\rho}_{xy}\right) = 1 - \mathrm{Tr}\left(\rho_x^2\right) - \mathrm{Tr}\left(\rho_y^2\right) + \mathrm{Tr}\left(\rho_{xy}^2\right) \geq 0 \tag{3.20}$$

for subsystems $x$ and $y$ having arbitrary Hilbert space dimensions. Under the assumption that $x$ and $y$ are in an overall pure state with a third subsystem $z$, Eq. (3.20) may be rewritten

$$\mathrm{Tr}\left(\rho_{xy}\widetilde{\rho}_{xy}\right) = 1 - \mathrm{Tr}\left(\rho_x^2\right) - \mathrm{Tr}\left(\rho_y^2\right) + \mathrm{Tr}\left(\rho_z^2\right) \geq 0, \tag{3.21}$$

where we have used the equality of the nonzero eigenvalue spectra of $\rho_{xy}$ and $\rho_z$. Then, by the observation [27] that for an arbitrary state of two qubits $A$ and $B$, the following upper bound on the tangle defined by Eq. (2.23) holds

$$\tau_2\left(\rho_{AB}\right) \leq \mathrm{Tr}\left(\rho_{AB}\widetilde{\rho}_{AB}\right), \tag{3.22}$$

and by Eq. (2.24) with $\nu_A = \nu_B = 1$, the inequality in Eq. (3.19) follows immediately.

Subtracting the terms on the left hand side of Eq. (3.19) from that on the right hand side yields a positive quantity referred to as the *residual tangle*

$$\tau_{ABC} \equiv \tau_{A(BC)} - \tau_{AB} - \tau_{AC}. \tag{3.23}$$

The residual tangle is interpreted as quantifying the inherent tripartite entanglement present in a system of three qubits, i.e., the entanglement that cannot be accounted for in terms of the various bipartite tangles. This interpretation is given further support by the observation that the residual tangle is invariant under all possible permutations of the subsystem labels [27].

We wish to generalize the residual tangle, defined for a system of three qubits, to apply to a $2 \otimes 2 \otimes D$ quantum system in an overall pure state so that we may study entanglement sharing in the two-atom TCM. Note that we already have all of the other tools needed for such an analysis. Specifically, from Section 3.3, we know the analytic forms for all of the different possible bipartite tangles in such a system.

Any proper generalization of the residual tangle must, at a minimum, be a positive quantity, and be equal to zero if and only if there is no tripartite entanglement in the system, i.e., if and only if all of the quantum correlations can be accounted for using only bipartite tangles. It should also reduce to the definition of the residual tangle in the case of three qubits. Further it is reasonable to require, if this is to be a true measure of irreducible three-body correlations, that symmetry under permutation of the subsystems be preserved, and that it remain invariant under local unitary operations. Finally, we conjecture that this quantity satisfies the requirements (2.2) and (2.3) for being an entanglement monotone [99, 35] under the set of stochastic local operations and classical communication (SLOCCs), or equivalently, under the set of invertible local operations [64]. We limit the monotonicity requirement to this restricted set of operations since, in the context of entanglement sharing, we are

*3.4. Entanglement Sharing and the Residual Tangle*

only concerned with LOCCs that preserve the local ranks of the marginal density operators such that all subsystem dimensions remain constant.

Let $A$ and $B$ again be qubits, and let $C$ now be a $D$-dimensional system with the composite system $ABC$ in an overall pure state. We note that, under these assumptions, we are still capable of evaluating each of the terms on the right hand side of Eq. (3.23) analytically using the results of Section 3.3. However, we cannot simply use the definition of the residual tangle (with $C$ now understood to represent a $D$-dimensional system) as the proper generalization for two reasons. First, since the three subsystems are no longer of equal dimension, symmetry under permutations of the subsystems is lost. However, as we will see, this problem is easily fixed by explicitly enforcing the desired symmetry. The second, and more difficult problem to overcome is the fact that the inequality given by Eq. (3.19) no longer holds for our generalized system because $\lambda_{min}$ in Eq. (2.26) can be negative, implying that Eq. (3.23) can also be negative.

The required permutation symmetry may be restored by taking our generalization of the residual tangle, dubbed the *I-residual tangle* [68] in reference to previous work, to be

$$\tau_{ABC} \equiv \frac{1}{3}\left[\tau_{A(BC)} + \tau_{B(AC)} + \tau_{C(AB)} - 2\left(\tau_{AB} + \tau_{AC} + \tau_{BC}\right)\right]. \quad (3.24)$$

The definition in Eq. (3.24) is obtained by averaging over all possible relabelings of the subsystems in Eq. (3.23). By inspection, it is obvious that Eq. (3.24) preserves permutation symmetry. However, it still suffers from the problem that its value can be negative. In order to deal with this difficulty, we make use of the arbitrary scale factors appearing in Eqs. (2.24) and (2.25).

Let $d$ be the smaller of the two 'dimensions' of two arbitrary dimensional subsystems $x$ and $y$, i.e., $d \equiv \min\{D_x, D_y\}$. Note that by dimension we do not necessarily mean the total Hilbert space dimension of the physical system under consideration, but only the number of different Hilbert space dimensions *that contribute to the formation* of the overall pure state of the system. This is a subtle but important point which automatically enforces insights like those due to Rungta, *et. al.*, [51] and Verstraete, *et. al.*, [100] which state that the scale chosen for a measure of entanglement must be invariant under the addition of extra, but unused, Hilbert space dimensions. The two-atom TCM provides one example of the relevant physics underlying these ideas.

Consider, for example, the bipartite partitioning of the TCM into a field subsystem with $D_F = \infty$, and an ensemble subsystem consisting of the two qubits with $D_{A_1 A_2} = 4$. Any entangled state of the overall system has a Schmidt decomposition with at most four terms, implying that the field effectively behaves like a four-dimensional system. Further, since the Tavis-Cummings Hamiltonian given by Eq. (3.3) does not induce couplings between the field and the singlet state of the atomic ensemble, i.e., the singlet state is a dark state, the field behaves effectively as a three-level system, or *qutrit*, in the context of the TCM. Accordingly, in any entangled state of the field with the ensemble, the field is considered to have a dimension no greater than three. We employ this revised definition of dimension throughout the remainder of the paper.

We now make the choice

$$\nu_A \nu_B = \frac{d}{2}, \qquad (3.25)$$

when calculating each of the bipartite tangles appearing on the right hand side of Eq. (3.24). This choice is made for several reasons. First of all, it is in complete agreement with the two qubit case, yielding $\nu_A \nu_B = 1$ as required. Indeed, when $A$, $B$, and $C$ are all qubits, the residual tangle given by Eq. (3.23) is recovered. Secondly, it takes differences in the Hilbert space dimensions of the subsystems into account when setting the relevant scale for each tangle. This is important since, in order to study the phenomenon of entanglement sharing, the tangles for each of the

*3.4. Entanglement Sharing and the Residual Tangle*

different bipartite partitions must be compared on a common scale. It is reasonable that this scale be a function of the smaller of the two subsystem dimensions since, for an overall pure state, it is this quantity that limits the number of terms in the Schmidt decomposition. Finally, it is conjectured that a proper rescaling of the various tangles will result in the positivity of Eq. (3.24).

Note that when applying the proposed rescaling to the terms on the right hand side of Eq. (3.24), the only term affected is $\tau_{C(AB)}$, which is rescaled by one-half of the smaller of the two subsystem dimensions $D_C$ and $D_{AB}$. Each of the other terms remains unaltered since, in each case, at least one of the two subsystems involved is a qubit. The net effect of this rescaling is to increase the 'weight' of the tangle between $C$ and $AB$ relative to that of the rest of the tangles. This is reasonable when one recognizes that both $AB$, a system of two qubits, and $C$, a $D$-dimensional system (in the case $D > 2$), have *entanglement capacities* [28] exceeding that of a single qubit.

The requirement that the I-residual tangle be invariant under local unitary operations follows trivially, since each term on the right hand side of Eq. (3.24) is known to satisfy this property individually. It is still an open question as to whether or not the proposed rescaling is sufficient to preserve positivity when generalizing the residual tangle, Eq. (3.23), to the I-residual tangle, Eq. (3.24). However, numerical calculations give strong evidence that this is the case. The I-residual tangle has been calculated for over two-hundred million randomly generated pure states of a $2 \otimes 2 \otimes 3$ system and of a $2 \otimes 2 \otimes 4$ system, the only nontrivial possibilities. In each instance the resulting quantity has been positive. We conjecture that the I-residual tangle satisfies the requirements of positivity and monotonicity under SLOCC not only for a $2 \otimes 2 \otimes D$ system, where closed forms currently exist for all of the terms on the right hand side of Eq. (3.24), but for the most general $D_A \otimes D_B \otimes D_C$ dimensional tripartite system in an overall pure state (with the proper scaling of each term again given by Eq. (3.25)). The I-residual tangle arising in the context of the two-atom TCM is shown by the blue curves in Figs. 3.1(a) - 3.3(a).

The residual tangle, as well as our proposed generalization of this quantity, may be interpreted as a measure of the irreducible tripartite entanglement in a system since it cannot be accounted for in terms of any combination of bipartite entanglement measures [27]. A slightly different, and possibly more enlightening interpretation is that the I-residual tangle quantifies the amount of freedom that a system has in satisfying the constraints imposed by the phenomenon of entanglement sharing. If the I-residual tangle of a tripartite system is zero, then each bipartite tangle is uniquely determined by the values of all of the other bipartite tangles. Alternatively, if $\tau_{ABC}$ is strictly greater than zero, then the bipartite tangles enjoy a certain latitude in the values that each may assume while still satisfying the positivity criterion. The larger the value of the I-residual tangle, the more freedom the system has in satisfying the entanglement sharing constraints. This reasoning highlights the relationship between entanglement sharing and the I-residual tangle.

Finally, we may interpret the I-residual tangle as the *average fragility* of a tripartite state under the loss of a single subsystem. That is, if one of the three subsystems is selected at random and discarded (or traced over), then the I-residual tangle quantifies the amount of three-body entanglement that is lost, on average. It is the existence of physically meaningful interpretations such as these which prompt us to postulate this new measure of tripartite entanglement for a $2 \otimes 2 \otimes D$ system in an overall pure state, rather than to rely on previously defined measures based on normal forms [100] or on the method of hyperdeterminants [101], for example. At this point it is unclear what, if any connection these entanglement monotones have to the entanglement that exists in different bipartite partitions of the system, a key ingredient in any discussion of entanglement sharing.

The constraint imposed by entanglement sharing on the values of the various

*3.4. Entanglement Sharing and the Residual Tangle*

bipartite tangles, each of which is known to be a positive function, is simply that Eq. (3.24) cannot be negative. It then follows that the strongest constraint of this form is placed on the two-atom TCM when the I-residual tangle is equal to zero. This occurs (to a good approximation) periodically in Fig. 3.1(a) for the initial condition given by Eq. (3.5a). It is at these points that each bipartite tangle is uniquely determined in terms of the values of all of the other bipartite tangles. Conversely, at one half of this period when the I-residual tangle achieves its maximum value, the various bipartite partitions enjoy their greatest freedom with respect to how entanglement may be distributed throughout the system while still satisfying the entanglement sharing constraints. The distribution of correlations is, of course, still determined by the initial state of the system and by the TCM time evolution, both of which we consider to be separate constraints.

Similarly, the dotted (blue) curves in Figs. 3.2(a) and 3.3(a) show the evolution of the residual tangle for the initial states given by Eqs. (3.5b) and (3.5c), respectively. Note how the more complicated behavior resulting from an initial coherent state field arises from a specific superposition of Fock states, the tangles of which all have a simple oscillatory evolution. This suggests that the phenomenon of entanglement sharing may offer a useful perspective from which to investigate the way in which the coherent state evolution results from a superposition of Fock state evolutions.

The fact that the TCM Hamiltonian leads to a nonzero I-residual tangle is interesting in its own right. Inspection of Eq. (3.3) shows that this model does not include a physical mechanism, e.g., a dipole-dipole coupling term enabling direct interaction between the two atoms in the ensemble, but only for coupling between the field and the atoms. Consequently, all interactions between the atoms are mediated by the electromagnetic field via the exchange of photons, and are in some sense indirect. This, however, turns out to be sufficient to allow genuine tripartite correlations to develop in the system as evidenced by values of the I-residual tangle that are strictly greater than zero.

## 3.5 Summary and Future Directions

The two-atom Tavis-Cummings model provides the simplest example of a collection of two-level atoms, or qubits, sharing a common coupling to the electromagnetic field. A detailed understanding of the evolution of entanglement in different bipartite partitions of this model is valuable for both fundamental theoretical investigations, and for accurately describing the behavior of certain nontrivial, yet experimentally realizable systems. Our proposed generalization of the residual tangle augments the current formalism, and enables one to analyze the irreducible three-body correlations that arise in a broader class of tripartite systems, providing a tool useful for studying the phenomenon of entanglement sharing in the context of a physically relevant and accessible system.

One possible extension of this work would be to generalize this analysis to include ensembles with an arbitrary number of atoms. This would entail further extensions of the tangle formalism in order to quantify both the entanglement in a mixed state of a bipartite system of arbitrary dimensions with rank greater than two, and the multipartite entanglement in a system with more than three subsystems. Completely characterizing multipartite entanglement is a difficult and as yet unsolved problem [27, 68, 46]. Thus, a comprehensive investigation of the phenomenon of entanglement sharing in the TCM when $N > 2$ is not currently feasible. However, using the results of Section 2.2 one may calculate a lower bound on the I-tangle of a mixed state of an arbitrary bipartite system that provides valuable information regarding the distribution of entanglement among the different bipartite partitions of the TCM when $N > 2$ [102].

Because it is an entanglement monotone, the quantity $\mathcal{L}_\tau(\rho)$ given by Eq. (2.64),

3.5. *Summary and Future Directions*

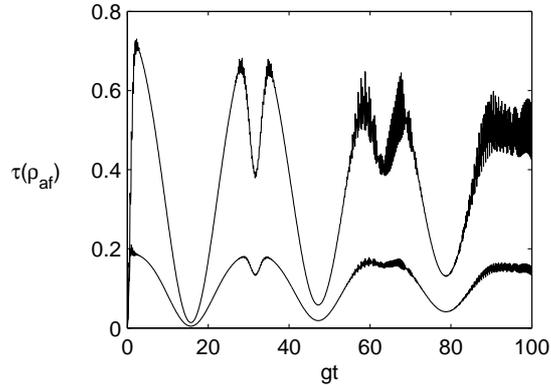

Figure 3.5: Analytic values of $\tau(\rho_{AF})$ (upper curve) and the corresponding lower bound $\mathcal{L}_\tau(\rho_{AF})$ (lower curve) in the two-atom TCM.

captures all of the qualitative features of the I-tangle. Fig. 3.5, which shows the I-tangle between a single atom and the field in the two atom TCM, as well as the corresponding lower bound, demonstrates this agreement. Thus, even though we may not, in general, be capable of analytically evaluating the I-tangles for certain bipartite partitions of the TCM when $N > 2$, we may still capture the qualitative features of the entanglement arising in these partitions by using $\mathcal{L}_\tau(\rho)$. Such calculations provide initial insight into the distribution of entanglement and the constraints imposed by entanglement sharing in the TCM with an arbitrary number of atoms.

Ultimately, one would hope to connect this analysis to the phenomenon of quantum backaction on individual particles when the whole ensemble is measured. The tradeoff between the information gained about a system and the disturbance caused to that system is certainly fundamental to quantum mechanics [69, 70, 71]. However, the relationship of this tradeoff to multiparticle entanglement is far from clear. Such an understanding would not only be a crucial step in designing protocols for the quantum control of ensembles, but would also provide deeper insight into the complementary relationship between correlations and correlated entities that we feel lies at the heart of quantum mechanics [103]. The next chapter places this intuition on a firm foundation by quantifying a tradeoff between bipartite entangled correlations and information about individual members of multi-qubit systems.

# Chapter 4

# Quantitative Complementarity Relations

## 4.1 Introduction

We now move from a discussion of entanglement and its properties to a discussion of complementarity which, we will argue, is an even more fundamental concept in quantum mechanics. Quantitative relations are introduced in this chapter which imply that many of the counterintuitive features of entangled quantum systems including: (i) the fact that maximal information about a composite quantum system does not necessarily imply maximal information about the component subsystems and (ii) the phenomenon of entanglement sharing, can be understood as specific consequences of complementarity in composite systems. This, in turn, has profound implications for the philosophical foundations of quantum mechanics.

Complementarity is perhaps the most important phenomenon distinguishing systems that are inherently quantum mechanical from those that may accurately be treated classically. Accordingly, a thorough understanding of this concept is of fun-damental importance in determining how to properly interpret quantum mechanics [1], as well as in studying the potential usefulness of quantum systems for enhancing specific information processing protocols.

Niels Bohr introduced the term complementarity to refer to the fact that information about a quantum object obtained under different experimental arrangements cannot always be comprehended within a single causal picture [4]. We identify the classical world with precisely those systems and processes for which it is possible to unambiguously combine the space-time coordinates of objects with the dynamical conservation laws that govern their mutual interactions. However, in the more general setting of quantum mechanics, complementarity precludes the existence of such a picture. It was this insight that led Bohr to consider complementarity to be the natural generalization of the classical concept of causality [4].

An alternative statement of complementarity, which makes no reference to experimental arrangements or measurements, states that a quantum system may possess properties that are equally real, but mutually exclusive. The wave-particle duality exhibited by a photon in a double-slit experiment [6] and the tradeoff between the uncertainties in the position and momentum of a subatomic particle governed by Heisenberg's relation [7] are two examples of complementarity in single quantum systems. The study of complementarity in composite systems has a fairly short history by comparison. Nevertheless important progress has been made, especially in the context of two-particle interferometers, where complementarity relations between single and two-particle fringe visibilities [104, 105], between distinguishability and visibility [106], and between the coherence and predictability [107] in a quantum eraser [92] are known and have been experimentally verified [108, 109, 110]. Additionally, Jaeger, *et al.*, [111] have recently derived a complementarity relation between multipartite entanglement and mixedness for specific classes of $N$-qubit systems.

Jakob and Bergou [112] took a major step forward by deriving a complementarity

*4.1. Introduction*

relation, valid for an arbitrary pure state of two qubits, which accounts for (and in some cases, generalizes) many of the main results put forward in [104, 105, 106, 107]. They showed that an arbitrary normalized pure state $|\psi\rangle$ of a two-qubit system satisfies the expression [112]

$$C_2^2(\psi) + \nu_k^2 + p_k^2 = 1. \tag{4.1}$$

The first term on the left hand side of Eq. (4.1) is the square of the concurrence given by Eq. (2.14), an inherently bipartite quantity which was shown to be equivalent to the two-qubit tangle in Section 2.1.3 .

The remaining two terms on the left hand side of Eq. (4.1) are the squares of single particle properties associated with qubit $k = 1, 2$. The first of these properties is the coherence $\nu_k$ of qubit $k$, which quantifies, e.g., the fringe visibility in the context of a two-state system incident on an interferometer. Defining the marginal density operator $\rho_k \equiv \text{Tr}_j(|\psi\rangle\langle\psi|)$ for $j \neq k$, the coherence is given by

$$\nu(\rho_k) \equiv 2\left|\text{Tr}\left(\rho_k \sigma_+^{(k)}\right)\right|, \tag{4.2}$$

where $\sigma_+^{(k)}$ is the raising operator acting on qubit $k$. Similarly, the predictability $p$ which quantifies the *a priori* information regarding whether qubit $k$ is in the state $|0\rangle$ or the state $|1\rangle$, e.g., whether it is more likely to take the upper or lower path in an interferometer, is given by

$$p(\rho_k) \equiv \left|\text{Tr}\left(\rho_k \sigma_z^{(k)}\right)\right|. \tag{4.3}$$

Here, $\sigma_z$ is a Pauli operator and $|0\rangle$ ($|1\rangle$) is the plus (minus) one eigenvector of $\sigma_z$.

Jakob and Bergou note that Eq. (4.1) becomes an inequality when applied to a mixed state of two qubits [112]. Here we generalize their result in two ways: (i) to apply to an arbitrary pure state of $N$ qubits, and (ii) to apply to an arbitrary state of two qubits, pure or mixed [103]. These expressions shed light on a wide range of topics in quantum information theory, including the highly investigated connection between entanglement and mixedness [66, 113, 114, 115, 116, 117] about which it was recently written that, "even for two qubits, the smallest nontrivial bipartite quantum system, the relation between entanglement and mixedness remains a fascinating open question" [117]. Our results lead to a complete characterization of the relationship between these two quantities in exactly this case.

The remainder of this chapter is as follows. In Section 4.2 we derive our two generalizations of Eq. (4.1) from a single common insight and interpret the various quantities involved. One immediate implication of our work is an explicit relationship between the residual tangle and the tangles of the different two-qubit marginals in a pure state of three qubits. The resulting expression illustrates a tradeoff between the various single particle properties, the bipartite entanglement, and the inherent three-body quantum correlations encoded in the tripartite quantum state, effectively reducing the phenomenon of entanglement sharing to a specific instance of complementarity in this system.

Next we derive a quantity, which we dub the separable uncertainty, that arises naturally in the context of our second complementarity relation, and show that it is a good measure of the uncertainty due to ignorance in a quantum state. The introduction of this quantity completes the two-qubit picture and enables a comprehensive analysis of the relationship between entanglement and mixedness in such systems. The section ends with several examples designed to illustrate the usefulness of our generalized relations. Finally, we discuss potential applications of our results, as well as some interpretational issues, in Section 4.3.

## 4.2 Multi-Qubit Complementarity Relations

Our two generalizations of Eq. (4.1) both hinge on the observation (which may be verified by direct computation) that the expression

$$M(\rho_k) = \frac{1}{2} - \overline{S^2}(\rho_k) \tag{4.4}$$

holds for an arbitrary state of a single qubit. Here, $M(\rho_k) \equiv 1 - \text{Tr}(\rho_k^2)$ is the marginal mixedness of qubit $k$, and $\overline{S^2}(\rho_k) \equiv 1/2\left[\nu^2(\rho_k) + p^2(\rho_k)\right]$ is the average of the squares of the single qubit properties defined by Eqs. (4.2) and (4.3). The mixedness, or linear entropy of a quantum state [118], varies continuously from zero for pure states to its maximum value of $1/2$ for the completely mixed state. The quantity $\overline{S^2}(\rho_k)$ is found to be invariant under local unitary operations performed on qubit $k$, and is therefore taken to be a measure of the single particle properties encoded in $\rho_k$. According to Eq. (4.4) then, the marginal mixedness quantifies our uncertainty regarding the properties possessed by a single qubit. Further, this expression leads to the relation

$$\sum_{k=1}^{N}\left[M(\rho_k) + \overline{S^2}(\rho_k)\right] = \frac{N}{2}, \tag{4.5}$$

when summed over an arbitrary system of $N$ qubits, and implies a tradeoff between the single particle properties in such a system and our uncertainty regarding these properties.

Consider now the case where the $N$ qubits are in an overall pure state. As we have already seen, the Schmidt decomposition theorem expressed by Eq. (2.8) implies that the marginal density operators describing the two subsystems resulting from a bipartite partitioning of this system have the same nonzero eigenvalue spectra. In particular, this holds for $\rho_k$ and $\rho_{\{R_k\}}$, where $\rho_k$ is the marginal state of the $k^{th}$ qubit and $\rho_{\{R_k\}}$ is the marginal state of the $N-1$ qubits in the set $R_k \equiv$ $\{1, 2, \ldots, k-1, k+1, \ldots, N-1, N\}$. From Eq. (2.24) with $\nu_A = \nu_B = 1$ it then follows that $2M(\rho_k) = \tau_{k\{R_k\}}$. Combined with Eq. (4.5), this yields

$$\sum_{k=1}^{N} \left[ \tau_{k\{R_k\}} + 2\overline{S^2}(\rho_k) \right] = N. \tag{4.6}$$

Equation (4.6) shows that a complementary relationship exists between the single particle properties $\overline{S^2}(\rho_k)$ and the bipartite quantum correlations $\tau_{k\{R_k\}}$.

As an interesting application of this expression, consider a pure state of three qubits. Expressing Eq. (4.6) (in the case $N = 3$) in terms of the symmetric form of the residual tangle given by Eq. (3.24) yields the relation

$$\tau_{123} + \frac{2}{3} \left[ \tau_{12} + \tau_{13} + \tau_{23} + \overline{S^2}(\rho_1) + \overline{S^2}(\rho_2) + \overline{S^2}(\rho_3) \right] = 1. \tag{4.7}$$

It then follows that the entanglement sharing behavior exhibited by this system, as described in Section 1.3, is essential for ensuring consistency with Eq. (4.7). This expression governs the underlying complementarity that exists between the potentially available information about individual subsystems, the bipartite entanglements in the marginal two-qubit states, and the irreducible tripartite quantum correlations. We conjecture that this connection between entanglement sharing and complementarity is a general feature of composite quantum systems.

The derivation of our second generalization of Eq. (4.1) makes use of Eq. (3.20), from which it immediately follows that

$$\mathrm{Tr}(\rho\tilde{\rho}) + M(\rho) = M(\rho_1) + M(\rho_2) \tag{4.8}$$

for an arbitrary state $\rho$ of two qubits. The quantity $\mathrm{Tr}(\rho\tilde{\rho}) + M(\rho)$ thus provides an alternative way of calculating the uncertainty regarding single particle properties encoded in such states. Substituting Eq. (4.8) into Eq. (4.5) in the case $N = 2$ yields

$$\mathrm{Tr}(\rho\tilde{\rho}) + M(\rho) + \overline{S^2}(\rho_1) + \overline{S^2}(\rho_2) = 1. \tag{4.9}$$

*4.2. Multi-Qubit Complementarity Relations*

The explicit role played by the sum of the first two terms in Eq. (4.9) is best illustrated by an example. The invariance under local unitary operations of both $\text{Tr}(\rho\tilde{\rho})$ and $M(\rho)$ allows us to consider, without loss of generality, density matrices $\rho_c$ in the computational basis of the form:

$$\rho_c = \begin{pmatrix} \omega_1 & a & a & e \\ a^* & \omega_2 & f & a \\ a^* & f^* & \omega_3 & a \\ e^* & a^* & a^* & \omega_4 \end{pmatrix}, \qquad (4.10)$$

where $0 \leq \omega_i \leq 1$ and $\sum_i \omega_i = 1$. Equation (4.10) is obtained by reducing the number of free parameters in an arbitrary density matrix from fifteen to nine via the six free parameters in a tensor product of two single-qubit unitary operators. In this representation the individual coherences for the two qubits are equal, i.e., $\nu\left(\rho_c^{(k)}\right) = 4|a|$, $k = 1, 2$. Accordingly, $\text{Tr}(\rho\tilde{\rho}) = \text{Tr}(\rho_c\tilde{\rho}_c) = 2\left(|e|^2 + |f|^2 - 4|a|^2 + \omega_2\omega_3 + \omega_1\omega_4\right)$, and $M(\rho) = M(\rho_c) = 1 - 2\left(|e|^2 + |f|^2 + 4|a|^2\right) - \sum_{i=1}^{4}\omega_i^2$. Some algebra then yields

$$\text{Tr}(\rho\tilde{\rho}) + M(\rho) = \sum_{i=1}^{4}\sigma_i^2 - 2C_{14} - 2C_{23} - \frac{1}{2}\sum_{k=1}^{2}\nu^2\left(\rho_c^{(k)}\right), \qquad (4.11)$$

where $\sigma_i^2 = \omega_i(1-\omega_i)$ is the variance of $\omega_i$ in the single trial frequencies that result from a measurement in the computational basis, $C_{ij} = -\omega_i\omega_j$ is the similarly defined covariance between $\omega_i$ and $\omega_j$, and $1/2\sum_{k=1}^{2}\nu^2\left(\rho_c^{(k)}\right)$ is the average of the squared coherences. The variances measure the spreads or uncertainties associated with the multinomial distribution $\{\omega_i\}$, the covariances are directly related to predictability information that is preserved by the spin-flip operation, and the average squared coherence quantifies the information encoded in the coherences of the individual qubits. Thus, $\text{Tr}(\rho\tilde{\rho}) + M(\rho)$ is the total uncertainty in the distribution $\{\omega_i\}$ minus the available information about properties possessed by the individual subsystems, in complete agreement with Eq. (4.8). The form of $\rho$ given by Eq. (4.10) makes this relationship readily apparent; however it holds for an arbitrary density operator due to the invariance of each term in Eq. (4.9) under local unitary operations.

Equation (4.9) also enables us to make a connection with the work of Jaeger, *et al.* [111] who showed that the following expression (in our notation) holds for an arbitrary state of $N$ qubits,

$$\text{Tr}(\rho\tilde{\rho}) + M(\rho) = I(\rho, \tilde{\rho}). \tag{4.12}$$

Here, $\tilde{\rho} \equiv \sigma_y^{\otimes N} \rho^* \sigma_y^{\otimes N}$ is the natural generalization of the spin-flip operation. The quantity $I(\rho, \tilde{\rho})$, referred to as the indistinguishability, is defined in terms of the Hilbert-Schmidt distance $D_{HS}(\rho - \rho') \equiv \sqrt{\frac{1}{2}\text{Tr}\left[(\rho - \rho')^2\right]}$ between two density matrices $\rho$ and $\rho'$ to be [111]

$$I(\rho, \tilde{\rho}) \equiv 1 - D_{HS}^2(\rho - \tilde{\rho}). \tag{4.13}$$

This quantity measures the indistinguishability of the state $\rho$ from the operator $\tilde{\rho}$, and thus serves as a measure of the spin-flip symmetry of the state. Further, Eqs. (4.8) and (4.12) imply that, at least in the two-qubit case, the indistinguishability also represents the total uncertainty in the quantum state regarding single particle properties.

Combining Eqs. (4.9) and (4.12) in the case $N = 2$ yields

$$I(\rho, \tilde{\rho}) + \overline{S^2}(\rho_1) + \overline{S^2}(\rho_2) = 1, \tag{4.14}$$

which implies a complementary relationship between information about properties possessed by the individual qubits and the spin-flip symmetry of the state. Substituting Eq. (4.13) into Eq. (4.14) then leads to the following relationship between single particle properties and the Hilbert-Schmidt distance between the density operator and its spin-flip,

$$D_{HS}(\rho - \tilde{\rho}) = \sqrt{\overline{S^2}(\rho_1) + \overline{S^2}(\rho_2)}. \tag{4.15}$$

*4.2. Multi-Qubit Complementarity Relations*

Equation (4.15) suggests a geometric picture in which the root mean square values $\sqrt{S^2(\rho_k)}$ of the single particle properties act like coordinates in the space of two-qubit density matrices, and shows that the distance between the quantum state and its spin-flip is determined solely by these local properties. Hence, our results yield a method of investigating the abstract space of two-qubit density matrices with simple Euclidean geometry.

Next, in order to determine the role played by entanglement in Eq. (4.9), we note that an arbitrary two-qubit density matrix may always be written in its unique optimal Lewenstein-Sanpera decomposition [119]

$$\rho = \lambda \rho_s + (1-\lambda) |\psi_e\rangle \langle \psi_e|, \tag{4.16}$$

where $\rho_s = \sum_i p_i \rho_1^{(i)} \otimes \rho_2^{(i)}$, $(0 \leq p_i \leq 1, \sum_i p_i = 1)$ is a separable density matrix, $|\psi_e\rangle$ is an entangled pure state, and $\lambda \in [0,1]$ is maximal. Calculating the quantity $\text{Tr}(\rho\tilde{\rho})$ using this representation, one finds that

$$\text{Tr}(\rho\tilde{\rho}) = \text{Tr}(\lambda^2 \rho_s \tilde{\rho}_s) + 2\lambda(1-\lambda)\text{Re}\langle \tilde{\psi}_e | \rho_s | \tilde{\psi}_e \rangle + (1-\lambda)^2 \left|\langle \psi_e | \tilde{\psi}_e \rangle\right|^2. \tag{4.17}$$

The first term in Eq. (4.17) quantifies that part of $\rho_s$ which is preserved under the spin-flip operation, and the second quantifies the (real) part of $|\tilde{\psi}_e\rangle$ that overlaps with $\rho_s$. Hence, neither of these terms involve entanglement. On the other hand, the last term in Eq. (4.17) is directly related to the quantum correlations in the system as we now demonstrate.

Written in the form of Eq. (4.16), all of the entanglement in the two-qubit state $\rho$ is concentrated in the pure state $|\psi_e\rangle$ as quantified by the expression [119]

$$C(\rho) = (1-\lambda)C(\psi_e). \tag{4.18}$$

Recalling the definition of the pure state concurrence given by Eq. (2.14) and that the squared concurrence is equivalent to the tangle for a system of two qubits, Eq. (4.18)

implies that

$$\tau(\rho) = (1-\lambda)^2 \left|\left\langle \psi_e \mid \tilde{\psi}_e \right\rangle\right|^2. \tag{4.19}$$

Thus, the last term in Eq. (4.17) represents the entanglement in the state $\rho$ as quantified by the mixed state tangle given in Eq. (2.25) with $\nu_A = \nu_B = 1$.

The two fundamental sources of uncertainty regarding single particle properties in composite quantum systems are: (i) ignorance of their values, and (ii) partial to total exclusion of these properties due to the presence of entanglement. Recalling from Eq. (4.8) that $\text{Tr}(\rho\tilde{\rho}) + M(\rho)$ is a measure of the total uncertainty regarding single particle properties in the quantum state, Eqs. (4.17) and (4.19) imply that the quantity

$$\eta(\rho) = \text{Tr}(\rho\tilde{\rho}) + M(\rho) - \tau(\rho), \tag{4.20}$$

$0 \le \eta(\rho) \le 1$, is a good measure of the *separable uncertainty*, or uncertainty due to ignorance (rather than to the presence of entanglement), in an arbitrary state of two qubits. For example, since $\text{Tr}\left(\psi\tilde{\psi}\right) = \tau(\psi)$ and $M(\psi) = 0$, we see from Eq. (4.20) that $\eta(\psi) = 0$, demonstrating that pure states contain no separable uncertainty. Similarly, $\eta(I/4) = 1$ for the completely mixed state, implying that the uncertainty in this case is maximal, and that this state encodes no information regarding either single particle properties or bipartite correlations. Finally, consider the maximally entangled states for fixed marginal mixednesses $\rho_m$ given by [117]

$$\rho_m = \begin{pmatrix} x_1 & 0 & 0 & \sqrt{x_1 x_2} \\ 0 & 0 & 0 & 0 \\ 0 & 0 & 1-x_1-x_2 & 0 \\ \sqrt{x_1 x_2} & 0 & 0 & x_2 \end{pmatrix}, \tag{4.21}$$

with $0 \le x_1, x_2 \le 1$ and $x_1 + x_2 \le 1$. We find that in this case $\eta(\rho_m) = M(\rho_m)$, i.e., the separable uncertainty is simply equal to the mixedness.

## 4.2. Multi-Qubit Complementarity Relations

Equation (4.20) completely characterizes the highly investigated connection between entanglement and mixedness in two-qubit systems by relating these quantities to the separable uncertainty and spin-flip invariance encoded in $\rho$. Likewise, combining Eqs. (4.8) and (4.20) yields

$$\eta(\rho) = M(\rho_1) + M(\rho_2) - \tau(\rho). \tag{4.22}$$

This alternative form of $\eta(\rho)$ quantifies the relationship between entanglement and the marginal mixednesses of the individual qubits.

Our second generalization of Eq. (4.1) is finally obtained by combining Eqs. (4.9) and (4.20), yielding

$$\eta(\rho) + \tau(\rho) + \overline{S^2}(\rho_1) + \overline{S^2}(\rho_2) = 1. \tag{4.23}$$

This expression shows that an arbitrary state of two qubits exhibits a complementary relationship between the amounts of separable uncertainty, entanglement, and information about single particle properties that it encodes. Further, it reduces to Eq. (4.1) for a pure state $|\psi\rangle$, and has the desirable property that each term is separately invariant under local unitary operations.

The following examples are adapted from [111] in order to highlight the additional insights provided by Eq. (4.23) over previous analyses. Consider the set of states for which $I(\rho, \tilde{\rho}) = 0$, or equivalently, for which $D_{HS}(\rho - \tilde{\rho}) = 1$. These states are maximally distinguishable from their spin-flips. From Eqs. (4.12) and (4.20) we see that $\tau(\rho) = M(\rho) = 0$ which implies that this class of states is equivalent to the set of separable pure states. Equation (4.23) confirms that in this case $\overline{S^2}(\rho_1) + \overline{S^2}(\rho_2) = 1$, i.e., states of this type possess maximal single particle properties and no bipartite correlations nor separable uncertainty.

Next, consider the class of states for which $M(\rho) = I(\rho, \tilde{\rho})$, i.e., for which the mixedness represents the total uncertainty about single particle properties. Equations (4.12) and (4.20) then imply that $\tau(\rho) + \eta(\rho) = M(\rho)$, which illustrates the precise relationship between entanglement and mixedness for states of this form. Further, from Eqs. (4.13) and (4.15) and the fact that the purity $P(\rho) = 1 - M(\rho)$, we surmise that $P(\rho) = \overline{S^2}(\rho_1) + \overline{S^2}(\rho_2)$, yielding an explicit geometric relationship between the purities and the allowable single particle properties of these states.

As a final example, we consider the states that possess perfect spin-flip symmetry. These are the states for which $\rho = \tilde{\rho}$, or by Eq. (4.13) for which $I(\rho, \tilde{\rho}) = 1$, and hence, $D_{HS}(\rho - \tilde{\rho}) = 0$. Equation (4.15) then implies that $\overline{S^2}(\rho_1) = \overline{S^2}(\rho_2) = 0$, yielding the result that no state with perfect spin-flip symmetry may encode any information about single particle properties.

A specific class of states satisfying these conditions are the Werner states $\rho_w$ [120]

$$\rho_w(\lambda) = \lambda |\text{Bell}\rangle \langle \text{Bell}| + \frac{1-\lambda}{4} I_2 \otimes I_2, \tag{4.24}$$

where $0 \leq \lambda \leq 1$, $|\text{Bell}\rangle$ represents one of the four Bell states, and $I_2$ is the identity operator for a single qubit. The Werner states vary continuously from the completely mixed state ($\lambda = 0$) to a maximally entangled state ($\lambda = 1$), and are known to be separable for $\lambda \leq 1/3$ [57]. It is a simple matter to show that the states given by Eq. (4.24) satisfy the condition that $\rho_w = \tilde{\rho}_w$. Equation (4.23) then implies that $\eta(\rho_w) = 1 - \tau(\rho_w)$ demonstrating, among other things, that all separable Werner states are associated with the maximum amount of separable uncertainty, independent of $\lambda$.

Jaeger, *et al.* claim that $\text{Tr}(\rho\tilde{\rho})$ is a good measure of multipartite entanglement, and therefore state that $\text{Tr}(\rho_w\tilde{\rho}_w) + M(\rho_w) = 1$ quantifies the relationship between entanglement and mixedness for the Werner states [111]. However, $\text{Tr}(\rho_w\tilde{\rho}_w)$ fails to satisfy the requirements for being an entanglement monotone [35], since it does not assign the same value to all of the separable Werner states. Indeed our results show that, when considering the class of Werner states the relevant tradeoff occurs not between entanglement and mixedness, but instead between entanglement and separable uncertainty.

## 4.3 Discussion and Future Directions

The examples presented above demonstrate that entanglement and separable uncertainty are quite similar in many respects. For instance, both quantities are related to information in the quantum state (or a lack thereof) which is preserved under the spin-flip operation. Further, both are invariant under local unitary operations, implying that they measure properties which are independent of the choice of local bases. Finally, Eq. (4.23) shows that both quantities share a complementary relationship with the properties of the local subsystems as well as with one another.

There are also important differences between entanglement and the separable uncertainty quantified by $\eta(\rho)$. First of all, the separable uncertainty vanishes for all pure states, while entanglement is both a pure and mixed state phenomenon. Further, as is well known, entanglement cannot be increased on average by local operations and classical communication (LOCC) [35], while this restriction does not hold for separable uncertainty where we are always allowed to throw away or 'forget' information. Finally, entanglement quantifies the information that we possess regarding the existence of quantum correlations, whereas separable uncertainty quantifies a lack of information about the individual subsystems.

We conclude from these observations that the entanglement $\tau(\rho)$, and the single particle properties $\overline{S^2}(\rho_1)$ and $\overline{S^2}(\rho_2)$, are the three fundamental and mutually complementary attributes of a two-qubit system about which we may possess information that does not depend on our choice of local bases. This, in turn, suggests an interpretation for the tangle as the fiducial measure of uncertainty regarding individual subsystem properties due to the presence of entanglement, rather than to our ignorance. Equivalently, because of the relationship between uncertainty and information [25, 16], the tangle also quantifies the amount of information directly encoded in the quantum correlations of the system.

When dealing with systems composed of more than two subsystems, entanglement sharing becomes possible. Equation (4.7) implies that, at least in the simplest case of a pure state of three qubits, this phenomenon also has its roots in complementarity; this time in terms of a tradeoff between the allowed single particle, bipartite, and tripartite information that such a system may encode. Verifying the conjecture that entanglement sharing in arbitrary composite systems is generally a consequence of complementarity requires the extension of relations such as Eq. (4.6) to multipartite systems with subsystems of arbitrary dimension. This in turn requires the identification of the appropriate multipartite generalization of the residual tangle as well as a determination of which of the possible partitions of such a system contribute to these relations and how to quantify them.

Because of these difficulties it is, at present, unclear how to further generalize our results. Fortunately, several potential applications of the complementarity relations presented here readily suggest themselves. Beyond fully investigating the relationship between entanglement and mixedness made explicit by Eq. (4.20), or, perhaps more interestingly, between entanglement and the individual subsystem mixednesses given by Eq. (4.22), our results also seem well-suited to formulating an information vs. disturbance tradeoff relation (see [71] and references therein) for two-qubit systems. The complementary behavior exhibited by these systems implies that, loosely speaking, a certain amount $\eta(\rho)$ of additional information regarding single particle properties may be obtained through observation without affecting the entanglement in the system. However, if one tries to obtain more information than this, then by Eq. (4.23) the entanglement must decrease. This behavior leads us to conjecture that complementarity between bipartite and single particle properties plays a fundamental role in the information-disturbance tradeoff phenomenon in composite

*4.3. Discussion and Future Directions*

systems.

A quantitative relation describing the information-disturbance tradeoff in an arbitrary state of two qubits would, e.g., prove useful for incorporating a measurement, feedback, and control loop into the two-atom Tavis-Cummings model analyzed in the previous chapter. Eventually, one would hope to generalize the complementarity relation in Eq. (4.23) to apply to arbitrary multipartite systems, and to extract from this a corresponding information-disturbance tradeoff relation governing these many-body systems. This would then enable the investigation of feedback and control on an arbitrarily large ensemble of two-level atoms in the context of the TCM. The main obstacle to such an extension of the current formalism once again appears to be the task of quantifying multipartite entanglement.

Finally, our generalized complementarity relations also suggest one possible way of thinking about the quantum state of a system from an information-theoretic point of view. Much has been written about the so-called Bayesian interpretation, which considers the quantum state to be a representation of our subjective knowledge about a quantum system [24]. One advantage of this interpretation is that the collapse of the wave function [22] is viewed not as a real physical process, but simply represents a change in our state of knowledge. However, it is unclear what this knowledge pertains to since, from this perspective, we are generally prohibited from associating objective properties with individual systems. This situation becomes even more confusing if one also contends that a qubit encodes in-principle information i.e. that information is physical [121], since the Bayesian interpretation fails to make a distinction between this type of information and the subjective knowledge of an observer. Equations (4.6) and (4.23) provide some insight regarding these observations, especially in the context of two-qubit systems.

We first assume, in agreement with the Bayesian interpretation, that the analysis of any such system must begin with our subjective human knowledge. Accordingly, we assign a quantum state to the system representing this knowledge. Associated with this quantum state assignment is a value for $\eta(\rho)$ which quantifies our subjective uncertainty regarding the in-principle information encoded by the two-qubit system. In this context Eq. (4.23) implies that, the smaller our separable uncertainty, the greater our ability to indirectly access and/or manipulate this in-principle information via the locally unitarily invariant bipartite correlations $\tau(\rho)$ and single particle properties $\overline{S^2}(\rho_k)$, about which we possess subjective information. However, even when we are able to assign a pure state to the quantum system such that $\eta(\psi) = 0$, Eq. (4.6) substantiates the observation that "maximal information is not complete and cannot be completed" [24]. This is a direct consequence of the complementary relationships that exist between (i) the single particle properties $\nu(\rho_k)$ and $p(\rho_k)$ of the individual subsystems, and (ii) between the total localized attributes $\overline{S^2}(\rho_k)$ of the subsystems and the inherently bipartite entangled correlations $\tau_{k\{R_k\}}$.

The second relationship above makes explicit the often stated fact that maximal information about a composite quantum system does not necessarily entail maximal information about the component subsystems. This, in turn, suggests that such systems possess the unique ability to encode information directly into entangled correlations. The next chapter investigates the possible importance of these directly encoded correlations in the performance of pure state quantum computation.

# Chapter 5

# Entanglement and Quantum Computation

## 5.1 Introduction

Bell's theorem [12] codifies the observation that entangled quantum-mechanical systems exhibit stronger correlations than are achievable with any local hidden-variable (LHV) model. As alluded to in Section 1.3, the ability to operate outside the constraints imposed by local realism serves as a resource for many information processing tasks such as communication [122] and cryptography [32].

The role of entanglement in quantum computation [2] is less clear, for the issue is not one of comparing quantum predictions to a *local* realistic description, but rather one of comparing a quantum computation to the *efficiency* of a realistic simulation. Nevertheless, various results indicate some connection between entanglement and computational power [123, 124]. Entanglement is a necessity if a pure-state quantum computer is to have scalable physical resources [125]. Moreover, systems with limited entanglement can often be efficiently simulated classically [126]. Jozsa and Linden [127] showed that if the entanglement in a quantum computer extends only to some fixed number of qubits, independent of problem size, then the computation can be simulated efficiently on a classical computer.

Despite these results, global entanglement is by no means sufficient for achieving an exponential quantum advantage in computational efficiency [128]. The set of Clifford gates (Hadamard, Phase, and CNOT) acting on a collection of $N$ qubits, each initialized to the state $|0\rangle$, can generate globally entangled states, yet according to the Gottesman-Knill (GK) theorem [2, 39], the outcomes of all measurements of products of Pauli operators on these states can be simulated with $O(N^2)$ resources [129] on a classical computer. The GK theorem is an expression of properties of the $N$-qubit Pauli group $\mathcal{P}_N$ [2], which consists of all products of Pauli operators multiplied by $\pm 1$ or $\pm i$: the allowed (Clifford) gates preserve $\mathcal{P}_N$, and the allowed measurements are the Hermitian operators in $\mathcal{P}_N$.

One approach to understanding the information processing capabilities of entangled states is to translate a quantum protocol involving entanglement into an equivalent protocol that utilizes only classical resources, e.g., the shared randomness of LHVs and ordinary classical communication. Toner and Bacon [130] showed that the quantum correlations arising from local projective measurements on a maximally entangled state of two qubits can be simulated exactly using a LHV model augmented by just a single bit of classical communication. Pironio [131] took this analysis a step further, showing that the amount of violation of a Bell inequality imposes a lower bound on the average communication needed to reproduce the quantum-mechanical correlations.

Working along these lines, Hardy [132] developed a local toy theory that allows for a nontrivial form of teleportation to occur. In this way, the ability to perform teleportation and the nonlocality exhibited by entangled quantum systems were shown to be two distinct phenomena. More generally, Spekkens developed a local model

*5.1. Introduction*

capable of reproducing several phenomena usually considered to be inherently quantum mechanical in nature including: interference, teleportation, dense coding, no cloning, no broadcasting, the noncommutativity of measurements, and many others [133]. Of course, due to the use of LHVs, neither of the above models is capable of reproducing all of the measurement predictions of quantum mechanics.

Taking a similar approach, we analyze the classical resources required to simulate measurements made on two important classes of globally entangled states: the $N$-qubit GHZ states [37] (also called "cat states"), and the (one- or two-dimensional) cluster states of $N$ qubits [38]. Specifically, we present a LHV model, augmented by classical communication, that simulates the quantum-mechanical predictions for measurements of arbitrary products of Pauli operators on these states. In each case the simulation is efficient since the required amount of communication scales linearly with the number of qubits.

The ability to perform such a simulation for an $N$-qubit GHZ state is surprising for a couple of reasons. First of all, using only a subset of the measurements in $\mathcal{P}_N$ on these states it is possible to demonstrate the incompatibility of the predictions of a LHV model with those of quantum mechanics deterministically, i.e., with only a single measurement [37]. Secondly, there exist Bell-type inequalities for $N$-qubit GHZ states which are violated by an amount that grows exponentially with $N$ [134]. Our results show that a simulation of the correlations that give rise to this exponential violation of local realism may be performed using solely classical resources that grow at most linearly with the number of qubits.

The results obtained for the simulation of the cluster states are even more enlightening. The structure of our model yields insight into the GK theorem, a result which goes a long way toward clarifying the role that global entanglement plays in pure state quantum computation. Specifically, we show that the correlations in the set of nonlocal hidden variables represented by the stabilizer generators [2, 39] that are tracked in the GK algorithm are captured by an appropriate set of local hidden variables augmented by $N-2$ bits of classical communication. This fact has profound consequences for our understanding of the necessary ingredients for achieving an exponential quantum advantage in computational efficiency. These implications are fully discussed towards the end of the chapter.

The remainder of this chapter is as follows. In Section 5.2 we briefly review Mermin's version of the three qubit GHZ argument which demonstrates the incompatibility of this state with the existence of a LHV model. Next, we introduce an efficient classical-communication-assisted local model in the context of this example, and show that it accounts for all of the quantum mechanical features considered by Mermin. We then generalize our result by summarizing the rules for simulating the creation of an $N$-qubit GHZ state, and for computing measurement predictions from the model. The section concludes with a proof showing that our simulation yields the correct quantum mechanical measurement predictions for all possible products of Pauli operators on these states.

Section 5.3 investigates the relationship between the Gottesman-Knill theorem and $N$-qubit cluster states that are subjected to single-qubit Pauli measurements. We begin by using our model to simulate the creation of an arbitrary cluster state in Section 5.3.1. The resulting ability to correctly predict the outcomes of all measurements in $\mathcal{P}_N$ on a two-dimensional cluster state of $N$ qubits makes it possible to efficiently simulate any GK circuit that will fit on the cluster. Section 5.3.2 elucidates the connection between Clifford gates implemented in the cluster state architecture and in our simulation procedure, and presents a consistent way of concatenating these simulated gates in order to model arbitrary GK circuits with our formalism. Finally, we discuss the implications of our results and suggest some directions for further research in Section 5.4.

## 5.2 Simulation of GHZ correlations

If we assume that locality is respected by quantum systems [8], then the violation of Bell-type inequalities demonstrates the in-principle failure of LHV models to account for all of the predictions of quantum mechanics. This violation is, however, a statistical phenomenon requiring multiple runs in order to generate the necessary statistics. On the other hand, a GHZ state of three or more qubits violates the assumption of the existence of LHVs deterministically [37]. The following is a brief review of Mermin's simplification of the GHZ argument [135], cast in the language of the stabilizer formalism, demonstrating this fact.

### 5.2.1 Deterministic violation of local realism

The GHZ state of three qubits $|\psi_3\rangle$ is given by

$$|\psi_3\rangle = \frac{1}{\sqrt{2}}\left(|000\rangle + |111\rangle\right), \tag{5.1}$$

where the logical basis state $|0\rangle$ ($|1\rangle$) represents the eigenvector of the Pauli $Z$ operator with eigenvalue $+1$ ($-1$). This state is uniquely specified by a complete set of commuting operators $g_{\psi_3}$, one choice for which is

$$g_{\psi_3} = \langle -XYY, -YXY, -YYX \rangle, \tag{5.2}$$

i.e., $|\psi_3\rangle$ is the unique eigenvector that yields the result $+1$ with certainty for each of these measurements. In the context of the stabilizer formalism [2, 39], the elements of the set $g_{\psi_3}$ are referred to as *stabilizer generators* of the three-qubit GHZ state.

Consider now the following attempt at capturing the behavior of this quantum system with a LHV model. We assume that there exist elements of reality which specify the outcomes for all measurements of the form $M_j$, where $M \in \{X, Y\}$ is a Pauli operator (we do not need to consider measurement of $Z$ for this argument), and $j \in \{1, 2, 3\}$ labels the qubit to be measured. We represent these six elements of reality by $m_x^j$ and $m_y^j$, each possessing either the value $+1$ or the value $-1$. The only constraints on the distribution of these values follow immediately from the stabilizer generators of $|\psi_3\rangle$; agreement between the LHV model and quantum mechanics requires that $m_x^1 m_y^2 m_y^3 = m_y^1 m_x^2 m_y^3 = m_y^1 m_y^2 m_x^3 = -1$. Multiplying these three quantities together, and using the fact that $\left(m_y^j\right)^2 = 1$ for all $j$, we find that $m_x^1 m_x^2 m_x^3 = -1$. That is, the LHV model predicts the result $-1$ with certainty for a measurement of the observable $XXX$. However, it is straightforward to check that the product of the three stabilizer generators in Eq. (5.2) is equal to $XXX$, such that quantum mechanics predicts the result $+1$ for this measurement with certainty. Thus, whereas Bell demonstrated that the elements of reality inferred from one group of measurements are incompatible with the statistics produced by a second group of measurements, requiring multiple runs for the generation of these statistics, the GHZ argument demonstrates the incompatibility of a LHV model with the predictions of quantum mechanics with just a single measurement [135].

The previous example shows that no LHV model can account for all of the quantum mechanical predictions for measurements performed on the three-qubit GHZ state, even when the set of allowed measurements is restricted to products of Pauli operators. We now present a LHV model, supplemented by classical communication between the qubits, which accounts for all of the features of the GHZ state considered by Mermin. We then extend this model to apply to a general $N$-qubit GHZ state, and show that it yields all of the correct quantum mechanical predictions for measurements of arbitrary products of Pauli operators.

### 5.2.2 Three-qubit GHZ simulation

The three-qubit GHZ state is generated by the quantum circuit shown in Fig. 5.1.[1] In the language of the GK theorem, the evolution of the state is tracked by the evolution of the stabilizer generators. The Hadamard gate $H$ transforms the Pauli operators $X, Y, Z$ according to

$$HXH^\dagger = Z, \qquad HYH^\dagger = -Y, \qquad HZH^\dagger = X. \tag{5.3}$$

Similarly, under the action of CNOT gate $(C)$, we have

$$\begin{aligned} C\,(XI)\,C^\dagger &= XX\,, & C\,(YI)\,C^\dagger &= YX\,, & C\,(ZI)\,C^\dagger &= ZI\,, \\ C\,(IX)\,C^\dagger &= IX\,, & C\,(IY)\,C^\dagger &= ZY\,, & C\,(IZ)\,C^\dagger &= ZZ\,, \end{aligned} \tag{5.4}$$

where the first qubit is the control, the second is the target, and $I$ represents the identity operator. The stabilizer generators evolve through the circuit in Fig. 5.1 as

$$\langle ZII, IZI, IIZ \rangle \xrightarrow{H_1} \langle XII, IZI, IIZ \rangle \xrightarrow{CNOT_{12}} \langle XXI, ZZI, IIZ \rangle$$
$$\xrightarrow{CNOT_{13}} \langle XXX, ZZI, ZIZ \rangle. \tag{5.5}$$

The full final stabilizer consists of all unique products of the last set of generators in Eq. (5.5), including the joint observables $-XYY$, $-YXY$, $-YYX$, and $XXX$ that form the basis of Mermin's argument.

The GK description provides an efficient method for simulating the outcome of a measurement of any product of Pauli operators on the globally entangled state $|\psi_3\rangle$, but it does so by keeping track of the nonlocal stabilizer generators specifying the state. We replace this nonlocal resource with a local description, augmented by classical communication, by constructing a LHV table where each row represents a qubit, and each column represents a measurement. Locality is enforced by only allowing changes in rows corresponding to qubits that participate in an interaction.

---

[1] The quantum circuits in this document were typeset using the LaTeX package `Qcircuit`, available at http://info.phys.unm.edu/Qcircuit/.

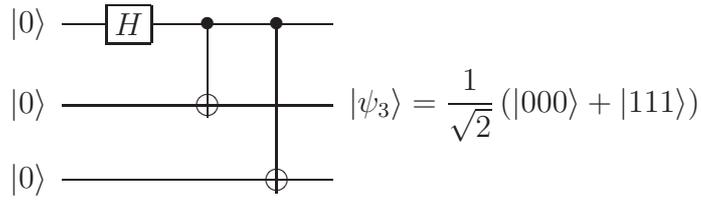

Figure 5.1: Circuit to generate the three-qubit GHZ state.

For the initial state $|000\rangle$, a measurement of $Z$ on any qubit yields $+1$ with certainty, and a measurement of $X$ or $Y$ yields $\pm 1$ with equal probabilities. Accordingly, the table

$$
\begin{array}{cccc}
 & X & Y & Z \\
\text{qubit 1} & R_1 & -iR_1 & 1 \\
\text{qubit 2} & R_2 & iR_2 & 1 \\
\text{qubit 3} & R_3 & iR_3 & 1
\end{array}
\tag{5.6}
$$

gives corresponding LHVs for this state, with $R_j$ denoting a classical random variable that returns $\pm 1$ with equal likelihood, and $j$ labeling the qubit to which the random variable refers. The reason for including a negative sign in the initial $Y$ entry for the first qubit will become clear in what follows.

The LHV table is read by choosing a measurement and multiplying the corresponding entries. The resulting product, with $i$ discarded whenever it appears, is the outcome predicted by the LHV model. For example, the measurement $Z \otimes Z \otimes X \equiv ZZX$ represents measuring $Z$ on the first two qubits and $X$ on the remaining qubit. Multiplying the corresponding entries in the above table, the resulting product is the fifty-fifty random result $R_3$, in agreement with the predictions of quantum mechanics.

The table given by Eq. (5.6) also contains a trivial column, which has not been shown, corresponding to leaving a qubit unmeasured (or, equivalently, to acting on

*5.2. Simulation of GHZ correlations*

a qubit with the identity operator $I$). Each entry in this column has the constant value one. This allows us to consider measurements on a subset of the qubits, e.g. $XXI$, yielding the result $R_1R_2$. Since the product of any number of $R$ variables with different subscripts is equivalent to a single random variable returning $\pm 1$ with equal probabilities, we find that this prediction is again consistent with quantum mechanics.

Now consider measurements involving entries in the $Y$ column, e.g. $ZYY$, which yields the result $-R_2R_3$. The minus sign that results from the product of the two imaginary phases in this example does nothing to change the fact that the result returned is still $\pm 1$ with equal likelihood, just as quantum mechanics requires. Similarly, suppose we were to measure the observable $YYY$. Discarding the imaginary phase that survives the multiplication of the entries in the $Y$ column, our model predicts the fifty-fifty random result represented by $R_1R_2R_3$, which is once again consistent with the predictions of quantum mechanics.

According to these rules, the LHV table in Eq. (5.6) yields the correct quantum-mechanical predictions for measurements of the $4^3 = 64$ products of Pauli operators on the state $|000\rangle$. The use of the imaginary phase $i$ in the model, apparently just a curiosity, actually plays a crucial role. It simulates some of the conflicting predictions of commuting LHVs and anticommuting quantum operators which form the basis of Mermin's GHZ argument [135]. In addition, our specific method for modeling each CNOT gate relies on the $X$ and $Z$ entries being real and the $Y$ entries being imaginary, as well as on the particular initial correlations that exist between the $X$ and $Y$ values for each qubit.

The first step in creating the three-qubit GHZ state is to apply the Hadamard gate to the first qubit. We extract rules for updating the LHV table from the transformations of the Pauli operators given in Eq. (5.3), which suggest that to simulate $H$ we should (i) swap the $X$ and $Z$ entries and (ii) flip the sign of the $Y$ entry of the transformed qubit. The resulting rules are given by

$$X^{\text{F}} = Z^{\text{I}}, \qquad Z^{\text{F}} = X^{\text{I}}, \qquad Y^{\text{F}} = -Y^{\text{I}}, \qquad (5.7)$$

where the superscripts 'I' and 'F', respectively, denote the initial and final values of the LHV table entry, before and after the application of a gate. Applying these rules to the first row of the table in Eq. (5.6) yields

$$\begin{array}{cccc} & X & Y & Z \\ \text{qubit 1} & 1 & iR_1 & R_1 \\ \text{qubit 2} & R_2 & iR_2 & 1 \\ \text{qubit 3} & R_3 & iR_3 & 1 \end{array} \qquad (5.8)$$

which returns the correct quantum-mechanical predictions for all measurements of Pauli products on the state $(|0\rangle + |1\rangle)|00\rangle/\sqrt{2}$. This is not surprising since the state remains a product state, and it is well known that a LHV model can be constructed for a single qubit [12]. The usefulness of our model only becomes apparent when we apply it to entangled states.

Applying the first CNOT gate in Fig. 5.1 yields the Bell entangled state

$$|\psi_2\rangle \otimes |0\rangle = \frac{1}{\sqrt{2}}(|00\rangle + |11\rangle)|0\rangle. \qquad (5.9)$$

In accordance with Eq. (5.4), we use the following rules to update the LHV table entries corresponding to the control $c$ and the target $t$ under the operation CNOT$[c, t]$:

$$\begin{aligned} X_c^{\text{F}} &= X_c^{\text{I}} X_t^{\text{I}}, & Y_c^{\text{F}} &= Y_c^{\text{I}} X_t^{\text{I}}, & Z_c^{\text{F}} &= Z_c^{\text{I}}, \\ X_t^{\text{F}} &= X_t^{\text{I}}, & Y_t^{\text{F}} &= Z_c^{\text{I}} Y_t^{\text{I}}, & Z_t^{\text{F}} &= Z_c^{\text{I}} Z_t^{\text{I}}. \end{aligned} \qquad (5.10)$$

Note that the update rules for $H$ and CNOT keep the $X$ and $Z$ entries real and the $Y$ entry imaginary, while the CNOT rule preserves the correlation $XYZ = i$ that holds for each qubit after the operation of the Hadamard gate. Applying the

## 5.2. Simulation of GHZ correlations

rules (5.10) to the first two rows of the table in Eq. (5.8) yields

$$
\begin{array}{cccc}
 & X & Y & Z \\
\text{qubit 1} & R_2 & iR_1R_2 & R_1 \\
\text{qubit 2} & R_2 & iR_1R_2 & R_1 \\
\text{qubit 3} & R_3 & iR_3 & 1
\end{array}
\tag{5.11}
$$

representing the state in Eq. (5.9).

Locality is enforced in Eq. (5.10) by allowing only those entries in the rows corresponding to the control and target qubits to change. Note that the updated values for the control (target) qubit are now allowed to depend on values previously associated with the target (control) qubit. This is valid since, in order to implement a CNOT gate, the qubits must be brought together and allowed to interact. Indeed, it has been shown that if it were possible to perform a CNOT on two spatially separated qubits, then this could be used to perform faster than light signaling [136] in violation of the assumption of locality.

The LHV rules (5.10) must be consistent with the fifteen transformations of nontrivial Pauli products under CNOT. For example, the transformation $C(XI)C^\dagger = XX$ requires that $X_c^{\text{I}} = X_c^{\text{F}} X_t^{\text{F}}$, which follows immediately from Eq. (5.10) since

$$X_c^{\text{F}} X_t^{\text{F}} = X_c^{\text{I}} \left(X_t^{\text{I}}\right)^2 = X_c^{\text{I}}. \tag{5.12}$$

More generally, the CNOT update rules are derived from the six transformations listed in Eq. (5.4), and because $C = C^\dagger$, these rules are automatically consistent with five additional transformations. Consistency with the remaining four transformations,

$$C(XY)C^\dagger = YZ, \tag{5.13}$$

$$C(XZ)C^\dagger = -YY, \tag{5.14}$$

and the inverse transformations, requires that

$$X_c^{\text{I}} Y_t^{\text{I}} = Y_c^{\text{F}} Z_t^{\text{F}} = Y_c^{\text{I}} Z_c^{\text{I}} Z_t^{\text{I}} X_t^{\text{I}} \tag{5.15}$$

and

$$X_c^{\text{I}} Z_t^{\text{I}} = -Y_c^{\text{F}} Y_t^{\text{F}} = -Y_c^{\text{I}} Z_c^{\text{I}} X_t^{\text{I}} Y_t^{\text{I}}. \tag{5.16}$$

The relations (5.15) and (5.16) do not hold in general, but they are satisfied whenever the LHV entries for both the control and target qubits are identically correlated according to either

$$XYZ = i \qquad \text{or} \qquad XYZ = -i, \tag{5.17}$$

with $X$ and $Z$ real and $Y$ imaginary. In all our applications of CNOT, this condition holds. In particular, it is for this reason that the initial sign of the $Y$ entry for the first qubit in the LHV table given by Eq. (5.6) (the only one on which a Hadamard is performed) must be opposite to that of all of the remaining qubits.

The table in Eq. (5.11) gives the correct quantum-mechanical predictions for all measurements of Pauli products on the Bell state $(|00\rangle + |11\rangle)\,|0\rangle/\sqrt{2}$. What is new are the correlations that have developed between the rows representing the first two qubits. For example, the single-qubit measurements $ZII$ and $IZI$ both return the random result $R_1$; the product of these outcomes always equals $+1$, the same as the outcome of a joint measurement of $ZZI$ on the first two qubits. In this context, the $i$'s in the correlated $Y$ entries now lead to a problem. The LHV model is designed to yield the correct predictions for all of the joint measurements, as well as the correct statistics for the local measurements. However, the different possible products of the local measurement results are not always equal to the corresponding joint measurement results as required by quantum mechanics.

For example, the single-qubit measurements $YII$ and $IYI$ both give the random result $R_1 R_2$, with product $+1$, inconsistent with the outcome $(iR_1R_2)(iR_1R_2) = -1$ of a joint measurement of $YYI$. [2] This problem persists throughout our analysis, occurring for joint measurements involving $Y$'s on some qubits and having outcomes that are certain (i.e., measurements of stabilizer elements). In fact, it is the reason our LHV model must be supplemented by classical communication.

At this point, the problem is restricted to the joint measurements $YYI$ and $YYZ$ and the corresponding local measurements. Thus, it can be corrected by flipping the sign of the outcome calculated from the LHV table in Eq. (5.11) whenever a local measurement of $Y$ is made on the first qubit, i.e., the model returns the random result $-R_1R_2$ for a measurement of $YII$. This sign flip fixes the required correlations and is irrelevant to other joint measurements that involve $Y$ on the first qubit, all of which have random results. Since the sign flip depends only on the measurement performed on the first qubit, it requires *no* communication between the qubits. Thus at this stage, with Bell-state entanglement, the LHV model gives correct quantum-mechanical predictions for all observables in $\mathcal{P}_3$ and their correlations. This simulation of correlations in a maximally entangled state of two qubits in terms of LHVs does not contradict the result of Toner and Bacon [130], which seems to imply that such a model must include at least a single bit of classical communication in order to succeed. The reason is that we are only considering measurements along a specific subset of the possible measurement directions. Indeed, any set of three measurement directions that are all oriented at ninety degrees with respect to one another, applied to a maximally entangled state of two qubits, was shown by Bell to give rise to correlations that can be obtained with a local hidden variable model [12, 87].

We complete the simulation of the creation of the GHZ state by performing the

---

[2] This problem can be traced back to the initial state: a measurement of $YXI$ has the result $-R_1R_2$, and a measurement of $XYI$ has the opposite result $R_1R_2$, yielding a product of $-1$, whereas a direct measurement of $ZZI$ yields the unequal result $+1$. For the initial state this is not seen as a problem because there is no *a priori* connection between these measurements in a LHV model.

CNOT between the first and third qubits, yielding

$$\begin{array}{cccc} & X & Y & Z \\ \text{qubit 1} & R_2R_3 & iR_1R_2R_3 & R_1 \\ \text{qubit 2} & R_2 & iR_1R_2 & R_1 \\ \text{qubit 3} & R_3 & iR_1R_3 & R_1 \end{array}. \quad (5.18)$$

This table yields correct quantum-mechanical predictions for all of the observables in $\mathcal{P}_3$, including those that form the basis of Mermin's GHZ argument [135], i.e., $XXX = 1$ and $XYY = YXY = YYX = -1$. As promised, the imaginary $Y$ entries make this agreement possible.

Consider now the scheme for ensuring consistency with local measurement predictions for the three-qubit GHZ state. The only local measurements that yield inconsistent results are those associated with stabilizer elements that contain $Y$'s; the joint measurements $XYY$, $YXY$, and $YYX$. Let Alice, Bob, and Carol each possess one of the qubits. If we put Alice in charge of ensuring compatibility, she should flip the sign of her outcome whenever she and/or Bob measures $Y$ locally. This sign flip fixes the local correlations associated with $XYY$, $YXY$, and $YYX$ and is irrelevant to other possible joint measurements that involve $Y$'s on the first two qubits, all of which have random outcomes. To implement this scheme, Bob must communicate to Alice one bit denoting whether or not he measured $Y$. For the three-qubit GHZ state, we thus have a LHV model, assisted by one bit of classical communication, that duplicates the quantum-mechanical predictions for all measurements in $\mathcal{P}_3$ and their correlations.

### 5.2.3   $N$-qubit GHZ simulation

The circuit that creates the general $N$-qubit GHZ state,

$$|\psi_N\rangle = \frac{1}{\sqrt{2}}\left(|00\ldots0\rangle + |11\ldots1\rangle\right), \quad (5.19)$$

*5.2. Simulation of GHZ correlations*

has the same topology as in Fig. 5.1: a Hadamard on the first qubit is followed by $N-1$ CNOT gates, with the leading qubit as the control and the remaining qubits serving successively as targets. The operator transformations (5.3) and (5.4) show that $|\psi_N\rangle$ is specified by the $N$ stabilizer generators

$$g_{\psi_N} = \langle X^{\otimes N}, ZZI^{\otimes(N-2)}, ZIZI^{\otimes(N-3)}, \ldots, ZI^{\otimes(N-2)}Z\rangle. \tag{5.20}$$

The full stabilizer consists of the $2^N$ observables in $\mathcal{P}_N$ that yield $+1$ with certainty, obtained by taking all possible unique products of the elements of $g_{\psi_N}$. It contains Pauli products that have (i) only $I$'s and an even number of $Z$'s and (ii) only $X$'s and an even number of $Y$'s, with an overall minus sign if the number of $Y$'s is not a multiple of 4. Of the remaining Hermitian observables in $\mathcal{P}_N$, $2^N$ are simply the negatives of the elements of the full stabilizer, and so return $-1$ with certainty, while the rest return $\pm 1$ with equal probability [2].

Following the same procedure as in the three-qubit case, one finds that the LHV table representing the $N$-qubit GHZ state is given by

$$\begin{array}{cccc}
 & X & Y & Z \\
\text{qubit 1} & R_2 R_3 \cdots R_N & i R_1 R_2 \cdots R_N & R_1 \\
\text{qubit 2} & R_2 & i R_1 R_2 & R_1 \\
\text{qubit 3} & R_3 & i R_1 R_3 & R_1 \\
\vdots & \vdots & \vdots & \vdots \\
\text{qubit } N & R_N & i R_1 R_N & R_1
\end{array}. \tag{5.21}$$

That this table gives the correct quantum-mechanical predictions for all measurements of Pauli products follows from the consistency of our LHV update rules, but it is nevertheless useful to check this directly. Suppose a Pauli product contains no $X$'s or $Y$'s, but consists solely of $I$'s and $Z$'s. Then it is clear from the table in Eq. (5.21) that the outcome is certain if and only if the number of $Z$'s in the product is even. Suppose now that the product has an $X$ or a $Y$ in the first position. Then it is apparent that to avoid a random variable in the overall product, all the other elements in the product must be $X$'s or $Y$'s and the number of $Y$'s must be even; the outcome is $+1$ if the number of $Y$'s is a multiple of 4 and $-1$ otherwise. Finally, suppose the Pauli product has an $X$ or a $Y$ in a position other than the first. Then the only way to avoid a random variable in the overall product is to have an $X$ or a $Y$ in the first position, and we proceed as before. This argument shows that the LHV table for the $N$-qubit GHZ state gives correct quantum-mechanical predictions for measurements of all Pauli products.

It remains to ensure that the products of the LHV predictions for local measurements are consistent with the corresponding joint measurement results. As before, the source of the inconsistency is the $i$ in the $Y$ table entries, the very thing that allows us to get all the Pauli products correct. Stationing Alice at the first qubit and putting her in charge of ensuring consistency, we see that what she needs to know is the number of $i$'s in the product for the corresponding joint measurement. In particular, letting $q_j = i$ if $Y$ is measured on the $j$th qubit and $q_j = 1$ otherwise, Alice can ensure consistency by changing the sign of her local outcome if the product $p_N = q_1 \cdots q_N$ is $-1$ or $-i$ and leaving her local outcome unchanged if $p_N$ is $+1$ or $i$. This scheme requires $N-1$ bits of communication as each of the other parties communicates to Alice whether or not they measured $Y$, but we can do a bit better. Alice's action is only important when $p_N$ is $+1$ or $-1$; when $p_N$ is $i$ or $-i$, the sign flip or lack thereof is irrelevant because the joint measurement outcome is random. As a result, Alice can get by with the truncated product $p_{N-1} = q_1 \cdots q_{N-1}$: she flips the sign of her local outcome if $p_{N-1}$ is $i$ or $-1$ and leaves the local outcome unchanged if $p_{N-1}$ is $-i$ or 1. The scheme works because whether $q_N$ is 1 or $i$, Alice flips when $p_N = -1$ and doesn't flip when $p_N = +1$, as required. This improved scheme requires $N-2$ bits of classical communication; it generalizes our previous results for the Bell state and the three-qubit GHZ state.

The consistency scheme generalizes trivially to the case of measurements made on $l$ disjoint sets of qubits. For each set $k$ chosen from the $l$ sets, the table yields a measurement product that is the predicted outcome multiplied by $q_k = i$ or $q_k = 1$. Letting Alice be in charge of the first set, all but the last of the other sets communicates $q_k$ to Alice, who computes the product $q_1 \cdots q_{l-1}$ and decides whether to flip her set's outcome just as before. Consistency with the corresponding joint measurement is thus ensured at the price of $l - 2$ bits of communication.

## 5.3 Cluster States and the Gottesman-Knill Theorem

The applicability of the LHV update rules (5.7) and (5.10) corresponding to the Hadamard and CNOT gates, respectively, is not limited to simulations involving $N$-qubit GHZ states. More generally, they may be used to model any circuit that: (i) is composed of sequences of only these two gates and (ii) is consistent with an initial choice of signs for the $Y$ column entries such that the LHVs representing the control and target qubits input to each CNOT are always identically correlated according to one of the two options in Eq. (5.17). An important example of a class of such circuits are those that generate the cluster states.

An $N$-qubit cluster state $|\Phi\rangle_{C(N)}$ is characterized by the set of eigenvalue equations [137]

$$K^{(a)} |\Phi\rangle_{C(N)} = |\Phi\rangle_{C(N)}, \qquad \forall a \in [1, \ldots, N] \qquad (5.22)$$

with the correlation operators

$$K^{(a)} = X^{(a)} \bigotimes_{b \in \text{ngbh}(a)} Z^{(b)}, \qquad (5.23)$$

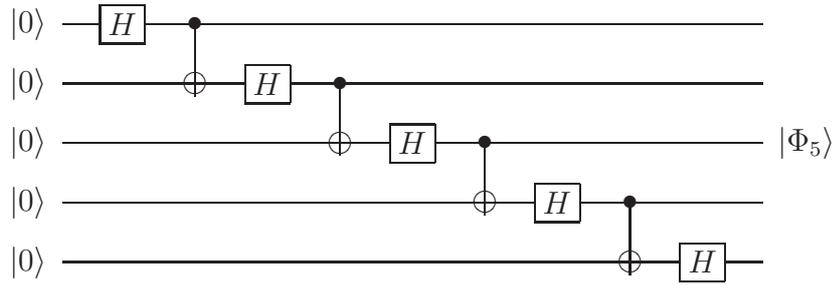

Figure 5.2: Circuit to generate the one-dimensional five-qubit cluster state.

where ngbh $(a)$ specifies the set of all neighboring qubits $b \in C(N)$ that interact with the qubit located at site $a$. Here we are using the convention that $|\Phi\rangle_{C(N)}$ is the unique eigenvector yielding $+1$ for each observable in Eq. (5.23) so that the complete set of commuting operators $\{K^{(a)}\}$ corresponds to the $N$ stabilizer generators of the state. The cluster states may be generated by a nearest-neighbor Ising-type interaction [38], or equivalently, by a specific sequence of Hadamard and CNOT gates. The general procedure to construct the circuit generating the cluster state with graph $C(N)$ is to (i) perform a Hadamard on the $a^{th}$ qubit and (ii) perform the gate CNOT$[a, b]$ for each $b \in$ ngbh $(a)$; repeating steps (i) and (ii) for each qubit $a \in [1, \ldots, N]$ [138]. For example, Fig. 5.2 depicts the quantum circuit that generates a one-dimensional cluster state of five qubits, which we denote by $|\Phi\rangle_5$.

The entanglement in the cluster states [139] provides a resource for universal quantum computation that is steadily consumed when subjected to adaptive single-qubit measurements [124, 137]. Further, any GK-type circuit, i.e. one that is (i) composed of qubits initially prepared in a computational basis state, (ii) acted upon by gates in the Clifford group, and (iii) subjected to measurements in the Pauli group, may be implemented on a cluster state of sufficient size with all measurements being performed simultaneously [137]. According to the GK theorem, any such circuit may be simulated efficiently by tracking the evolution of the stabilizer generators for the measured cluster state.

*5.3. Cluster States and the Gottesman-Knill Theorem*

The results presented in Chapter 4 on extending the formalism of complementarity to composite quantum systems imply that entanglement between subsystems possesses the unique ability to encode correlations directly, without the need to fully represent the logical entities to which these correlations refer. Further, this ability seems to be one of the vital requirements for achieving an exponential quantum advantage in computational efficiency [123]. However, at first glance the GK theorem appears to be at odds with this assertion. The nonlocal stabilizer generators tracked in the GK simulation represent global information about joint measurement results, but encode no information at all about local properties [2, 39]. Thus, they are a specific example of so-called correlations without correlata [21], the potential existence of which follows from the $N$-qubit complementarity relation given by Eq. (4.6). Specifically, one finds that $\overline{S^2}\left(\rho_k^{\Phi_{C(N)}}\right) = 0$ and $\tau_{k\{R_k\}}^{\Phi_{C(N)}} = 1, \forall k \in [1, \ldots, N]$ for an arbitrary cluster state $\left|\Phi_{C(N)}\right\rangle$ of $N$ qubits. This demonstrates that all of the information in a cluster state is encoded in the entangled correlations, not in the individual qubits. Nevertheless, when the set of possible measurements is restricted to $\mathcal{P}_N$, these entangled correlations may be classically simulated in an efficient manner. The GK theorem therefore seems to imply that the ability of composite quantum systems to encode information directly in entangled correlations is not an important ingredient for performing pure state quantum computation.

Our results yield an alternative perspective on the GK theorem, and demonstrate that we may replace the nonlocal hidden variables represented by the stabilizer generators with LHVs and an amount of classical communication that scales efficiently with the size of the problem. This is a general feature of quantum circuits obeying the constraints of the GK theorem since, as our model illustrates, such circuits do not utilize the full capabilities of the available entanglement in the probability distributions that they generate. Accordingly, one cannot rule out that the ability to directly encode correlations is important for performing truly quantum computation on the basis of the GK theorem. Rather, our model demonstrates that an exponen-tial quantum computational advantage can only be achieved when the entanglement available in the cluster state is utilized in a way that precludes the existence of an efficient, local realistic description of the process, even when supplemented by an efficient amount of nonlocal but classical communication.

### 5.3.1 Simulating cluster state correlations

The creation of an arbitrary cluster state, e.g., the state generated by the circuit in Fig. 5.2, can be cast in the language of the GK theorem, where the stabilizer generators evolve according to Eqs. (5.3) and (5.4), just as was done for the $N$-qubit GHZ states. According to these rules, the stabilizer generators for the initial state $|0\rangle^{\otimes 5}$

$$g_{0_5} = \langle ZIIII, IZIII, IIZII, IIIZI, IIIIZ \rangle \tag{5.24}$$

evolve through the circuit in Fig. 5.2 to the set

$$g_{\Phi_5} = \langle XZIII, ZXZII, IZXZI, IIZXZ, IIIZX \rangle, \tag{5.25}$$

which uniquely determines the state $|\Phi\rangle_5$. Fig. 5.3 illustrates the corresponding evolution of the LHV table simulating the creation of this state.

It is easily verified that, at each step in the circuit of Fig. 5.2, the corresponding LHV table yields the correct joint measurement predictions for all products of Pauli operators. In particular, the last table in Fig. 5.3 yields the correct quantum mechanical predictions for all joint measurements in $\mathcal{P}_5$ on the state $|\Phi\rangle_5$. Additionally, one sees from the first table in Fig. 5.3 that our model requires an initial sign distribution among the $Y$ column entries such that no two neighboring qubits start with the same sign. This is a generic feature of our model when used to simulate the creation of a cluster state; it is required in order to satisfy condition (5.17). This initialization is different from that required to create an $N$-qubit GHZ state. We

5.3. *Cluster States and the Gottesman-Knill Theorem*

$$
\begin{array}{ccc}
X & Y & Z \\
R_1 & -iR_1 & 1 \\
R_2 & iR_2 & 1 \\
R_3 & -iR_3 & 1 \\
R_4 & iR_4 & 1 \\
R_5 & -iR_5 & 1
\end{array}
\xrightarrow{H_1}
\begin{array}{ccc}
X & Y & Z \\
1 & iR_1 & R_1 \\
R_2 & iR_2 & 1 \\
R_3 & -iR_3 & 1 \\
R_4 & iR_4 & 1 \\
R_5 & -iR_5 & 1
\end{array}
\xrightarrow{C_{12}}
\begin{array}{ccc}
X & Y & Z \\
R_2 & iR_1R_2 & R_1 \\
R_2 & iR_1R_2 & R_1 \\
R_3 & -iR_3 & 1 \\
R_4 & iR_4 & 1 \\
R_5 & -iR_5 & 1
\end{array}
$$

$$
\xrightarrow{\cdots}
\begin{array}{ccc}
X & Y & Z \\
R_2 & iR_1R_2 & R_1 \\
R_1R_3 & -iR_1R_2R_3 & R_2 \\
R_2R_4 & iR_2R_3R_4 & R_3 \\
R_3R_5 & -iR_3R_4R_5 & R_4 \\
R_5 & -iR_4R_5 & R_4
\end{array}
\xrightarrow{H_5}
\begin{array}{ccc}
X & Y & Z \\
R_2 & iR_1R_2 & R_1 \\
R_1R_3 & -iR_1R_2R_3 & R_2 \\
R_2R_4 & iR_2R_3R_4 & R_3 \\
R_3R_5 & -iR_3R_4R_5 & R_4 \\
R_4 & iR_4R_5 & R_5
\end{array}
$$

Figure 5.3: Evolution of the LHV model simulating the creation of the one-dimensional five-qubit cluster state. Note that, for compactness, not all of the intermediate LHV tables have been included.

conjecture that this difference is fundamental, and gives rise to the different types of entanglement that exist in these two classes of states. This idea is discussed more fully at the end of the chapter.

In general, the LHV table representing an arbitrary two-dimensional cluster state of $N$ qubits, constructed according to the above rules, yields the correct joint predictions for all measurements in the set $\mathcal{P}_N$ as we now show; the proof for the one-dimensional case then follows immediately. Consider the stabilizer generators for the two-dimensional $3 \times 3$ cluster state, written so that each entry in a given generator appears in the lattice position of the corresponding qubit,

$$g_{\Phi_{(3\times3)}} = \left\langle \begin{pmatrix} X & Z & I \\ Z & I & I \\ I & I & I \end{pmatrix} - \begin{pmatrix} Z & X & Z \\ I & Z & I \\ I & I & I \end{pmatrix} - \begin{pmatrix} I & Z & X \\ I & I & Z \\ I & I & I \end{pmatrix} \right. \\ \left. \begin{pmatrix} Z & I & I \\ X & Z & I \\ Z & I & I \end{pmatrix} - \begin{pmatrix} I & Z & I \\ Z & X & Z \\ I & Z & I \end{pmatrix} - \begin{pmatrix} I & I & Z \\ I & Z & X \\ I & I & Z \end{pmatrix} \right. \\ \left. \begin{pmatrix} I & I & I \\ Z & I & I \\ X & Z & I \end{pmatrix} - \begin{pmatrix} I & I & I \\ I & Z & I \\ Z & X & Z \end{pmatrix} - \begin{pmatrix} I & I & I \\ I & I & Z \\ I & Z & X \end{pmatrix} \right\rangle. \quad (5.26)$$

One may extract the following insights into the structure of our model from Eq. 5.26. First, the distribution of $Y$'s that result from the multiplication of the various stabilizer generators depends solely on the multiplication of generators that neighbor one another in the above representation, i.e., only on those that are joined by a red horizontal or vertical line. [3] The overlap of an $X$ with a $Z$ in the entries of the generators of the stabilizer occurs only when this is true. Furthermore, the number of $Y$'s that result from any product of generators is always even, which implies that no imaginary phases survive the multiplication of LHV table elements corresponding to the measurement of an entry in the full stabilizer. Finally, only three unique patterns (appearing in different orientations and identifiable by the yellow shading) occur in the above set of generators. These are (i) the "corner" pattern in, e.g., position #1, (ii) the "edge" pattern in, e.g., position #2, and (iii) the "center" pattern in position #5. For larger lattices no new patterns nor nontrivial overlaps arise in the various possible products of the associated generators. Thus, the success of our model for an arbitrary two-dimensional lattice follows immediately from its success for the $3 \times 3$ lattice, which has been verified by the computer algorithm included in appendix B for the measurement of all $4^9$ products of Pauli operators on this state. We conjecture that a similar argument holds for three-dimensional cluster states. However, we have been unable to verify this even for the $3 \times 3 \times 3$ lattice, the simplest non-trivial extension to three dimensions, since using this method of proof would require calculating the quantum and LHV predictions for $4^{27}$ possible measurements.

---

[3]Just as in the corresponding two-dimensional lattice of qubits, diagonally adjacent generators are not considered neighbors.

The above argument demonstrates that our LHV model automatically yields the correct quantum mechanical predictions for all joint measurements in $\mathcal{P}_N$ on an arbitrary $N$-qubit cluster state. Similar to what was done for the $N$-qubit GHZ states, consistency between the joint measurement predictions and the products of the corresponding local measurements may once again be ensured by performing the procedure outlined at the end of Section 5.2.3, at a cost of $N-2$ bits of classical communication.

### 5.3.2 Simulating Gottesman-Knill circuits

In general, the number of physical qubits needed to create a cluster state large enough to implement a GK circuit with $m$ Clifford gates acting on $n$ logical qubits scales as $N = O(mn)$. The results of the previous section demonstrate that our model yields the correct predictions for both joint and local measurements associated with the set of observables in $\mathcal{P}_N$ on any such $N$-qubit cluster state $|\Phi\rangle_{C(N)}$ with only $N-2$ bits of classical communication overhead. A subset of these observables will correspond to the set of GK circuits that fit on $|\Phi\rangle_{C(N)}$. For example, the Hadamard gate is implemented in the cluster state architecture by performing the measurement pattern

$$H = X_1 Y_2 Y_3 Y_4 \square_5 \tag{5.27}$$

on four neighboring cluster qubits [137], where the positions of the qubits in the lattice are represented by subscripts. This has the effect of performing $H$ on the qubit in the first position and teleporting the result to the output position denoted by $\square$. Similarly, the measurement

$$P = X_1 X_2 Y_3 X_4 \square_5 \tag{5.28}$$

implements the phase gate $P$, and

$$\text{CNOT} = \begin{matrix} X_1 & Y_2 & Y_3 & Y_4 & Y_5 & Y_6 & \square_7 \\ Z_8 & Z_9 & Z_{10} & Y_{11} & Z_{12} & Z_{13} & Z_{14} \\ X_{15} & X_{16} & X_{17} & Y_{18} & X_{19} & X_{20} & \square_{21} \end{matrix} \tag{5.29}$$

implements the CNOT between the control qubit in position #1 and the target qubit in position #15 [137]. Strung together in different combinations, with the output position(s) of one Clifford gate measurement pattern overlapping the input position(s) of another, these different configurations give rise to the set of GK circuits that will fit on a cluster state of a given size.

Our objective is to use the LHV model presented in Section 5.3.1 to simulate the cluster state implementation of the Clifford gates. Taking this approach, one finds that simply calculating the LHV predictions for the given measurement patterns does not work. Consider, for example, a measurement of the four-qubit pattern given by Eq. (5.27) on the state $|\Phi_5\rangle$. The result of this operation is to perform a Hadamard on the first qubit (initially stabilized by $X$) with the result being teleported to the output qubit position. A subsequent measurement of $Z$ on the output qubit is then expected to yield the result $+1$ with certainty. In terms of our simulation, however, multiplying the appropriate LHV entries from the last table in Fig. 5.3 yields

$$X_1 Y_2 Y_3 Y_4 = (R_2)(-iR_1R_2R_3)(iR_2R_3R_4)(-iR_3R_4R_5) = -iR_1R_2R_3R_5. \tag{5.30}$$

Since the LHV table entry for a measurement of $Z_5$ on the state $|\Phi_5\rangle$ is $R_5$, it is easily verified that the prediction of our model for this measurement (using Eq. (5.30) and

*5.3. Cluster States and the Gottesman-Knill Theorem*

discarding the imaginary factor) is the completely random result $-R_1R_2R_3$, which is inconsistent with the cluster state result.

It is not surprising that this straightforward application of our LHV model fails to simulate the evolution of a single logical qubit in the cluster state architecture, which we may think of as being teleported through a Hadamard gate in the above example, since such a process relies crucially on the entanglement possessed by the cluster state. The procedure outlined in Section 5.3.1 is designed not to simulate the effects of these quantum correlations on one or more logical qubits, but to predict the outcomes of measurements in the $N$-qubit Pauli group performed on the physical qubits; a task which our model carries out successfully.

Nevertheless, a method also exists for simulating the evolution of the logical qubits with our LHV model. The connection between the Clifford gates and our model is properly made by referring to [137], which identifies three sets of $2l$ eigenvalue equations associated, respectively, with the gates $H$, $P$, and CNOT, where $l$ is the number of logical qubits on which the gate in question operates. For example, the five qubit cluster state $|\Phi\rangle_5$ obeys the two eigenvalue equations

$$\begin{aligned}
|\Phi\rangle_5 &= K^{(1)}K^{(3)}K^{(4)}|\Phi\rangle_5 \\
&= \boxdot\square\boxdot\square\square|\Phi\rangle_5 \\
|\Phi\rangle_5 &= X_1 I_2 Y_3 Y_4 Z_5 |\Phi\rangle_5,
\end{aligned} \tag{5.31}$$

and

$$\begin{aligned}
|\Phi\rangle_5 &= K^{(2)}K^{(3)}K^{(5)}|\Phi\rangle_5 \\
&= \square\boxdot\boxdot\square\boxdot|\Phi\rangle_5 \\
|\Phi\rangle_5 &= Z_1 Y_2 Y_3 I_4 X_5 |\Phi\rangle_5.
\end{aligned} \tag{5.32}$$

The middle expressions above are written in correlation center notation [137] where,

e.g., the pattern ⊡□⊡⊡□ represents the stabilizer element that results from the product of the generators $K^{(1)}$, $K^{(3)}$, and $K^{(4)}$. This notation will prove useful in the upcoming discussion of how to combine Clifford gates in the LHV simulation.

Equations (5.31) and (5.32) play a fundamental role in the context of the cluster state architecture, since they are used to determine the measurement pattern given by Eq. (5.27) to implement the Hadamard gate [137]. Further, these elements of the stabilizer act as a generating set for the complete Pauli algebra of $H$. The product of the stabilizer elements in Eqs. (5.31) and (5.32) yields the eigenvalue equation

$$\begin{aligned} |\Phi\rangle_5 &= K^{(1)}K^{(2)}K^{(4)}K^{(5)} |\Phi\rangle_5 \\ &= \boxdot\boxdot\square\boxdot\boxdot |\Phi\rangle_5 \\ |\Phi\rangle_5 &= Y_1 Y_2 I_3 Y_4 Y_5 |\Phi\rangle_5, \end{aligned} \quad (5.33)$$

which governs the evolution of the stabilizer entry $Y$ under the operation of the Hadamard. The three intermediate stabilizer components, i.e., those that are not associated with either the input or the output qubit in the above eigenvalue equations, all commute with the single-qubit measurements performed on the intermediate qubits in Eq. (5.27). Such a relationship was shown to preserve the correlations $K^{(i)}, i \in [1,\ldots,N]$ possessed by a given state [137]. The ability of our model to capture these correlations is sufficient to simulate the evolution of the logical qubits, implemented by measuring single-qubit patterns on the cluster state, as we now show.

The mapping between our model and the cluster state implementation of a Clifford gate is made by taking the products of the LHV table entries corresponding to the stabilizer appearing in the appropriate eigenvalue equation as determined by the input qubit. This procedure ensures that the measurement pattern simulated by our LHV model corresponds to the unique stabilizer element that preserves the same set of correlations respected by the cluster state measurement pattern. For example, if a

5.3. *Cluster States and the Gottesman-Knill Theorem*

Hadamard gate is implemented in the cluster state architecture via the single-qubit measurement pattern shown in Eq. (5.27) on an input qubit stabilized by $X$, then the connection to the LHV model is made via Eq. (5.31). Specifically, the product of the elements in the last table of Fig. 5.3, corresponding to a measurement of $X_1 I_2 Y_3 Y_4$, is given by

$$(R_2)(1)(iR_2 R_3 R_4)(-iR_3 R_4 R_5) = R_5 = Z_5, \tag{5.34}$$

demonstrating that $X_1 \xrightarrow{H} Z_5$ as expected. Similarly, from Eqs. (5.32) and (5.33) we find that

$$Z_1 Y_2 Y_3 I_4 = R_4 = X_5 \implies Z_1 \xrightarrow{H} X_5 \tag{5.35}$$

and

$$Y_1 Y_2 I_3 Y_4 = -i R_4 R_5 = -Y_5 \implies Y_1 \xrightarrow{H} -Y_5, \tag{5.36}$$

respectively, all in perfect agreement with the stabilizer algebra of the $H$ gate.

A similar analysis may be performed in order to connect the $P$ and CNOT gates to the LHV formalism. The generating set of eigenvalue equations for the phase gate is [137]

$$\begin{aligned}
|\Phi\rangle_5 &= K^{(1)} K^{(3)} K^{(4)} K^{(5)} |\Phi\rangle_5 \\
&= \boxdot\square\boxdot\boxdot\boxdot\, |\Phi\rangle_5 \\
|\Phi\rangle_5 &= -X_1 I_2 Y_3 X_4 Y_5 |\Phi\rangle_{5,}
\end{aligned} \tag{5.37}$$

$$\begin{aligned}
|\Phi\rangle_5 &= K^{(2)} K^{(4)} |\Phi\rangle_5 \\
&= \square\boxdot\square\boxdot\square\, |\Phi\rangle_5 \\
|\Phi\rangle_5 &= Z_1 X_2 I_3 X_4 Z_5 |\Phi\rangle_{5,}
\end{aligned} \tag{5.38}$$

which leads to the additional equation

$$|\Phi\rangle_5 = K^{(1)}K^{(2)}K^{(3)}K^{(5)}|\Phi\rangle_5$$

$$= \boxdot\boxdot\boxdot\square\boxdot\,|\Phi\rangle_5 \tag{5.39}$$

$$|\Phi\rangle_5 = -Y_1 X_2 Y_3 I_4 X_5 |\Phi\rangle_5.$$

These equations lead, respectively, to the LHV results

$$-X_1 I_2 Y_3 X_4 = -iR_4 R_5 = -Y_5 \implies X_1 \xrightarrow{P} Y_5, \tag{5.40}$$

$$Z_1 X_2 I_3 X_4 = R_5 = Z_5 \implies Z_1 \xrightarrow{P} Z_5, \tag{5.41}$$

$$-Y_1 X_2 Y_3 I_4 = R_4 = X_5 \implies Y_1 \xrightarrow{P} -X_5. \tag{5.42}$$

Finally, the generating set of equations for the CNOT gate is given by

$$|\Phi\rangle_{(3\times7)} = K^{(1)}K^{(3)}K^{(4)}K^{(5)}K^{(7)}K^{(11)}K^{(19)}K^{(21)}|\Phi\rangle_{(3\times7)}$$

$$= \begin{pmatrix} \boxdot & \square & \boxdot & \boxdot & \boxdot & \square & \boxdot \\ \square & \square & \square & \boxdot & \square & \square & \square \\ \square & \square & \square & \square & \boxdot & \square & \boxdot \end{pmatrix} |\Phi\rangle_{(3\times7)} \tag{5.43}$$

$$|\Phi\rangle_{(3\times7)} = -\begin{pmatrix} X_1 & I_2 & Y_3 & Y_4 & Y_5 & I_6 & X_7 \\ Z_8 & I_9 & I_{10} & Y_{11} & Z_{12} & I_{13} & I_{14} \\ I_{15} & I_{16} & I_{17} & I_{18} & X_{19} & I_{20} & X_{21} \end{pmatrix} |\Phi\rangle_{(3\times7)},$$

## 5.3. Cluster States and the Gottesman-Knill Theorem

$$|\Phi\rangle_{(3\times 7)} = K^{(2)}K^{(3)}K^{(5)}K^{(6)} |\Phi\rangle_{(3\times 7)}$$

$$= \begin{pmatrix} \square & \cdot & \cdot & \square & \cdot & \cdot & \square \\ \square & \square & \square & \square & \square & \square & \square \\ \square & \square & \square & \square & \square & \square & \square \end{pmatrix} |\Phi\rangle_{(3\times 7)} \qquad (5.44)$$

$$|\Phi\rangle_{(3\times 7)} = \begin{pmatrix} Z_1 & Y_2 & Y_3 & I_4 & Y_5 & Y_6 & Z_7 \\ I_8 & Z_9 & Z_{10} & I_{11} & Z_{12} & Z_{13} & I_{14} \\ I_{15} & I_{16} & I_{17} & I_{18} & I_{19} & I_{20} & I_{21} \end{pmatrix} |\Phi\rangle_{(3\times 7)},$$

$$|\Phi\rangle_{(3\times 7)} = K^{(15)}K^{(17)}K^{(19)}K^{(21)} |\Phi\rangle_{(3\times 7)}$$

$$= \begin{pmatrix} \square & \square & \square & \square & \square & \square & \square \\ \square & \square & \square & \square & \square & \square & \square \\ \cdot & \square & \cdot & \square & \cdot & \square & \cdot \end{pmatrix} |\Phi\rangle_{(3\times 7)} \qquad (5.45)$$

$$|\Phi\rangle_{(3\times 7)} = \begin{pmatrix} I_1 & I_2 & I_3 & I_4 & I_5 & I_6 & I_7 \\ Z_8 & I_9 & Z_{10} & I_{11} & Z_{12} & I_{13} & Z_{14} \\ X_{15} & I_{16} & X_{17} & I_{18} & X_{19} & I_{20} & X_{21} \end{pmatrix} |\Phi\rangle_{(3\times 7)},$$

$$|\Phi\rangle_{(3\times 7)} = K^{(5)}K^{(6)}K^{(11)}K^{(16)}K^{(18)}K^{(20)} |\Phi\rangle_{(3\times 7)}$$

$$= \begin{pmatrix} \square & \square & \square & \square & \cdot & \cdot & \square \\ \square & \square & \square & \cdot & \square & \square & \square \\ \square & \cdot & \square & \cdot & \square & \cdot & \square \end{pmatrix} |\Phi\rangle_{(3\times 7)} \qquad (5.46)$$

$$|\Phi\rangle_{(3\times 7)} = \begin{pmatrix} I_1 & I_2 & I_3 & I_4 & Y_5 & Y_6 & Z_7 \\ I_8 & Z_9 & Z_{10} & Y_{11} & I_{12} & I_{13} & I_{14} \\ Z_{15} & X_{16} & I_{17} & Y_{18} & I_{19} & X_{20} & Z_{21} \end{pmatrix} |\Phi\rangle_{(3\times 7)},$$

corresponding, respectively, to the four transformations

$$C(XI)C^\dagger = XX, \quad C(ZI)C^\dagger = ZI,$$
$$C(IX)C^\dagger = IX, \quad C(IZ)C^\dagger = ZZ, \tag{5.47}$$

in Eq. (5.4). Taking the different possible products of the above generating set yields the complete set of eigenvalue equations corresponding to the sixteen possible input combinations to the CNOT operation. The connection between the cluster state architecture and the LHV model for the CNOT gate is found to be analogous to that for the Hadamard and Phase gates.

We now demonstrate how to simulate the behavior of single-qubit measurement patterns that correspond to connected Clifford gates in the cluster state picture. The general procedure is to create a correlation center diagram representing concatenated gates by (i) identifying the stabilizer element for each gate that corresponds to the given input, (ii) concatenating the patterns by letting the output position(s) of earlier gates overlap the input position(s) of later gates, and (iii) assigning a correlation center to the overlap position(s) if and only if the concatenated patterns contain a correlation center in that position. For simplicity, we restrict our examples to concatenated single-qubit gates. However, the procedure works in general, and allows for the efficient local simulation of any GK-type circuit that will fit on the cluster being modeled.

Consider, for example, the single-qubit gate sequence

$$X \xrightarrow{H} Z \xrightarrow{H} X. \tag{5.48}$$

Since an $X$ is input to the first Hadamard gate and a $Z$ is input to the second, the relevant correlation center patterns are given by the middle expressions in Eqs. (5.31) and (5.32), respectively. Combining these patterns according to the above rule yields the stabilizer element

$$\boxdot\square\boxdot\square\boxdot\square\boxdot\square\boxdot = X_1 I_2 Y_3 Y_4 I_5 Y_6 Y_7 I_8 X_9, \tag{5.49}$$

5.3. Cluster States and the Gottesman-Knill Theorem

|  | X | Y | Z |
|---|---|---|---|
| qubit 1 | $R_2$ | $iR_1R_2$ | $R_1$ |
| qubit 2 | $R_1R_3$ | $-iR_1R_2R_3$ | $R_2$ |
| qubit 3 | $R_2R_4$ | $iR_2R_3R_4$ | $R_3$ |
| qubit 4 | $R_3R_5$ | $-iR_3R_4R_5$ | $R_4$ |
| qubit 5 | $R_4R_6$ | $iR_4R_5R_6$ | $R_5$ |
| qubit 6 | $R_5R_7$ | $-iR_5R_6R_7$ | $R_6$ |
| qubit 7 | $R_6R_8$ | $iR_6R_7R_8$ | $R_7$ |
| qubit 8 | $R_7R_9$ | $-iR_7R_8R_9$ | $R_8$ |
| qubit 9 | $R_8$ | $iR_8R_9$ | $R_9$ |

Figure 5.4: LHV table representing the one-dimensional nine-qubit cluster state.

of the one-dimensional nine-qubit cluster state $|\Phi\rangle_9$. The LHV table corresponding to $|\Phi\rangle_9$ is shown in Fig. 5.4 and yields the result

$$X_1 I_2 Y_3 Y_4 I_5 Y_6 Y_7 I_8 = R_8 = X_9 \implies X_1 \xrightarrow{H^2} X_9, \qquad (5.50)$$

in complete agreement with Eq. (5.48).

As a final example, consider the gate sequence

$$Y \xrightarrow{P} -X \xrightarrow{H} -Z. \qquad (5.51)$$

Combining the correlation center patterns in Eqs. (5.39) and (5.31) yields

$$\boxdot\boxdot\boxdot\boxdot\boxdot\boxdot\boxdot\boxdot\boxdot = -Y_1 X_2 Y_3 I_4 X_5 I_6 Y_7 Y_8 Z_9. \qquad (5.52)$$

The LHV table in Fig. 5.4 then implies that

$$Y_1 X_2 Y_3 I_4 X_5 I_6 Y_7 Y_8 = -R_9 = -Z_9 \implies Y_1 \xrightarrow{PH} -Z_9 \qquad (5.53)$$

as required. The method outlined above is completely general since this procedure always results in a valid stabilizer element that corresponds to evolving the given input to the output appropriate to the simulated gate sequence.

## 5.4 Summary and Future Directions

Using local hidden variables and an efficient amount of classical communication, we have shown that it is possible to simulate the correlations that arise when measuring arbitrary products of Pauli operators on two classes of globally entangled states: the $N$-qubit GHZ states and the $N$-qubit cluster states. In each case the procedure to do this consists of constructing a LHV table yielding all of the correct quantum mechanical predictions for the allowed set of joint measurements, and ensuring that the products of the local measurement results are consistent with these joint predictions. This second step was shown to require an amount of classical communication that scales linearly with the number of qubits in the state.

The $N$-qubit GHZ states yield deterministic (as well as exponentially increasing) violations of local realism, while the $N$-qubit cluster states provide an entanglement resource for performing universal quantum computation. Nevertheless, the probability distributions arising from measurements of Pauli products on these states are essentially trivial, in every case being either certainty or binary randomness. This property is shared by all states produced by GK circuits, suggesting that the inability of these circuits to generate nontrivial probability distributions is, at least in part, responsible for the existence of an efficient classical simulation procedure. Conversely, allowing just one additional nontrivial measurement, say of the observable $(X+Y)/\sqrt{2}$, leads to correlations for which our simulation will no longer work. In the cluster state architecture, adding this single-qubit measurement is equivalent to replacing the phase gate $P$ in the Clifford group with the so-called $T$ gate [2] which corresponds to a $\pi/4$-rotation about the $Z$ axis, and yields a universal gate set for quantum computation [140]. We anticipate that under this more general measurement scheme, the amount of classical communication required to make any LHV model work grows exponentially in the number of qubits.

*5.4. Summary and Future Directions*

Our results yield a new perspective on the GK theorem by demonstrating that, when restricted to observables in the Pauli group, the allowed class of single-qubit measurements does not make full use of the entanglement available in the cluster states. The success of our simulation provides strong evidence that the power of quantum computation arises not directly from entanglement, but rather from the nonexistence of an efficient, local realistic description of the computation, even when supplemented by an efficient amount of nonlocal, but classical communication. Further, it supports the conjecture that the unique ability of quantum systems to encode correlations directly, without the need for the underlying correlata to possess physically meaningful values [21], (an immediate consequence of the complementary relationship between entanglement and individual subsystem properties derived in Chapter 4) is a necessary ingredient in truly quantum computation. Accordingly, this work constitutes further progress towards quantifying the classical resources required to simulate the correlations arising in an arbitrary quantum circuit in order to gain insight into the roles played by complementarity and entanglement in achieving an exponential quantum advantage in computational efficiency.

We have outlined procedures for efficiently simulating both individual Clifford gates and arbitrary GK circuits with our model using only classical resources. One possible path for further research in this area would be to try and quantify (in an asymptotic sense) the maximum number of $T$ gates (implemented in the cluster state architecture by allowing measurements of the observable $(X+Y)/\sqrt{2}$) that may be added to an arbitrary GK circuit such that a LHV model capable of simulating the circuit efficiently, i.e., with communication overhead that scales at most polynomially in the number of qubits, still exists. This would provide further insight into the specific way in which the available entanglement in the cluster states must be utilized in order to yield an exponential speedup.

It would also be interesting if one could prove or disprove the optimality of our procedure to implement the GK circuits and identify the minimal resources required for such a simulation. Two approaches readily suggest themselves. First, one could allow classical, but nonlocal communication between the qubits containing not only information about the measurements being made on the individual qubits, but also about the actual values possessed by the LHVs in order to calculate lower bounds on the required amount of communication. A second approach, which need not be taken independently of the first, would be to try and find a consistent set of rules to implement the phase gate directly, as was done for the $H$ and CNOT gates, and so rid our model of the overhead of having to use several cluster state qubits to implement gates on each logical qubit in the original GK circuit. The main obstacle to this type of extension of our model seems to be the identification of a consistent set of update rules for the phase gate that also preserves the correlations in Eq. (5.17).

Finally the structure of our model, in particular the roles played by the initial conditions encoded in the LHV tables, suggests the possibility of obtaining insights of a more fundamental nature. For example, the success of our simulation procedure relies heavily on the initial correlations that exist between the LHVs corresponding to noncommuting observables on individual qubits. This seems to be a generic feature of any such simulation, not an artifact of our specific implementation.

The initial distribution of signs in the $Y$ column entries for each qubit also appears to play a fundamental role in determining the specific entanglement properties of the state being generated. For example, we have seen that simulating the creation of an $N$-qubit GHZ state requires that the qubit on which the Hadamard is to act starts with a sign in its $Y$ column entry opposite that of all other qubits, while the condition for the generation of a cluster state is that all neighboring qubits begin with opposite signs in their $Y$ columns. These conditions can be made to overlap in our LHV implementation in the cases $N = 2$ and $N = 3$, precisely when these two classes of states are equivalent to one another (up to local unitaries) [38]. Alternatively, for

*5.4. Summary and Future Directions*

$N \geq 4$, when the initial LHV conditions for the simulation of the creation of the GHZ and cluster states are no longer compatible with one another, these two classes of states are found to possess very distinct types of entanglement [38, 139]. We feel that both of these features of our model warrant more detailed study in the hope that they will lead to new insights into the foundations of quantum mechanics and the nature of entanglement.

# Chapter 6

# Summary and Conclusions

The research presented in this thesis is focused on determining the role played by entanglement in certain information-theoretic processes, as well as on understanding various characteristics of these quantum mechanical correlations in terms of a broader framework based on the phenomenon of complementarity in composite systems. Our results concerning the dynamical evolution of entanglement in the different partitions of the two-atom Tavis-Cummings model demonstrate how the discussion of entanglement sharing can be extended to increasingly complex systems of both theoretical and experimental significance. Additionally, we show how one can gain initial insight into the entanglement generated in larger atomic ensembles by making use of one specific member of our new class of computable entanglement monotones. Achieving a thorough understanding of the constraints governing the distribution of entanglement in multipartite systems of this sort has important implications for designing quantum feedback and control protocols with potential applications in the field of quantum computing.

Our multi-qubit complementarity relations imply that entanglement sharing, as well as other unique features of entanglement, e.g. the fact that maximal information about a composite quantum system does not necessarily imply maximal information about the component subsystems, can be understood as specific consequences of the complementary relationships that exist between different types of information that one may simultaneously possess about such systems. As we have seen, this tradeoff suggests an interpretation of the tangle as the fiducial measure of uncertainty about individual qubits due to the presence of entanglement, rather than to our ignorance. Equivalently, because of the relationship between uncertainty and information, the tangle also quantifies the amount of information directly encoded in the quantum correlations of the system.

Our result concerning the efficient communication-assisted local simulation of the Gottesman-Knill circuits is fully consistent with multi-qubit complementarity, and supports the conclusion that the unique ability of entangled quantum systems to directly encode information in correlations is a necessary ingredient in performing truly quantum computation. The fundamental tradeoff between the information that a quantum state may encode about entangled correlations and about the correlated subsystems appears necessary to ensure that any process yielding an exponential quantum advantage in computational efficiency does not have an efficient, local realistic description, even when supplemented by an efficient amount of nonlocal, but classical communication.

The above results all imply that a constructive approach to better understanding quantum mechanics would be to merge the Bayesian, Ithaca, and Copenhagen interpretations into a single interpretation that retains certain key features of each. Proceeding in this way, we find that the lesson learned from the Bayesian approach is that probabilities, even those that are derived from a pure state wavefunction [24], are subjective quantities that represent our uncertainty, or equivalently, our information about a quantum system. According to this view, the quantum state itself must be considered to be nothing more than a concise encapsulation of our information about the system, and not a description of its objective properties. This is not to say that physical quantum systems do not possess objective properties of fundamental importance, only that to identify them one must look elsewhere in the theory.

The mantra of the Ithaca interpretation of quantum mechanics is that "correlations have physical reality; that which they correlate does not [21]." While we agree that probabilities should apply to individual systems and not be defined merely in terms of either real or hypothetical ensembles, we must disagree with the contention that quantum mechanical probabilities are objective features of the world as argued in [21]. Rather, we see the main contribution of the Ithaca interpretation to be the recognition of the fundamental importance of information that is directly encoded in entangled correlations. In our hybrid interpretation this information is to be treated on the same footing as information that we possess about individual subsystems, but at the same time is recognized to have a different character from the information normally associated with classical correlations. This distinction highlights the 'primacy' of information stored in entangled correlations which cannot be inferred, even in principle, from information about the subsystems to which the correlations refer; all in marked contrast to the 'secondary' nature of classical correlations.

At first glance, Mermin's argument that correlations should be granted physical reality at the expense of the reality of the correlated entities, due to the observed mutual consistency of all of the different correlations (or joint probability distributions) encoded in the quantum state [21], is quite compelling. The appeal of this point of view stems from the fact that each of the different possible joint probability distributions derivable from the quantum state is associated with a certain conceptual division of the system into subsystems; a distinction that nature simply does not make. Therefore the consistency of the correlations, even in the face of such arbitrary observer-induced distinctions, strongly suggests that granting physical reality to the correlations will alleviate many of the conceptual difficulties normally associated with quantum mechanics.

Nonetheless, the proposed objectivity of these correlations is convincingly refuted by an argument going back to Einstein [141] and reiterated in [24], where it is noted that the measurement of an observable performed on one part $A$ of a composite system $AB$ in an entangled pure state, e.g., two qubits in the maximally entangled singlet state, allows one to immediately write down a pure state for the unobserved system $B$. Depending on what observable is measured, the new state of system $B$ may come from one of a number of disjoint sets. Further, this is true no matter how far apart the two systems are from one another. These observations led Einstein to conclude that quantum states (and the correlations which they encode) cannot be real states of affairs since, whatever the real, objective state of affairs at $B$ is, it should not depend upon the choice of measurement made at $A$. Of course, according to the Bayesian point of view there is no difficulty; the changing state assignment simply reflects a corresponding change in our state of knowledge due to information obtained via the measurement performed on subsystem $A$ and inferred from our knowledge of the entangled correlations.

From the perspective afforded by our hybrid interpretation, the observed mutual consistency of the different correlations encoded in the quantum state becomes a statement about the noncontradictory nature of the different types of information that we may simultaneously possess about a system. It is not surprising that the various forms that our subjective information may take should in part be determined by the assumptions that we make about, or the conceptual subdivisions that we impose on, the system being described. The logical implications of quantum mechanics are always fully consistent as long as we base our predictions only on information that we actually possess, and not on information that we contemplate having about the system, e.g., through postulated but unperformed measurement-based or Hamiltonian evolution. 'Contradictions' only arise when we reason counterfactually, i.e., when we attempt to compare valid quantum predictions with predictions made by assuming quantum state assignments which we know to be incorrect or, at the very least, for which we have no evidence.

Our extension of the phenomenon of complementarity, which forms the heart of the Copenhagen interpretation, makes explicit the distinction between information about individual subsystems and information about the correlations that exist between these subsystems. The multi-qubit complementarity relations that we derive demonstrate that this distinction is fundamental to the theory, since possessing knowledge of one type limits the in-principle availability of information of a complementary type. Specifically, we have shown that the complementarity that precludes simultaneous knowledge about the outcomes of noncommuting measurements performed on individual systems [6] may be extended to include information encoded in bipartite, tripartite, and (we conjecture) higher order quantum correlations.

In the course of our investigation of entanglement sharing in the context of multi-qubit complementarity, we showed that a pure state of three qubits is associated with two distinct relations given, respectively, by Eq. (4.6) with $N = 3$ and by Eq. (4.7). These expressions quantify tradeoffs between complementary types of potentially available information corresponding to two different conceptualizations of the system. The first concerns information associated with the individual qubits, as well as information stored in the correlations between each of these qubits and the remainder of the system. The second expression involves information encoded in the irreducible three-body correlations, as well as in the individual qubits. In this context, Eq. (4.7) shows that additional information may also be obtained regarding the correlations that exist in the various two-party marginal states.

The three qubit example shows that regardless of which exhaustive set of correlations we choose to consider, the amount of information encoded by the individual qubits remains unaffected; we only see changes in the forms and amounts of bipar-tite and tripartite entanglements. We conjecture that this 'fungibility' of correlation information is a general feature of multipartite quantum systems, and is related to the fact that all possible joint and marginal probability distributions encoded in a density operator are mutually consistent with one another. One possible path for further research along these lines is therefore to try and quantify the relationship between the tangles (and multipartite generalizations thereof) and the associated joint probability distributions corresponding to all possible valid conceptual identifications of 'subsystems' in a multi-dimensional quantum system. As a concrete example, identifying all of the different complementarity relations associated with a four-qubit pure state should yield insight into the observed differences in the types of entanglement encoded by the four-qubit GHZ and cluster states [38, 139], and therefore into the different possible joint probability distributions that these states can yield as part of some information processing protocol.

The existence of multiple distinct complementarity relations governing the types and amounts of potentially available information encoded in multipartite pure states suggests, more generally, that a unique relation exists for every valid conceptual partitioning of a given system into subsystems and all possible sets of multipartite correlations between these subsystems. That is, the set of complementarity relations applicable to a given quantum system constrains the possible choices as to what may be consistently treated as a subsystem in the theory. The different quantities appearing in such generalized complementarity relations then play fundamental roles, to the degree that each exists, since together they exhaust the possible types of information that the quantum state may encode.

In this context, it becomes important to identify the objective properties of a physical system that determine the number and forms of such relations. One property that seems to be important in this respect is the dimension of the Hilbert space. Indeed, the Hilbert space dimension of the full system seems to determine, or at least to constrain, both the allowed conceptual divisions of the system into subsystems and the associated multipartite correlations. Further the designated subsystems, associated with specific Hilbert space dimensions themselves, yield quantities which remain invariant under different manifestations of multipartite complementarity as long as these designations remain unchanged. Thus, our results provide new evidence in support of the conjecture put forward in [142] that Hilbert space dimension is one of the objective properties of a physical system that plays a fundamental role in quantum mechanics, a connection which we feel deserves further investigation.

To summarize, our hybrid interpretation views the quantum state as a concise encapsulation of the information that we possess about a quantum system, and recognizes that the various forms that this information can take depends on our conceptual identification of component subsystems. The Hilbert space dimension of the system appears to be the key objective property that constrains the spectrum of valid conceptual decompositions. Each of these decompositions is, in turn, thought to correspond to the existence of one or more complementarity relationships between the designated subsystems and the various possible multipartite correlations that may exist between them. Unlike in classical systems, information is not restricted to be encoded solely in the individual subsystems, but can also be directly encoded in correlations; all the while being subject to tradeoffs governed by the relevant complementarity relations. From this point of view, complementarity may be identified as that part of quantum theory where objectivity (Hilbert space dimension) and subjectivity (the arbitrary conceptual distinctions which we impose on the system and the information that we possess about them) intersect.

Ultimately, we find that the conceptual framework that we impose on nature plays a fundamental role in this hybrid interpretation, where the subjective and objective intermingle to give rise to the phenomenon of complementarity. The reason for this, and perhaps some inkling of how to proceed from here, can be found in the writings of Chuangtse who recognized that "The disadvantage of regarding things in their separate parts is that when one begins to cut up and analyze, each (part) tries to be exhaustive ... Only one who can imagine the formless in the formed can arrive at the truth."

# Appendices

*Chapter 6. Summary and Conclusions*

# Appendix A

# Numerical Evolution of Entanglement in the Two-Atom TCM

This program calculates the atomic inversion, the tangles in all of the bipartite partitions, and the generalized residual tangle for the two-atom TCM as functions of time.

## Initialization

### Directives

```
In[1]:= Off[General :: spell]
        Off[General :: spell1]
```

### Additional Packages

```
In[2]:= << LinearAlgebra`MatrixManipulation`
```

## Global Variables

```
In[3]:= $HistoryLength = 0;
```

```
In[4]:= (* Coherent state amplitude *)
        α = √100 ;
```

```
In[5]:= (* Atom - Field coupling constant *)
        g = 1;
```

```
In[6]:= (* Initial Time of TCM evolution *)
        InitTime = 0;
```

```
In[7]:= (* Final Time of TCM evolution *)
        FinalTime = 100;
```

```
In[8]:= (* Time step for TCM evolution *)
        Δ = 1;
```

```
In[9]:= (* Minimum number of photons in the Poisson distribution to take
        into account. Should usually be set to about n - 3√n = α² - 3α.
        Zero causes problems so don't let S_min be less than one. *)
        S_min = Max[Floor[α² - 3α], 0];
```

```
In[10]:= (* Maximum number of photons in the Poisson distribution
         to take into account. *)
         S_max = Ceiling[α² + 3α];
```

```
In[11]:= NumFockStates = S_max - S_min + 1;
```

# Functions Needed to Calculate the Various Tangles

*In[12]:=* (* Time - dependent "angle". "n" is the number of photons and "x" is an integer offset which results from the δ functions between the Fock basis states when the field is traced out. *)

$\delta[n\_, x\_, t\_] := \delta[n, x, t] = g t \sqrt{2(2(n + x) + 3)}$

*In[13]:=* (* Time - dependent probability amplitudes *)

$p[n\_, x\_, t\_] := p[n, x, t] = \frac{1}{2(n + x) + 3}((n + x) + 2 + ((n + x) + 1)\cos[\delta[n, x, t]])$

*In[14]:=* $q = \text{Compile}\left[\{n, x, t\}, -i\sqrt{\frac{(n + x) + 1}{2(n + x) + 3}}\sin[\delta[n, x, t]]\right];$

*In[15]:=* $r[n\_, x\_, t\_] := r[n, x, t] = -\sqrt{\frac{(n + x) + 2}{(n + x) + 1}}(1 - p[n, x, t])$

*In[16]:=* (* Coherent state expansion coefficients used for the two - atom and single atom/field marginal density operators, respectively. *)

$c[n\_, x\_] := c[n, x] = e^{-\alpha^2}\frac{\alpha^n \alpha^{n+x}}{\sqrt{n!(n+x)!}}$

*In[17]:=* $c_{nm}[n\_, m\_] := c_{nm}[n, m] = e^{-\alpha^2}\frac{\alpha^n \alpha^m}{\sqrt{n!m!}}$

## Plot of Atomic Inversion

```mathematica
In[18]:= Inversion =
          Plot[ Sum[c[n, 0] (p[n, 0, t]^2 - r[n, 0, t]^2), {n, Smin, Smax}],
             {t, InitTime, FinalTime},
             PlotRange -> {{InitTime, FinalTime}, {-1, 1}}];

In[19]:= (* Two-atom marginal density operator. *)
         ρAA[t_] :=
          Sum[ {{c[n, 0] p[n, 0, t]^2,
                 c[n, 1] p[n, 1, t] Conjugate[q[n, 0, t]],
                 c[n, 2] p[n, 2, t] r[n, 0, t], 0},
                {c[n, 1] q[n, 0, t] p[n, 1, t],
                 c[n, 0] Abs[q[n, 0, t]]^2,
                 c[n, 1] q[n, 1, t] r[n, 0, t], 0},
                {c[n, 2] r[n, 0, t] p[n, 2, t],
                 c[n, 1] r[n, 0, t] Conjugate[q[n, 1, t]],
                 c[n, 0] r[n, 0, t]^2, 0}, {0, 0, 0, 0}},
               {n, Smin, Smax}]

In[20]:= (* Calculate the value of ρ_A[t] for each time step in the
          given interval. *)
         Do[RhoList[t] = ρAA[t], {t, InitTime, FinalTime, Δ}]
```

## Field - Ensemble Tangle

```
In[21]:= (* The tangle between the field and ensemble is given
         by 2 * purity of ρ_A^2. The factor of 3/2 comes from choosing
         the scale factor of each tangle to be m/2 where m is the
         smaller of the two dimensions. *)
         Do[FieldEnsTangle[t] =
           N[(2(1 - Tr[RhoList[t].RhoList[t]]))],
         {t, InitTime, FinalTime, Δ}]
```

```
In[22]:= (* Construct a list for plotting *)
         FieldEnsTangleList =
           Table[{t, FieldEnsTangle[t]},
                 {t, InitTime, FinalTime, Δ}];
```

```
In[23]:= (* Marginal density operator for a single atom. *)
         ρA1[t_] :=
           $\sum_{n=S_{min}}^{S_{max}}$ {{c[n, 0] (p[n, 0, t]^2 + 1/2 Abs[q[n, 0, t]]^2),
              $\frac{1}{\sqrt{2}}$ c[n, 1] *
                 (p[n, 1, t] Conjugate[q[n, 0, t]] + q[n, 1, t] r[n, 0, t])},
             {$\frac{1}{\sqrt{2}}$ c[n, 1] *
                 (q[n, 0, t] p[n, 1, t] + Conjugate[q[n, 1, t]] r[n, 0, t]),
              c[n, 0] (1/2 Abs[q[n, 0, t]]^2 + r[n, 0, t]^2)}}
```

```
In[24]:= (* Calculate the matrix at different times. *)
         Do[OneAtomRhoList[t] = ρA1[t], {t, InitTime, FinalTime, Δ}]
```

## One Atom - Remainder Tangle

```
In[25]:= (* Take a single atom as one subsystem and the remainder
         (the other atom and the field) as the second subsystem
         and calculate the tangle between them. *)
         Do[AtomRestTangle[t] =
           2(1 - Tr[OneAtomRhoList[t].OneAtomRhoList[t]]),
         {t, InitTime, FinalTime, Δ}]

In[26]:= (* Construct a list for plotting. *)
         AtomRestTangleList =
           Table[{t, AtomRestTangle[t]},
             {t, InitTime, FinalTime, Δ}];

In[27]:= (* Tensor product of σ_Y's on subsystems A and B in computational
         basis. *)
         SigYSigY = {{0, 0, 0, -1}, {0, 0, 1, 0},
                     {0, 1, 0, 0}, {-1, 0, 0, 0}};

In[28]:= (*Tranformation matrix to basis of ρ_A*)
         U = {{1, 0, 0, 0}, {0, 1/√2, 0, 1/√2},
              {0, 1/√2, 0, -1/√2}, {0, 0, 1, 0}};

In[29]:= (* SigYSigY in basis of ρ_A *)
         NewSigY = Conjugate[Transpose[U]].SigYSigY.U;
```

```
In[30]:= (* Calculate the product of ρ_A with it's spin-flip *)
        Do[Overlap[t] = RhoList[t].NewSigY.
                            Conjugate[RhoList[t]].NewSigY,
          {t, InitTime, FinalTime, Δ}]

In[31]:= (* Find the eigenvalues of ρ_A (ρ̃)_A *)
        Do[λ[t] =
          Reverse[Sort[Sqrt[Chop[Eigenvalues[N[Overlap[t]]]]]]],
          {t, InitTime, FinalTime, Δ}]
```

## Wootters' Tangle

```
In[32]:= Do[AtomAtomTangle[t] =
          Max[0, λ[t][[1]] - λ[t][[2]] - λ[t][[3]] - λ[t][[4]]]^2,
          {t, InitTime, FinalTime, Δ}]

In[33]:= (* Construct a list for plotting *)
        AtomAtomTangleList =
          Table[{t, AtomAtomTangle[t]},
            {t, InitTime, FinalTime, Δ}];
```

## Osborne's Tangle

```
In[34]:= (* Dimension of the single atom and field system. The basis being
        used is {|g, S_min >, ..., |g, S_max + 2 >, |e, S_min >, ..., |e, S_max + 2 >} *)
        D_A1 = 2;  (* Qubit *)
        D_F = (S_max + 2) - S_min + 1;  (* Truncated Field *)
        Dim = D_A1 * D_F;
```

# Appendix A. Numerical Evolution of Entanglement in the Two-Atom TCM

```
In[35]:= (* This next block calculates the single atom/field marginal
          density operator by populating an initially empty matrix with
          the correct matrix elements. *)
```

```
In[36]:= Entry1 = Compile[{{element, _Complex}, x, y, t},
                    element + c_{nm}[x,y]p[x,0,t]p[y,0,t]];
         Entry2 = Compile[{{element, _Complex}, x, y, t,
                    {qm, _Complex}},
                    element + \frac{1}{\sqrt{2}}c_{nm}[x,y]p[x,0,t]Conjugate[qm]];
         Entry3 = Compile[{{element, _Complex}, x, y, t,
                    {qn, _Complex}},
                    element + \frac{1}{\sqrt{2}}c_{nm}[x,y] * qn * p[y,0,t]];
         Entry4 = Compile[{{element, _Complex}, x, y, t,
                    {qn, _Complex}, {qm, _Complex}},
                    element + \frac{1}{2}c_{nm}[x,y]qn * Conjugate[qm]];
         Entry5 = Compile[{{element, _Complex}, x, y, t,
                    {qn, _Complex}},
                    element + \frac{1}{\sqrt{2}}c_{nm}[x,y] * qn * r[y,0,t]];
         Entry6 = Compile[{{element, _Complex}, x, y, t,
                    {qm, _Complex}},
                    element + \frac{1}{\sqrt{2}}c_{nm}[x,y]r[x,0,t]Conjugate[qm]];
         Entry7 = Compile[{{element, _Complex}, x, y, t},
                    element + c_{nm}[x,y]r[x,0,t]r[y,0,t]];
```

```
In[37]:=  (* Calculate ρ_AF for each time *)
          For[t = InitTime, t ≤ FinalTime, t+ = Δ,
            (*Create an initially empty matrix*)
            ρAF = Table[0, {Dim}, {Dim}];
            (* Loop through each possible value for n and m, and insert the
               appropriate matrix element. *)
            For[n = 1, n ≤ NumFockStates, n++,
              For [m = 1, m ≤ NumFockStates, m++,

                nOff = n + S_min - 1;
                mOff = m + S_min - 1;

                ρAF = ReplacePart[ρAF,
                    Entry1[N[ρAF[[n + D_F, m + D_F]]],nOff,mOff,t],
                    {n + D_F, m + D_F}];
                ρAF = ReplacePart[ρAF,
                    Entry2[N[ρAF[[n + D_F, m + 1]]],nOff,mOff,
                      t, q[mOff, 0, t]], {n + D_F, m + 1}];
                ρAF = ReplacePart[ρAF,
                    Entry3[N[ρAF[[n + 1, m + D_F]]],nOff,mOff,
                      t, q[nOff, 0, t]], {n + 1, m + D_F}];
                ρAF = ReplacePart[ρAF,
                    Entry4[N[ρAF[[(n + 1) + D_F, (m + 1) + D_F]]],
                      nOff,mOff,t, q[nOff, 0, t], q[mOff, 0, t]],
                    {(n + 1) + D_F, (m + 1) + D_F}];
```

```
In[38]:=     ρAF = ReplacePart[ρAF,
                Entry4[N[ρAF[[n + 1, m + 1]]],nOff,mOff,
                  t, q[nOff,0,t], q[mOff,0,t]],{n + 1,m + 1}];
             ρAF = ReplacePart[ρAF,
                Entry5[N[ρAF[[(n + 1) + D_F, m + 2]]],nOff,
                  mOff,t, q[nOff,0,t]], {(n + 1) + D_F,m + 2}];
             ρAF = ReplacePart[ρAF,
                Entry6[N[ρAF[[n + 2, (m + 1) + D_F]]],nOff,
                  mOff,t, q[mOff,0,t]], {n + 2, (m + 1) + D_F}];
             ρAF = ReplacePart[ρAF,
                Entry7[N[ρAF[[n + 2, m + 2]]],nOff,mOff,t],
                {n + 2,m + 2}]];

             (* Find the eigenvectors of ρ_AF *)
             EvectList[t] = Eigenvectors[Chop[N[ρAF]]]

In[39]:=  (* Form the four "γ" matrices corresponding to the outer products
             of the different combinations of the two eigenvectors with
             nonzero eigenvalues. *)
             Do[γ[t, i, j] = Outer[Times, Chop[EvectList[t][[i]]],
                Conjugate[Chop[EvectList[t][[j]]]]],
             {t, InitTime, FinalTime, Δ}, {i, 1, 2}, {j, 1, 2}]
```

*In[40]:=* (* Form the four $\gamma_F$ matrices corresponding to tracing over the remaining atom in each $\gamma$. The trace over the atom picks out all terms of the form |g,n><g,m| and |e,n><e,m|, i.e., the [n,m] entry in the $\gamma_F$ matrix is given by the sum of the [n,m] and [2n, 2m] terms in the original $\gamma$ matrix. *)

```
For[t = InitTime, t ≤ FinalTime, t+ = Δ,
  For[i = 1, i ≤ 2, i++,
    For[j = 1, j ≤ 2, j++,
      γF[t,i,j] = Table[0, {D_F}, {D_F}];
      For[n = 1, n ≤ D_F, n++,
        For[m = 1, m ≤ D_F, m++,
          γF[t,i,j] = ReplacePart[γF[t,i,j],
            γF[t,i,j][[n,m]] + γ[t,i,j][[n,m]] +
              γ[t,i,j][[n + D_F, m + D_F]], {n,m}]]]]]]
```

## Appendix A. Numerical Evolution of Entanglement in the Two-Atom TCM

```
In[41]:=  (* Form the four γ_A1 matrices corresponding to tracing over the field
          in each γ. The trace over the field picks out all of the terms of the
          form |g(e),n><g(e),n|. *)
          For[t = InitTime, t ≤ FinalTime, t+ = Δ,
            For[i = 1, i ≤ 2, i++,
              For[j = 1, j ≤ 2, j++,
                γA1[t,i,j] = Table[0, {D_{A1}}, {D_{A1}}];
                For[n = 1, n ≤ D_F, n++,
```
$$(* \text{ The } |g><g| \text{ entry is given by } \sum_{n=S_{min}}^{S_{max+2}} \gamma[n,n]. *)$$
```
                  γA1[t,i,j] = ReplacePart[γA1[t,i,j],
                      γA1[t,i,j][[1,1]] + γ[t,i,j][[n,n]], {1,1}];
```
$$(* \text{ The } |e><e| \text{ entry is given by } \sum_{n=S_{min}}^{S_{max+2}} \gamma[n+D_F, n+D_F]. *)$$
```
                  γA1[t,i,j] = ReplacePart[γA1[t,i,j],
                      γA1[t,i,j][[2,2]] + γ[t,i,j][[n+D_F, n+D_F]],
                      {2,2}];
```
$$(* \text{ The } |g><e| \text{ entry is given by } \sum_{n=S_{min}}^{S_{max+2}} \gamma[n, n+D_F]. *)$$
```
                  γA1[t,i,j] = ReplacePart[γA1[t,i,j],
                      γA1[t,i,j][[1,2]] + γ[t,i,j][[n, n+D_F]],
                      {1,2}];
```
$$(* \text{ The } |e><g| \text{ entry is given by } \sum_{n=S_{min}}^{S_{max+2}} \gamma[n+D_F, n]. *)$$
```
                  γA1[t,i,j] = ReplacePart[γA1[t,i,j],
                      γA1[t,i,j][[2,1]] + γ[t,i,j][[n+D_F, n]],
                      {2,1}]]]]]
```

```
In[42]:=  (* Create identity matrices needed to find γ̃ *)

          I_A1 = IdentityMatrix[D_A1];

          I_F = IdentityMatrix[D_F];

          I_A1,F = IdentityMatrix[Dim];

In[43]:=  (* This next block calculates the field marginal density operator

          by populating an initially empty matrix with the correct matrix

          elements. *)

In[44]:=  (* Form ρF corresponding to tracing over the remaining atom. The

          trace over the atom picks out all terms of the form |g,n><g,m|

          and |e,n><e,m|, i.e., the [n,m] entry in the ρF matrix is given

          by the sum of the [n,m] and [n + D_F, m + D_F] terms in the original

          ρAF matrix. For testing only. *)
          (* For[t = InitTime, t ≤ FinalTime, t+ = Δ,

            (*Create an initially empty matrix*)
            ρF[t] = Table[0, {D_F}, {D_F}];

            (* Loop through each possible value for n and m,
            and insert the appropriate matrix element. *)
            For[n = 1, n <= D_F, n++,
              For[m = 1, m <= D_F, m++,
                ρF[t] = ReplacePart[ρF[t],
                  ρF[t][[n,m]] + N[ρAF[t][[n,m]]] +
                    N[ρAF[t][[n + D_F, m + D_F]]], {n, m}]]]] *)
```

```
In[45]:= (* Calculate the γ̃ matrices for each of the four γ matrices in

          Osborne's paper. *)

        For[t = InitTime, t ≤ FinalTime, t+ = Δ,

          For[k = 1, k ≤ 2, k++,

            For[l = 1, l ≤ 2, l++,

              (* Take the adjoints of γ, γ_A1, and γ_F *)

              β[t,k,l] = Transpose[Conjugate[γ[t,k,l]]];

              βA1[t,k,l] = Transpose[Conjugate[γA1[t,k,l]]];

              βF[t,k,l] = Transpose[Conjugate[γF[t,k,l]]];

              (* Compute the necessary tensor products *)

              TempA1 = Outer[Times, βA1[t,k,l], I_F];

              TempF = Outer[Times, I_A1, βF[t,k,l]];

              For[row = 1, row ≤ D_A1, row++,

                For[col = 1, col ≤ D_A1, col++,

                  MatListA1[row, col] = TempA1[[row]][[col]];

                  MatListF[row, col] = TempF[[row]][[col]]]];

              (γ̂)_{A1,I_F} [t,k,l] =
                BlockMatrix[{{MatListA1[1,1], MatListA1[1,2]},

                    {MatListA1[2,1], MatListA1[2,2]}}];

              (γ̂)_{I_A1,F} [t,k,l] =
                BlockMatrix[{{MatListF[1,1], MatListF[1,2]},

                    {MatListF[2,1], MatListF[2,2]}}];

              γflip[t,k,l] = Tr[β[t,k,l]]I_{A1,F} -
                    (γ̂)_{A1,I_F} [t,k,l] - (γ̂)_{I_A1,F} [t,k,l] + β[t,k,l]]]]
```

```mathematica
In[46]:= (* Calculate the tensor (T) obtained by taking the trace of
       each γ with it's "spin - flip". *)
      Do[T[t, i, j, k, l] = Tr[γ[t, i, j].γflip[t, k, l]],
       {t, InitTime, FinalTime, Δ}, {i, 1, 2}, {j, 1, 2}, {k, 1, 2},
       {l, 1, 2}]

In[47]:= (* Form the real symmetric matrix elements (M) from the elements
       of the tensor (T). *)

In[48]:= Do[M[t, 1, 1] =
        Chop[1/4 T[t, 1, 2, 2, 1] + 1/2 T[t, 1, 1, 2, 2] +
          1/4 T[t, 2, 1, 1, 2]],
        {t, InitTime, FinalTime, Δ}]

In[49]:= Do[M[t, 1, 2] = M[t, 2, 1] =
        Chop[i/4 (T[t, 1, 2, 2, 1] - T[t, 2, 1, 1, 2])],
        {t, InitTime, FinalTime, Δ}]

In[50]:= Do[M[t, 1, 3] = M[t, 3, 1] =
        Chop[1/4 (T[t, 1, 1, 2, 1] - T[t, 2, 1, 2, 2] +
          T[t, 1, 1, 1, 2] - T[t, 1, 2, 2, 2])],
        {t, InitTime, FinalTime, Δ}]

In[51]:= Do[M[t, 2, 2] =
        Chop[-1/4 T[t, 1, 2, 2, 1] + 1/2 T[t, 1, 1, 2, 2] -
          1/4 T[t, 2, 1, 1, 2]], {t, InitTime, FinalTime, Δ}]
```

```mathematica
In[52]:= Do[M[t, 2, 3] = M[t, 3, 2] =
          Chop[i/4 (T[t,1,1,2,1] - T[t,1,1,1,2] +
              T[t,2,1,2,2] - T[t,1,2,2,2])],
        {t, InitTime, FinalTime, Δ}]

In[53]:= Do[M[t, 3, 3] =
          Chop[1/4 T[t,1,1,1,1] - 1/2 T[t,1,1,2,2] +
              1/4 T[t,2,2,2,2]],
        {t, InitTime, FinalTime, Δ}]

In[54]:= Do[CurveMat[t] = {{M[t,1,1], M[t,1,2], M[t,1,3]},
                           {M[t,2,1], M[t,2,2], M[t,2,3]},
                           {M[t,3,1], M[t,3,2], M[t,3,3]}},
         {t, InitTime, FinalTime, Δ}]

In[55]:= (* Find the minimum eigenvalue of the matrix consisting of the
          elements (M) at each time. *)
         Do[λ_min[t] = Min[Eigenvalues[CurveMat[t]]],
         {t, InitTime, FinalTime, Δ}]

In[56]:= (* The tangle between and single atom and the field may be shown to
          be equal to 1/3 τ_{F(A1,A2)} + λ_min * τ_{A1(A2,F)} using the scale factor m/2 = 3/2
          for the field with the ensemble. *)
         Do[τ_AF[t] = 1/2 FieldEnsTangle[t] + λ_min[t] AtomRestTangle[t],
         {t, InitTime, FinalTime, Δ}];
```

*In[57]:=* **(\* Construct a list for plotting. \*)**

**AtomFieldTangleList =**

**Table[{t, $\tau_{AF}$[t]}, {t, InitTime, FinalTime, Δ}];**

*In[58]:=* **(\* Calculate the three tangle using one of the two nontrivial**

**possible symbol permutations. \*)**

**Do[ThreeTangle1[t] =**

$\frac{1}{3}$ **(2 AtomRestTangle[t] + $\frac{3}{2}$FieldEnsTangle[t] −**

**2 AtomAtomTangle[t] − 4$\tau_{AF}$[t]),**

**{t, InitTime, FinalTime, Δ}];**

*In[59]:=* **(\* Construct a list for plotting. \*)**

**ThreeTangle1List =**

**Table[{t, ThreeTangle1[t]},**

**{t, InitTime, FinalTime, Δ}];**

*Appendix A. Numerical Evolution of Entanglement in the Two-Atom TCM*

# Appendix B

# LHV and Quantum Measurement Algorithm

This program simulates the operation of the Hadamard and CNOT gates, and computes both the quantum mechanical and LHV predictions for the measurements of all possible products of Pauli operators on the generated N-qubit state.

## Initialization

### Directives

```
In[60]:= Off[General :: spell]
         Off[General :: spell1]
```

### Additional Packages

```
In[61]:= << LinearAlgebra`MatrixManipulation`
```

# Global Variables

```
In[62]:= (* Number of qubits *)
        n = rows = 5;

In[63]:= (* Number of columns I, X, Y, Z *)
        cols = 4;

In[64]:= (* Column labels *)
        Id = 1;
        X = 2;
        Y = 3;
        Z = 4;

In[65]:= (* Table of indices for all possible measurements *)
        Index = Flatten[Table[{i, m_i}, {i, n}]];

In[66]:= Indices = Flatten[Table[Index, {m_1, cols}, {m_2, cols}, {m_3, cols},
                                {m_4, cols}, {m_5, cols}], n - 1];

In[67]:= (* Total number of possible measurement results *)
        NumResults = 4^n;
```

```mathematica
In[68]:= (* Table of local gate induced phases *)
        Phases = Table[1, {n}, {cols}];

        (* Since Y = ⅈXZ initialize all Y phases to ⅈ *)
        For[i = 1, i ≤ n, i++,
          Phases[[i, Y]] = (-1)^i * ⅈ];

In[69]:= (* Table of "Local Hidden Variables" *)
        counter = 0;

In[70]:= LHV = Table[0, {rows}, {cols}];

In[71]:= For[i = 1, i ≤ rows, i++,

          counter = counter + 1;
          For[j = 1, j ≤ cols, j++,

            (* If the current column is the I column or the Z column *)
            If[j == 1 || j == Z,

              (* then insert the empty list *)
              LHV[[i, j]] = List[],

              (* else if j == X insert an $R_i$ representing either a plus one
              or a minus one with equal probabilities *)
              If[j == X || j == Y,
                LHV[[i, j]] = List[R_counter]]]]];

In[72]:= (* Table of stabilizer generators in symplectic notation *)
        SymS = Table[0, {n}, {2n}];
```

```
In[73]:= (* Table of stabilizer generators composed of products of Pauli
         operators *)
         PauliS = Table[Id, {n}, {n}];
```

```
In[74]:= (* Initialize SymS and PauliS =< ZII..., IZI..., ...IIZ >, for
         n qubits initially in the state |0>. *)
         For[i = 1, i ≤ n, i++,
           SymS[[i]] = ReplacePart[SymS[[i]], 1, n+i];
           PauliS[[i]] = ReplacePart[PauliS[[i]], Z, i]];
```

## Useful Functions

```
In[75]:= View[A_] := A//MatrixForm
```

```
In[76]:= (* This function removes all pairs of the form R_i, R_j for i = j.
         This corresponds to the multiplication of two identical R's
         which always yields a one. *)
         RemovePairs[L_] :=
           (Result = {};
            For[i = 1, i ≤ Length[L], i++,
              L1 = L[[i]];
              Result = Complement[Union[Result, L1],
                                  Intersection[Result, L1]]];
            Result)
```

```
In[77]:= (* This function returns the input measurement result list with
         all of the R outcomes deleted *)
         RemoveR[L_] := Delete[L, Position[L, R]];
```

# Gates

## Hadamard

```mathematica
In[78]:= H[i_] :=

            (* Swap X and Z entries in row i of LHV *)
            ({LHV[[i,X]], LHV[[i,Z]]} = {LHV[[i,Z]], LHV[[i,X]]};

            (* Swap the local X and Z phases and flip the local Y phase for
            the ith bit *)
            {Phases[[i,X]], Phases[[i,Z]]} =
                    {Phases[[i,Z]], Phases[[i,X]]};
            Phases[[i,Y]] = -Phases[[i,Y]];

            (* Update the symplectic and Pauli stabilizer generators by
            changing all X's in the ith position to Z's and vice versa. *)
            For[g = 1, g ≤ n, g++,
              {SymS[[g,i]], SymS[[g,i+n]]} =
                    {SymS[[g,i+n]], SymS[[g,i]]};

              PauliSign = Sign[PauliS[[g,i]]];

              If[Abs[PauliS[[g,i]]] == X,
                PauliS[[g,i]] = PauliSign * Z,

                If[Abs[PauliS[[g,i]]] == Y,
                  PauliS[[g]] = -PauliS[[g]],

                  If[Abs[PauliS[[g,i]]] == Z,
                    PauliS[[g,i]] = PauliSign * X]]]; )
```

## Phase Gate

```
In[79]:= P[i_] :=

            (* Swap X and Y entries in row i of LHV *)
            ({LHV[[i, X]], LHV[[i, Y]]} = {LHV[[i, Y]], LHV[[i, X]]};

            (* Read table as : X := -Y and Y := X *)
            {Phases[[i, X]], Phases[[i, Y]]} =
                {-Phases[[i, Y]], Phases[[i, X]]};

            (* Update the stabilizer generators by changing all X's in the
            ith position to Y's and vice versa *)
            For[g = 1, g ≤ n, g++,
              If[SymS[[g, i]] == 1,
                SymS[[g, i + n]] = Mod[SymS[[g, i + n]] + 1, 2]];

              PauliSign = Sign[PauliS[[g, i]]];

              If[Abs[PauliS[[g, i]]] == X,
                PauliS[[g, i]] = PauliSign * Y,

                If[Abs[PauliS[[g, i]]] == Y,
                  PauliS[[g, i]] = PauliSign * X;
                  PauliS[[g]] = -PauliS[[g]]]]]; )
```

## Pauli X

```
In[80]:= GateX[i_] :=

            (* Flip the local Y and Z phases for the ith bit *)
            (Phases[[i, Y]] = -Phases[[i, Y]];
            Phases[[i, Z]] = -Phases[[i, Z]];

            For[g = 1, g ≤ n, g++,
              If[Abs[PauliS[[g, i]]] == Y || Abs[PauliS[[g, i]]] == Z,
                PauliS[[g]] = -PauliS[[g]]]]; )
```

## Pauli Y

```
In[81]:= GateY[i_] :=

            (* Flip the local X and Z phases for the ith bit *)
            (Phases[[i, X]] = -Phases[[i, X]];
             Phases[[i, Z]] = -Phases[[i, Z]];

             For[g = 1, g ≤ n, g++,
               If[Abs[PauliS[[g, i]]] == X || Abs[PauliS[[g, i]]] == Z,
                 PauliS[[g]] = -PauliS[[g]]]]; )
```

## Pauli Z

```
In[82]:= GateZ[i_] :=

        (* Flip the local X and Y phases for the ith bit *)
        (Phases[[i, X]] = -Phases[[i, X]];
        Phases[[i, Y]] = -Phases[[i, Y]];

        For[g = 1, g ≤ n, g++,
          If[Abs[PauliS[[g, i]]] == X || Abs[PauliS[[g, i]]] == Y,
            PauliS[[g]] = -PauliS[[g]]]];)
```

## CNOT

```
In[83]:= (* Table that contains the updates under CNOT that the various
         products of Pauli operators undergo *)
        CNOTUpdates = {{{Id, Id}, {Id, X}, {Z, Y}, {Z, Z}},
                       {{X, X}, {X, Id}, {Y, Z}, {Y, Y}},
                       {{Y, X}, {Y, Id}, {X, Z}, {X, Y}},
                       {{Z, Id}, {Z, X}, {Id, Y}, {Id, Z}}};
```

```
In[84]:=  CNOT[i_, j_] :=
           (* Update LHV table by multiplying the appropriate entries
           together and removing any pairs of R_i's with the same index i,
           corresponding to a multiplicative factor of one. *)
          (LHV[[i, X]] = RemovePairs[{LHV[[i, X]], LHV[[j, X]]}];
           LHV[[j, Z]] = RemovePairs[{LHV[[i, Z]], LHV[[j, Z]]}];
           LHV[[i, Y]] = RemovePairs[{LHV[[i, X]], LHV[[i, Z]]}];
           LHV[[j, Y]] = RemovePairs[{LHV[[j, X]], LHV[[j, Z]]}];

           (* Update the local Y phases *)
           (*If[Phases[[i, Y]] == -i,
              Phases[[j, Y]] = -Phases[[j, Y]]]; *)

           (* Update the symplectic stabilizer generator by taking
           the XOR (addition mod 2) of the appropriate entries. *)
           For[g = 1, g ≤ n, g++,
            SymS[[g, j]] = BitXor[SymS[[g, j]], SymS[[g, i]]];
            SymS[[g, i + n]] = BitXor[SymS[[g, i + n]], SymS[[g, j + n]]];
            PauliSign = Sign[PauliS[[g, i]]];

            (*Update the Pauli stabilizer generator by looking up the
            appropriate product of Pauli's in the CNOTUpdates table *)
            {PauliS[[g, i]], PauliS[[g, j]]} =
              CNOTUpdates[[Abs[PauliS[[g, i]]], Abs[PauliS[[g, j]]]]];

            If[PauliSign == -1,
             PauliS[[g, i]] = -PauliS[[g, i]];
             PauliS[[g, j]] = -PauliS[[g, j]]];

            If[(Abs[PauliS[[g, i]]] == X && Abs[PauliS[[g, j]]] == Z) ||
               (Abs[PauliS[[g, i]]] == Y && Abs[PauliS[[g, j]]] == Y),
                PauliS[[g]] = -PauliS[[g]]]];)
```

# GK Measurements

## Classical Predictions

```
In[85]:= (* This function constructs a list containing the local values
         contained in LHV for the given measurement specified by Index *)
         BuildLHVList[Index_] := (LHVList = {};

         For[i = 1, i ≤ 2n, i+ = 2,
           (* The ith entry in Index gives the qubit number and the (i + 1)th
           entry gives the Pauli operator to be measured on the ith qubit *)
           LHVList = Append[LHVList, LHV[[Index[[i]], Index[[i + 1]]]]];)

In[86]:= (* This function determines whether the LHV model predicts a plus
         one or a minus one *)
         FlipSign[Index_] :=
           (TotPhase = 1;

           (* Calculate the product of all relevant local phases *)
           For[i = 1, i ≤ n, i + +,
             TotPhase = Phases[[i, Index[[2i]]]] * TotPhase];

           (* If the total phase is equal to -1 or to (-𝚤) *)
           If[TotPhase == -1||TotPhase == -𝚤,
             (* then the LHV model predicts -1 *)
             Return[True],
             (* else the LHV model predicts 1 *)
             Return[False]];)
```

```
In[87]:= (* This function constructs a list containing the LHV predictions
         for every possible GK measurement that may be performed on the
         current state of the system *)
         LHVMeasure[Index_] :=
           (* Build a list of all of the local values contained in LHV for
           the given measurement.  The classical prediction is then given
           by the product of all of these values *)
           (BuildLHVList[Index];

              (* If all random variables square to one *)
             If[RemovePairs[LHVList] == {},

                (*then if Flipsign is True *)
                If[FlipSign[Index],
                   (* then the LHV predicts - 1 *)
                   ClassPred = Append[ClassPred, -1],

                   (* otherwise the LHV predicts + 1 *)
                   ClassPred = Append[ClassPred, 1]],

                (* else the LHV predicts a random plus or minus one with
                50 - 50 probability *)
                ClassPred = Append[ClassPred, R]]; )
```

## Quantum Predictions

```
In[88]:= (* This function constructs the symplectic representation for
         the n qubit measurement given by Index *)
         BuildSymplecticList[Index_] :=
           (SymplecticList = Table[0, {2n}];

           (* Symplectic notation uses a 2n bit string divided in half
           with a zero in the ith and (i + n)th position representing an I,
           a one in the ith and a zero in the (i + n) position representing
           an X, a one in both positions representing a Y, and a zero in the
           ith and a one in the (i + n) position representing a Z *)
             For[i = 1, i ≤ n, i++,
               If[Index[[2i]] == X,
                 SymplecticList = ReplacePart[SymplecticList, 1, i],

               If[Index[[2i]] == Y,
                 SymplecticList =
                     ReplacePart[SymplecticList, 1, i];
                 SymplecticList =
                     ReplacePart[SymplecticList, 1, i + n],

               If[Index[[2i]] == Z,
                 SymplecticList =
                     ReplacePart[SymplecticList, 1, i + n]]]]; )
```

**Create the symplectic inner product matrix**

```
In[89]:= SigX = {{0, 1}, {1, 0}};
         I_n = IdentityMatrix[n];
         I_sym = Outer[Times, SigX, I_n];

         For[i = 1, i ≤ 2, i++,
           For[j = 1, j ≤ 2, j++,
             TempMat[i, j] = I_sym[[i]][[j]]]];

         I_sym = BlockMatrix[{{TempMat[1, 1], TempMat[1, 2]},
                             {TempMat[2, 1], TempMat[2, 2]}}];

In[90]:= (* This function returns True if the given measurement
         anticommutes with at least one stabilizer generator and
         False otherwise *)
         NotinStabilizer[SymplecticList_] :=
           (For[i = 1, i ≤ n, i++,
              (* If the symplectic inner product between the
              measurement and the current stabilizer generator is not
              equal to zero then these two operators anticommute *)
              If[Mod[SymS[[i]].I_sym.SymplecticList, 2] ≠ 0,
                Return[True]]];
            Return[False]; )
```

```
In[91]:= (* This function constructs the Pauli representation of
          the measurement corresponding to Index *)
         BuildPauliM[Index_] :=
           (PauliM = {};
             (* For each qubit pull out the entry representing the
             measurement to be performed on that qubit *)
             For[i = 1, i ≤ n, i++,
               PauliM = Append[PauliM, Index[[2i]]]];
             Return[PauliM];)
```

```
In[92]:= (* This function determines whether quantum mechanics predicts
          a plus one or a minus one for the measurement specified by Index *)
         DetermineSign[Index_] :=
           (* Construct the Pauli representation of the current
           measurement *)
           (PauliM = BuildPauliM[Index];

             (* If the current measurement is a member of the stabilizer *)
             If[MemberQ[S, PauliM],
               (* then quantum mechanics predicts plus one *)
               Return[1],
               (* otherwise, quantum mechanics predicts minus one *)
               Return[-1]];)
```

```
In[93]:= (* This function constructs a list containing the quantum
         predictions for every possible GK measurement that may be
         performed on the current state of the system *)
         QMeasure[Index_] :=

           (* Construct the symplectic representation of the current
           joint GK measurement *)
           (BuildSymplecticList[Index];

             (* If the current measurement does not commute with all of
             the stabilizer generators *)
             If[NotinStabilizer[SymplecticList],

               (* then QM predicts a random outcome *)
               QMPred = Append[QMPred, R],

               (* else QM predicts either a plus or minus one *)
               QMPred = Append[QMPred, DetermineSign[Index]]]; )
```

```
In[94]:= (* Table containing the outcomes for all possible products of
         two Pauli operators *)
         PauliProducts = {{Id, X, Y, Z}, {X, Id, i Z, -i Y},
                         {Y, -i Z, Id, i X}, {Z, i Y, -i X, Id}};
```

```
In[95]:=  (* This function returns the operator product of two n qubit
          measurements, M1 and M2 *)
          MeasurementProduct[M1_, M2_] :=
            (Result = {};

              (* For each qubit, calculate the product of the two
              measurements on the ith qubit by looking up the result in
              the PauliProducts table *)
              For[i = 1, i ≤ n, i++,
                Result = Append[Result, PauliProducts[[Abs[M1[[i]]],
                                                      Abs[M2[[i]]]]]]];

              (* If the sign of the stabilizer generator times the sign of
              the new stabilizer element is -1 *)
              If[Sign[M1[[1]]] * Sign[M2[[1]]]
                *Sign[Apply[Times, Result]] == -1,
                  (* then -g is a member of the stabilizer *)
                  Return[-Abs[Result]],
                  (* else g is a member of the stabilizer *)
                  Return[Abs[Result]]]; )
```

```mathematica
In[96]:= (* This function computes the list consisting of all products
           of the n qubit measurement M with all elements of S containing
           no duplicates *)
         ProductWithS[M_] :=
            (Temp = {};
            
             (* For each entry in S, add the appropriate product to the
                temporary list *)
             For[x = 1, x ≤ Length[S], x++,
               Temp = Append[Temp, MeasurementProduct[M, S[[x]]]]];
             
             (* Update S and remove duplicates *)
             S = Union[S, Temp];)
             
             
           (* This function generates the full stabilizer for the current
              state *)
           GenerateStabilizer :=
             (* Append the identity operation to the list of Pauli
                generators *)
             (S = Append[PauliS, Table[Id, {n}]];
             
              (* For each of the Pauli generators (g), calculate the
                 product of g with each entry of S *)
              For[g = 1, g ≤ n, g++,
                ProductWithS[PauliS[[g]]];)
```

```
In[97]:= (* This function performs a measurement by calculating both the
         classical and quantum predictions for all possible 4^n products
         of Pauli operators for the current state of the system *)
         Measurement :=
           (* Empty the lists which will contain the classical and quantum
           predictions *)
           (ClassPred = {};
            QMPred = {};

            (* Generate the full n element stabilizer from the Log_2 n
            stabilizer generators *)
            GenerateStabilizer;

            (* For each possible GK measurement, calculate the classical
            and quantum predictions *)
            For[x = 1, x ≤ NumResults, x + +,
              LHVMeasure[Indices[[x]]];
              QMeasure[Indices[[x]]];])
```

## Sample GK Circuit (1-D five qubit cluster state)

```
In[98]:= H[1]

In[99]:= CNOT[1, 2]

In[100]:= H[2]

In[101]:= CNOT[2, 3]

In[102]:= H[3]

In[103]:= CNOT[3, 4]
```

```
In[104]:= H[4]

In[105]:= CNOT[4,5]

In[106]:= H[5]

In[107]:= Measurement;

In[108]:= ClassPred == QMPred
Out[108]= True

In[109]:= View[Phases]
```

$$Out[109]= \begin{pmatrix} 1 & 1 & i & 1 \\ 1 & 1 & -i & 1 \\ 1 & 1 & i & 1 \\ 1 & 1 & -i & 1 \\ 1 & 1 & i & 1 \end{pmatrix}$$

```
In[110]:= View[LHV]
```

$$Out[110]= \begin{pmatrix} \{\} & \{R_2\} & \{R_1, R_2\} & \{R_1\} \\ \{\} & \{R_1, R_3\} & \{R_1, R_2, R_3\} & \{R_2\} \\ \{\} & \{R_2, R_4\} & \{R_2, R_3, R_4\} & \{R_3\} \\ \{\} & \{R_3, R_5\} & \{R_3, R_4, R_5\} & \{R_4\} \\ \{\} & \{R_4\} & \{R_4, R_5\} & \{R_5\} \end{pmatrix}$$

*Appendix B. LHV and Quantum Measurement Algorithm*